\RequirePackage{lineno}
\documentclass{elsart}

\usepackage{epsfig}
\usepackage{graphicx}
\usepackage{xspace}

\def\D0{D0}

\def\vetmis {\mbox{${\hbox{${\bf E}$\kern-0.6em\lower-.1ex\hbox{/}}}_T$ }}
\def\etmisv {\mbox{${\hbox{${\vec E}$\kern-0.6em\lower-.1ex\hbox{/}}}_T$ }}
\def\etmiss {\mbox{${\hbox{$E$\kern-0.6em\lower-.1ex\hbox{/}}}_T$ }}
\def\etmis  {\mbox{${\hbox{$E$\kern-0.6em\lower-.1ex\hbox{/}}}_T$ }}
\def\ptmiss {\mbox{${\hbox{$p$\kern-0.6em\lower-.1ex\hbox{/}}}_T$ }}
\def\ptmis  {\mbox{${\hbox{$p$\kern-0.6em\lower-.1ex\hbox{/}}}_T$ }}

\def\Zeemis 
    {\mbox{${\hbox{$Z\rightarrow ee$\kern-0.4em\lower-.1ex\hbox{/}}}$ }}
\def\ifmath#1{\relax\ifmmode #1\else $#1$\fi}%
\def\GeV{\ifmmode {\mathrm{ Ge\kern -0.1em V}}\else
                   \textrm{Ge\kern -0.1em V}\fi}%
\def\MeV{\ifmmode {\mathrm{ Me\kern -0.1em V}}\else
                   \textrm{Me\kern -0.1em V}\fi}%
\def\keV{\ifmmode {\mathrm{ ke\kern -0.1em V}}\else
                   \textrm{ke\kern -0.1em V}\fi}%
\def\eV{\ifmmode  {\mathrm{ e\kern -0.1em V}}\else
                   \textrm{e\kern -0.1em V}\fi}%
\def\GeVc{\ifmmode {\mathrm{ \GeV/c}}\else
                    \textrm{Ge\kern -0.1em V/c}\fi}%
\def\GeVcc{\ifmmode {\mathrm{ \GeV/c^2}}\else
                   \textrm{Ge\kern -0.1em V/c$^2$}\fi}%

\newcommand{\degC}    {\ensuremath{^\circ\rm C}\xspace}

\newcommand{\nope}[1]{}
\begin{document}
\setpagewiselinenumbers
\modulolinenumbers[2]

\thispagestyle{empty}
\begin{center}

\vspace{1.5in}
\leftline{\quad}
\rightline{\today } 
\leftline{\quad}
\leftline{\quad}
\leftline{\quad}
\leftline{\quad}
\Large{\bf{ The \D0 Silicon Microstrip Tracker } }
\vskip 2.0cm 
\normalsize
{\begin{flushleft}
\author{ \small
	S.N.~Ahmed,$^{i}$
	R.~Angstadt,$^{v}$
	M.~Aoki, $^{v}$
	B.~{\AA}sman,$^{k}$
	S.~Austin,$^{v}$
	L.~Bagby,$^{v}$	
}
\author{ \small
	E.~Barberis,$^{ad}$		
	P.~Baringer,$^{aa}$
	A.~Bean,$^{aa}$
	A.~Bischoff,$^{s}$
% ...
}

% ...
\address{$^{i}$Radboud University Nijmegen/NIKHEF, Nijmegen, The 
		Netherlands} 
% ...
\address{$^{k}$Stockholm University, Stockholm, Sweden} 
% ...
\address{$^{s}$University of California Riverside, Riverside, California 92521, USA}
% ...
\address{$^{v}$Fermi National Accelerator Laboratory, Batavia, 
		Illinois 60510, USA} 
% ...
\address{$^{aa}$University of Kansas, Lawrence, Kansas 66045, USA} 
% ...
\address{$^{ad}$Northeastern University, Boston, Massachusetts 02115, USA} 
% ...

\end{flushleft}

     \quad            \\
     \quad            }

\vskip 1.1cm 
    \rule{6.0cm}{1.0pt}
    \vskip 5.0pt
    \rule{3.0cm}{0.5pt} 
\rm
\small
\vspace{1.5cm}

\parbox{5.5in}{
\hspace{5mm}
{\large Abstract \\}
 
This paper describes the mechanical design, the readout chain, the production, 
testing and the installation of the Silicon Microstrip Tracker of the D0 
experiment at the Fermilab Tevatron collider. In addition, description of 
the performance of the detector during the experiment data collection 
between 2001 and 2010 is provided.

}
\vspace{5mm}

\end{center}
\vfill 
\pagebreak
\newpage

\pagenumbering{arabic} 
\tableofcontents
\newpage
\pagebreak

\pagenumbering{arabic} 

\section{Introduction}

The \D0 detector~\cite{d0nim-old} is one of the two detectors at the
Tevatron accelerator at Fermilab. The detector performed extraordinarily 
well in Run~I (1992-1996), as demonstrated by the discovery of the top quark 
\cite{top} and many other published physics results~\cite{manyph}. 

During Run~I, the Tevatron operated using six bunches each of 
protons and anti-protons with 3500\,ns between bunch crossings.
In Run~II,started in 2001, it is operated with 36 bunches of 
protons and anti-protons with a bunch spacing of 396\,ns. 
The instantaneous luminosity exceeds $4\times10^{32}$ cm$^{-2}$s$^{-1}$, 
and more than 12 fb$^{-1}$ of data are expected to be delivered in Run~II.
The center-of-mass energy is 1.96\,TeV in Run~II compared to 1.8\,TeV 
in Run~I.  

With the expected increase in the Tevatron luminosity and the experience 
gained in operating the \D0 detector and in analyzing data from Run~I, 
the \D0 collaboration upgraded the detector~\cite{d0nim} to ensure the full 
exploitation of the physics opportunities in Tevatron Run~II. 

\begin{figure}[h]
\begin{center}
    \epsfxsize=12.0cm 
    \epsfbox{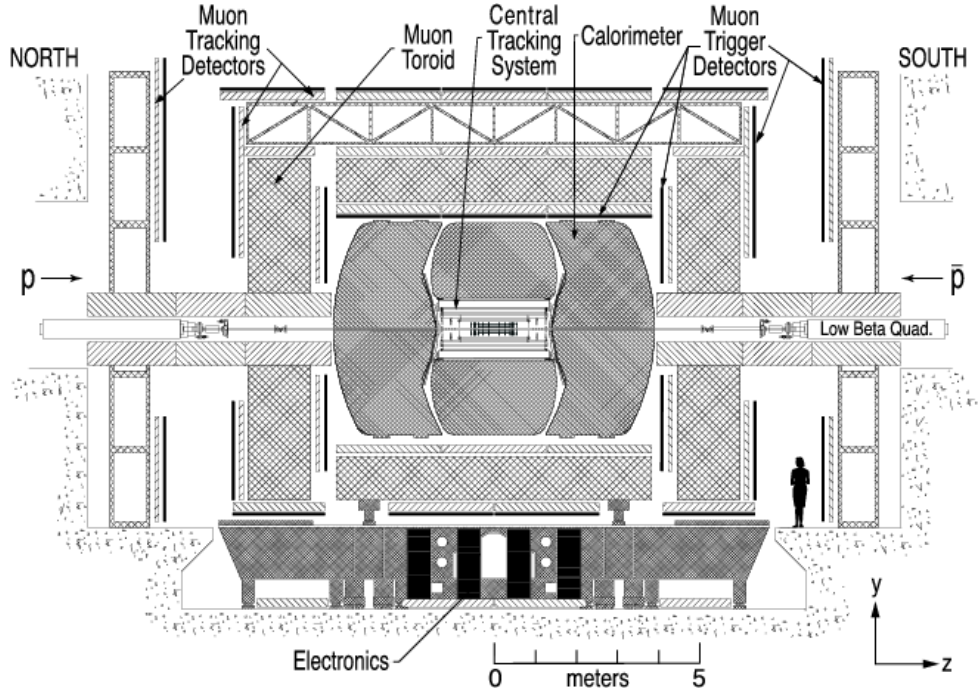}
    \caption{Cross sectional view of the Run II \D0 detector.}
\label{fig:dzero_2a}
\end{center}
\end{figure}

\begin{figure}
\begin{center}
    \epsfxsize=12.0cm 
    \epsfbox{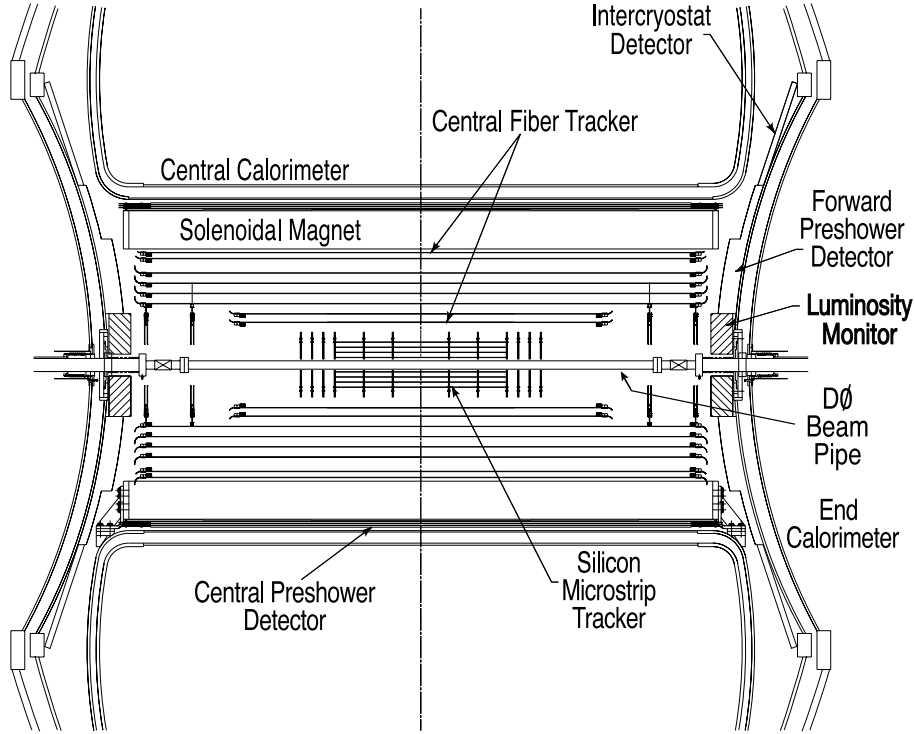}
    \caption{Cross sectional view of the Run IIa \D0 tracking system. }
\label{fig:dzero_tracker}
\end{center}
\end{figure}

Figure~\ref{fig:dzero_2a} shows a cross sectional view of  
the Run~II \D0 detector. This upgrade includes faster 
electronics to match the reduced bunch crossing interval, upgrades
of the muon system~\cite{muon}, and new central and forward preshower 
detectors to improve $e/ \gamma$ identification. Most important is 
a completely new central tracking system, illustrated in 
Fig.~\ref{fig:dzero_tracker}. A 2\,T axial magnetic field is provided 
by a $\approx 2.6$\,m long superconducting solenoid magnet with 
$\approx 0.5$\,m inner radius. The solenoid encloses a scintillating 
fiber tracker and a silicon microstrip tracker. The Central Fiber 
Tracker (CFT) consists of eight concentric barrels of scintillating 
fibers with axial and $\pm 3^0$ stereo read out. The Silicon Microstrip 
Tracker (SMT) is situated inside the CFT.     

Since Run~II has a higher instantaneous luminosity than Run~I, the trigger 
system was also upgraded. This new system is formed by three distinct 
levels. The first stage, Level~1, comprises a collection of 
hardware trigger elements which reduces the input rate of 2\,MHz
to approximately 2\,kHz. The Level~2 trigger system reduces the rate
further by a factor of two. In this second stage, hardware engines 
and embedded  microprocessors associated with specific subdetectors 
provide information to a global processor which  constructs a trigger 
decision based on individual objects as well as object correlations.
The SMT is one of the subdetectors taking part in the Level~2 trigger.
This so called Silicon Track Trigger (STT) is further described 
in Ref.~\cite{STT}. Events that passed Level~1 and Level~2 are sent to 
the Level~3 trigger farm for real time reconstruction, reducing the 
rate to 50\,Hz. 

The purpose of this paper is to provide a description of the 
\D0 Silicon Microstrip Tracker built for Run~II. In 2006 the dector 
was upgraded with an extra layer of silicon which was added close 
to the beam pipe. This is called layer 0 and is described in 
Ref.~\cite{layor0}. Furthermore the most forward and backwards 
disks seen in Fig.~\ref{fig:dzero_tracker} where removed.
The period before this upgrade is referred to as Run~IIa and the 
period after as Run~IIb. In the following section a general overview 
of the detector is given. Section~3 provides an overview of the 
mechanical aspects of the tracker, including a description of the cooling. 
This is followed by a section on the readout system. In Sec.~5, 
silicon wafer production and testing are described. The two following 
sections are about the production and quality control of the various 
detector components. The detector assembly is described in Sec.~8, and 
in Sec.~9, the installation is described. The paper concludes by 
summarizing some aspects of the performance of the detector.

In the description of the detector a right-handed coordinate system is 
used. The $z$-axis is along the proton beam direction and the $y$-axis
is upward, as illustrated in Fig.~\ref{fig:dzero_2a}. The angles 
$\phi$ and $\theta$ are the azimuthal and polar angles, respectively. 
The $r$ coordinate denotes the perpendicular distance from the $z$
axis. 

\newpage
\pagebreak
\section{General Overview}

The design goal of the SMT was to provide both tracking and 
vertexing over the full pseudorapidity ($\eta$)\footnote 
{The pseudorapidity is defined as 
{$\eta = -\ln\left[ \tan (\frac{1}{2}\theta)\right]$}where {$\theta$} 
is the polar angle relative to the proton beam axis.}  
coverage of the calorimeter and muon systems. 
 
\begin{figure}[ht]
\begin{center}
    \epsfxsize=10.0cm
    \epsfbox{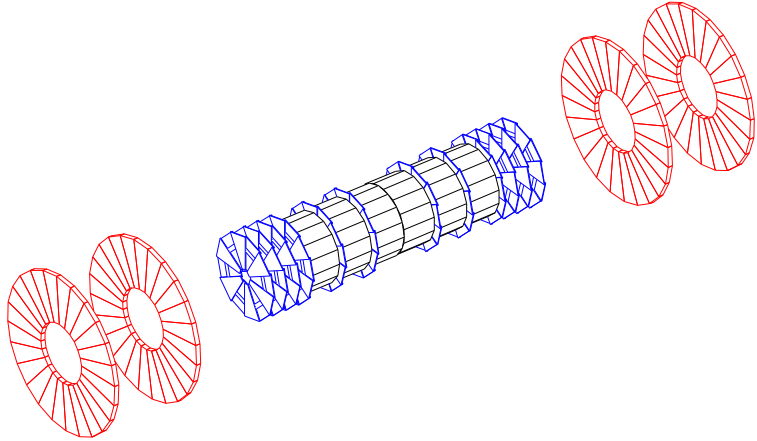}
    \caption{Isometric view of the Run~IIa \D0 silicon tracker.}
    \label{fig:smt_iso}
\end{center}
\end{figure}

Several of the Tevatron machine parameters had a large effect on the 
SMT design. The luminosity sets a scale for the radiation damage 
expected over the life of the detector, which in turn dictates the 
operating temperature. The bunch spacing sets the design parameters 
for the electronics and readout, as well as the probability that 
multiple interactions occur in a single crossing. The length of the 
interaction region sets the length scale of the device. With a long 
interaction region of about 25\,cm rms, it is difficult to deploy 
detectors such that the tracks are generally perpendicular to detector 
surfaces for all $\eta$. This feature led to the hybrid system shown 
in Fig.~\ref{fig:smt_iso}, with barrel detectors measuring primarily 
the $r$-$\phi$ coordinate and disk detectors which measure $r$-$z$ as 
well as $r$-$\phi$. Thus verticies for high $|\eta|$ particles are 
reconstructed in three dimensions by the disks, and verticies of particles 
at small values of $|\eta|$ are measured in the barrels. This design 
poses difficult mechanical challenges in the arrangement of the detector 
types to provide space for cooling and cables while minimizing dead areas.
 
The design of the SMT was also influenced by the requirements of minimal 
mass, a precise alignment between sensors and support structures,
and adequate thermal performance of the silicon modules in both the 
barrels and the disks. The barrel/disk design made it necessary to have 
the front-end electronics attached to the sensor structures. All these 
design constraints of the barrel silicon modules, called the ladders, 
led to a layout of the basic detector unit with approximately rectangular 
beryllium substrates with one (single-sided sensors) or two 
(double-sided sensors) silicon surfaces glued to them. The beryllium 
substrates conduct heat from sensors and their on-board readout
to cooling and support bulkheads. They also include features which 
provide precise alignment of ladder sensors to the bulkhead. A printed 
flex circuit, called the High Density Interconnect (HDI), carrying the 
readout chips and passive electronics components, was laminated onto 
the beryllium. A long flexible tail, called the pigtail, of the 
HDI allowed easy routing of the data and supply lines through the 
barrel structures. The readout chip, called the SVXIIe chip~\cite{svx2} 
provides 128 channels, each including a preamplifier, a 32 cell deep 
analog pipeline, and an 8 bit ADC. 

\begin{figure}[ht]
\begin{center}
    \epsfxsize=15.0cm
    \epsfbox{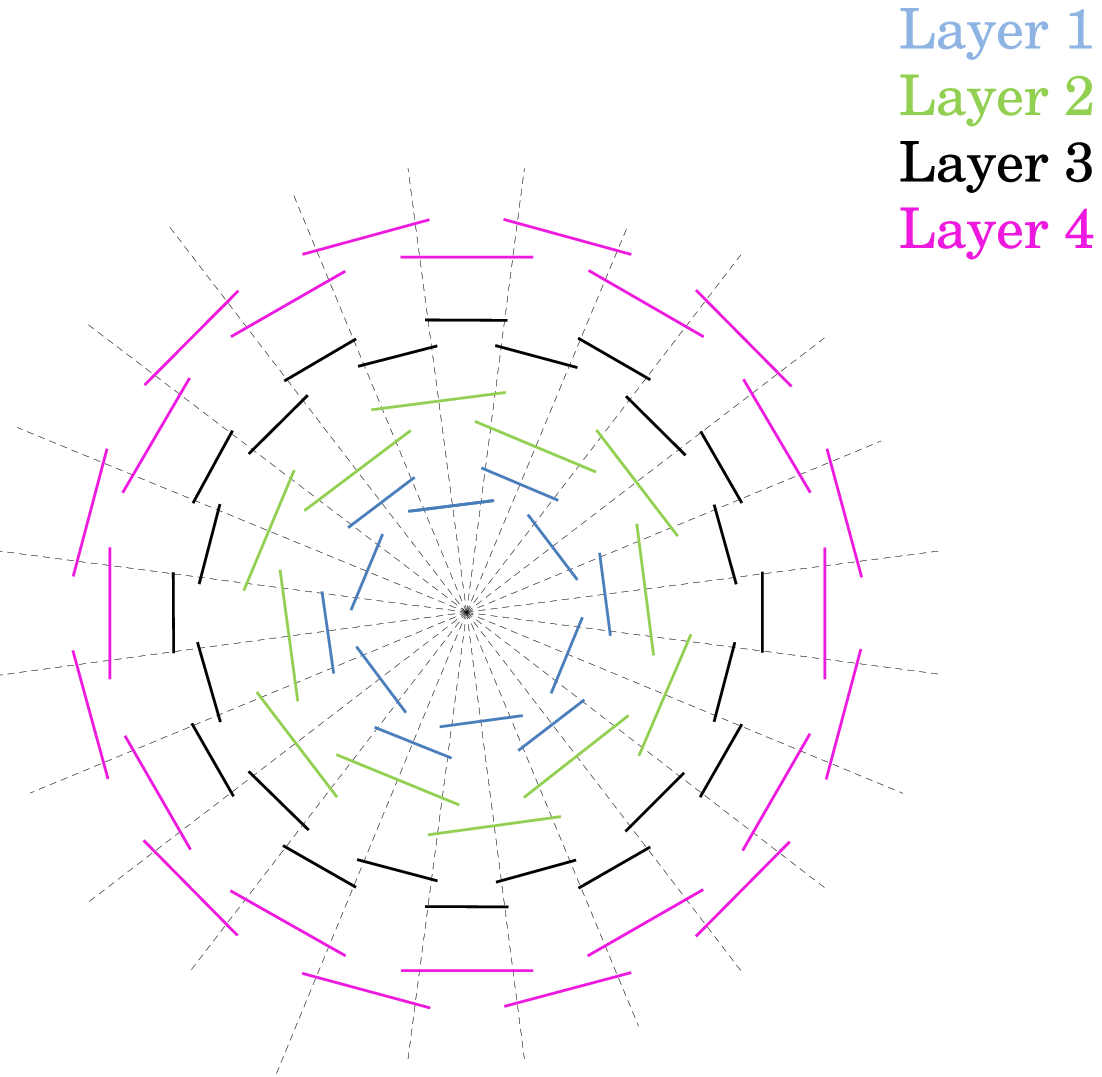}
    \caption{Arrangement of barrel modules in $r-\phi$.}
    \label{mech_F4}
\end{center}
\end{figure}

\begin{figure}
\begin{center}
\centerline{\epsfig{figure=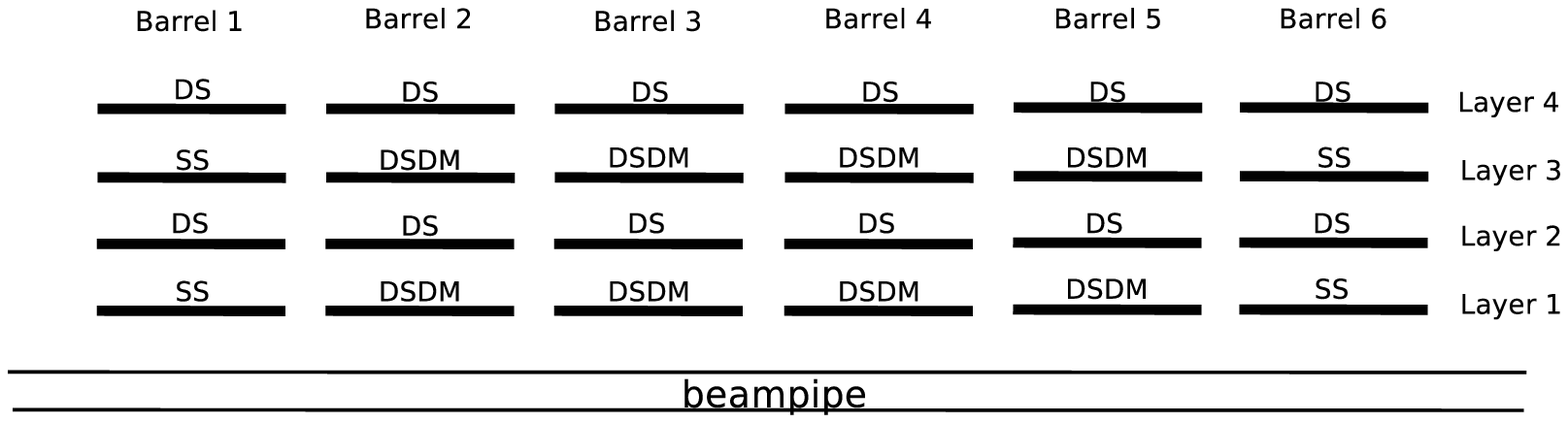,height=2.0in}}
\end{center}
\caption{The locations of the different types of ladders.}  
\label{fig:sensors}
\end{figure}

Each of the six barrels is 12\,cm long and has 72 ladders arranged in 
four layers, with each layer having two sub-layers at slightly different 
radii to provide azimuthal overlap, as illustrated in Fig.~\ref{mech_F4}. 
The two outer barrels have 36 single-sided (SS) and 36 double-sided 
$2^\circ$~stereo (DS) ladders. The four inner barrels have 36 double-sided 
double-metal (DSDM) $90^\circ$~stereo and 36 double-sided $2^\circ$~stereo  
ladders. Figure~\ref{fig:sensors} shows the locations of the different
sensors in the six barrels. In the text the ladders are often referred to as
3-chip (SS), 6-chip (DSDM), and 9-chip (DS) ladders according to the 
number of SVXIIe chips mounted on them. The ladders are mounted between 
two precision-machined beryllium bulkheads. The bulkhead that supports 
the sides of the ladders carrying the readout electronics is equipped 
with cooling channels.   

Each barrel is capped with a disk of wedge detectors, called the F-disks.
The F-disks comprise twelve wedges made of double-sided silicon wafers
with trapezoidals shapes. The stereo angle of the F-wedges is $30^\circ$.
The n-side is connected to 6 readout chips and the p-side to 8 chips. To
provide further coverage at intermediate $|\eta|$, the central system is 
completed with a set of three F-disks on each end of the barrel assembly. 
Each disk is rotated by $7.5^\circ$ with respect to its neighboring disk.

In both the far-forward and far-backward region, two large diameter disks,
called the H-disks, provide tracking information. The H-disks are made of 
24 pairs of single-sided detectors glued back to back, giving a stereo 
angle of $15^\circ$. Each detector is read out by 6 chips. Since these 
disks give the last track-measuring points before the end calorimeters, 
the mass constraints on these detectors are more relaxed than for the 
F-disks or barrel ladders.  This allowed the use of back-to-back 
single-sided detectors in a less compact package than the F-disks. The 
H-disks were designed to improve momentum resolution for tracks up to 
$|\eta|=3$. 

For both the F-disks and the H-disks, the wedges are mounted and aligned 
on beryllium rings which include cooling channels. The barrels together 
with the F-disks are precisely mounted on two carbon fiber cylinders which 
meet at the nominal interaction point in the \D0 detector. The four H-disks 
are individually mounted on carbon fiber cylinders. 

Table ~\ref{tab:radii} summarizes the SMT design parameters.

\begin{table}
\caption{SMT design parameters (module refers to both ladders and
 wedges).\label{tab:radii}}
\vspace{0.8cm}
\begin{center}
\footnotesize
\begin{tabular}{|l|l|l|l|}
\hline
             & \bf {Barrels}   & \bf {F-disks}   & \bf {H-disks}  \\
\hline
Layers/Discs &    4      &    12      &    4     \\
Channels     &  387,072   &  258,048   &  147,456  \\
Modules      &  432       &    144     &  96 pairs\\
Si area      &  1.4\,m$^2$  &  0.4\,m$^2$   &  1.2\,m$^2$\\
Inner radius &  2.7\,cm    &  2.6\,cm    &   9.6\,cm\\
Outer radius &  9.4\,cm    &  10.5\,cm   &    23.6\,cm\\
\hline
\end{tabular}
\end{center}
\vspace{0.6cm}
\end{table}

\newpage
\pagebreak

\section{Mechanical Structure and Environment}
\par
\subsection{Introduction}
\par
The SMT requires mechanical structures and enclosures that provide 
accurate positioning of the detector elements and an acceptable
operating environment for the silicon.  To allow optimal reconstruction 
of trajectories, the positions of detector elements must be known well 
enough that uncertainties in their positions contribute negligibly to 
track reconstruction inaccuracies. This placed requirements upon the 
position stability and upon the accuracy with which positions need to 
be known.  

Each barrel is 120\,mm long and 201\,mm in diameter. The ladders are 
parallel to  the beam pipe. The wedges are held in disks in which the 
planes of the wedges are normal to the beam line. The two types of 
disks, F-disks and H-disks, have active regions with outer diameter 
of  206.4\,mm and of 470.1\,mm, respectively, measured along the 
center-lines of the sensors. Six of the F-disks are interleaved with 
and attached to barrels to form disk/barrel modules. The remaining 
disks are located at the ends of the barrel region, three F-disk and 
two H-disks at either end. The fiducial length of the central 
disk-barrel region is approximately 766\,mm, the fiducial length 
including the F-disks is approximately 1066\,mm and that including 
the H-disks is approximately 2432\,mm. The central silicon detector, 
including cabling and mechanical structures, fits into a 360\,mm 
diameter region. Including readout structures and protective enclosures 
but excluding mount structures, the total diameter of the H-disks is 
approximately 488\,mm.

To limit the effects of multiple scattering and the production of 
secondary particles and showers, it is important that the amount of 
material, in the mechanical structure measured in radiation lengths, 
is minimized.  This implies the use of large radiation length materials 
and the minimization of the mass of the support structure in a way 
consistent with other design requirements.  

Both the disk-barrel modules and the end F-disks are supported from two 
carbon-fiber-laminate cylinders of combined length approximately 1660\,mm.
The two cylinders are butted against one another but electrically insulated 
from each another at $z = 0$.  In turn, they are supported from the inner 
surface of the innermost CFT barrel.  Removable portions of the cylinders 
allowed modules to be installed.  The cylinders also provide support for 
cables, cable connections, and coolant distribution manifolds. Each H-disk 
structure is independently supported from the inner surface of the 
innermost CFT barrel at large $z$ via leaf-spring, ball-and-cone 
kinematic mounts.

In the SMT region, the beam pipe consists of a beryllium cylinder with
an outer diameter of 38.1\,mm and an overall length of 2378\,mm. 

To control noise and leakage currents in the high radiation environment, 
sensors are indirectly cooled by forced flow of an ethylene glycol-water 
mixture.  The coolant temperature is typically $-8^\circ$C. 

\subsection{Ladders and Barrels}
\label{s:mech_barrel}
Barrel silicon sensors are mounted on ladders which  
serve four purposes:

\begin{enumerate}
\item They fix the relative positions of pairs of silicon sensors and provide
          features which allow the sensors to be accurately positioned in the
          barrel.
\item They aid in flattening the sensors and in maintaining flatness.
\item They hold the HDIs.
\item They provide a path by which the silicon and HDI components can be
          cooled.
\end{enumerate}

As mentioned in Sec.~2, three different types of ladders are used: 
3-chip ladders, 6-chip ladders which have the same width as the 3-chip 
ladders, and 9-chip ladders which are wider. While the sensors for the 
6-chip ladders are 120\,mm long, the 3-chip and 9-chip ladders are made 
of two 60\,mm long rectangular silicon sensors mounted end-to-end. A 
silicon barrel comprises four ladder layers at different radii; each 
layer consists of an inner and an outer sub-layer. A summary of ladder 
dimensions is given in Table~\ref{mech_T1}. 

\begin{table}[ht]
\begin{center}
\caption{Ladder dimensions (mm)}\label{mech_T1}
\vspace{0.6cm}
    \begin{tabular}{|l|r|r|r|}
\hline
\bf {Ladder type}        &  \bf {3-chip} &  \bf {6-chip} &    \bf {9-chip} \\
\hline
Single/double sided       &       S     &	D	&     D    \\ \hline
Stereo Angle (degrees)	  &      none   &      90	&	2    \\ \hline
Silicon thickness         &      0.300  &      0.300  &    0.300 \\ \hline

Length of each sensor      &     60.0  &     120.0  &   60.0 \\ \hline
Overall length            &    120.125  &    120.125 	&  120.125 \\ \hline
Length of active portions &    116.7  &    116.7  &  116.7 \\ \hline

Silicon width             &     21.2  &     21.2  &   34.0 \\ \hline
Active width              &     19.2  &     19.2  &   32.0 \\ \hline
Maximum HDI width         &     25.2  &     25.2  &   38.0 \\ \hline

\end{tabular}
\end{center}
\end{table}

\begin{figure}
\begin{center}
\centerline{\hspace{-.1in}\psfig{figure=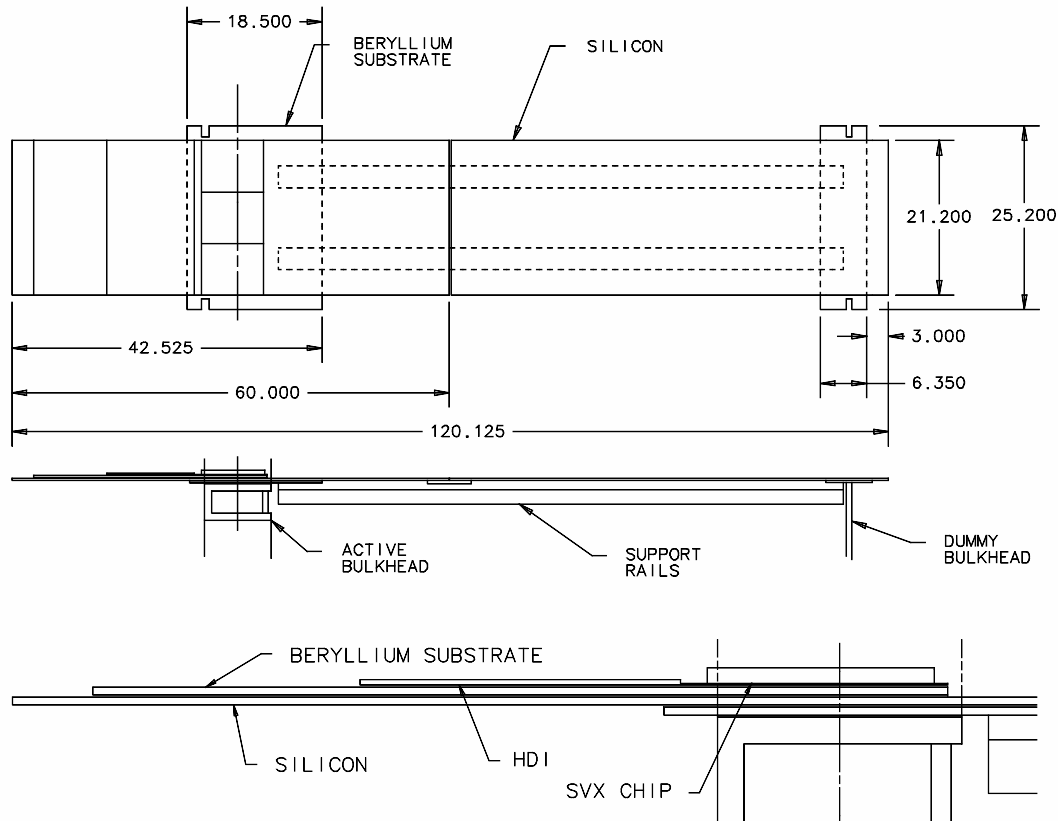,height=4.5in}}
  \caption{Mechanical structure of the 3-Chip ladder with dimensions 
given in mm.
The bottom picture is a blow-up of the HDI area shown in the middle
picture. The bulkheads are part of the support
structure and not part of the ladder.}\label{mech_F2a}
\end{center}
\end{figure}

The mechanical structure of a 3-chip ladder is shown in Fig.~\ref{mech_F2a}. 
Two rails, each made of a fiber composite-Rohacell foam sandwich, 
interconnect and support the silicon sensors.  A layer of carbon-boron 
hybrid fibers plus a layer of carbon fibers was used for each fiber composite 
to closely match the thermal expansion of the silicon sensors. At the HDI 
end, the rails were bonded to the 0.40\,mm thick beryllium substrate which 
provides reinforcement in the HDI region and aids in conducting heat from 
the HDI components. This beryllium piece, along with another piece near the 
opposite end of the ladder, also provides features used in positioning and 
holding the ladder. 

\begin{figure}
\begin{center}
\centerline{\psfig{figure=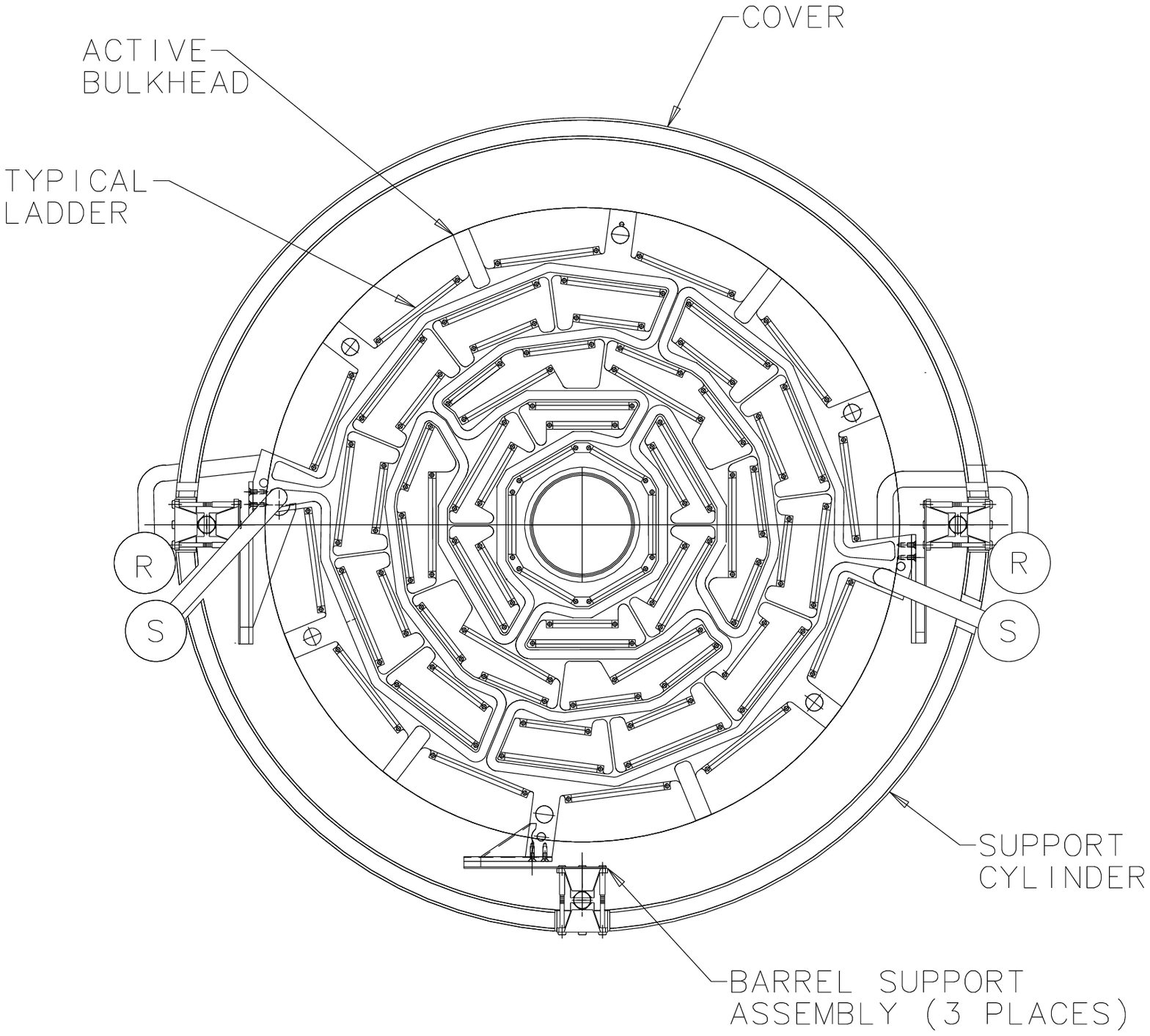, height=6.0in,angle=270.}} 
\caption{Ladder positions on the active bulkhead. Four manifolds,
two supply and two return, are indicated in the figure with 
S and R.}\label{mech_F3} 
\end{center}
\end{figure}

The ladders were positioned between two bulkheads, a cooled, active,
bulkhead and an uncooled, passive, bulkhead. Radial positions of the 
silicon are shown in Table~\ref{mech_T2} and Fig.~\ref{mech_F3}.  The 
ladders are grouped into 24 equal $\phi$ towers centered on the layer 3 
ladders as shown in Fig.~\ref{mech_F4}. Layers 1 and 2, the two innermost 
layers, have half as many ladders as layer 3. The $\phi$ positions of 
ladders in layers 1 and 2 have been chosen to cover two towers. Each tower 
includes portions of 4 ladders. 

\begin{table}[ht]
\begin{center}
\caption{Ladder silicon radial positions along each 
ladders centerline.}\label{mech_T2}
\vspace{0.6cm}
    \begin{tabular}{|c|c|c|c|}
\hline
\bf {Layer} & \bf {Sub-layer} & \bf {Ladders/}& \bf {R (mm)} \\
       &             &  \bf {sub-layer}   &     \\
\hline
  1  &     Inner   &   6     &  27.15            \\
     &     Outer   &   6     &  36.45            \\ \hline
  2  &     Inner   &   6     &  45.50            \\
     &     Outer   &   6     &  55.54            \\ \hline
  3  &     Inner   &  12     &  67.68            \\
     &     Outer   &  12     &  75.82             \\ \hline
  4  &     Inner   &  12     &  91.01              \\
     &     Outer   &  12     & 100.51              \\ \hline
\end{tabular}
\end{center}
\vspace{0.6cm}
\end{table}

Several considerations drove the choice of ladder location on the
bulkheads. 
\begin{enumerate}
\item Adequate overlap between adjacent ladders was provided to ensure
complete $\phi$ coverage within each layer.  This places restrictions on the
maximum average radius of a layer, given the active widths and numbers of
ladders.  

\item Sufficient space was provided between layers to permit ladder
installation.  The overall radial space for installing ladders is
indicated in Tables~\ref{mech_T4} and~\ref{mech_T5}.
The minimum bulkhead layer-to-layer radial clearance is 10.25\,mm,  
allowing 1.73\,mm clearance on both surfaces of an installed ladder,
assuming a ladder already in place on the opposite sub-layer surface. 

\begin{table}[ht]
\begin{center}
\caption{Total radial height of a ladder 
from the bottom up.}\label{mech_T4}
\vspace{0.6cm}
    \begin{tabular}{|l|c|}
\hline
\bf {Element} &  \bf {Radial} \\
               & \bf {height (mm)}  \\
\hline
Rail                                         &  2.757  \\
Adhesive                                     &  0.075  \\
Silicon sensor                               &  0.300  \\
Adhesive                                     &  0.075  \\
Beryllium                                    &  0.406  \\
Adhesive                                     &  0.075  \\
HDI under chips                              &  0.117  \\
Adhesive                                     &  0.075  \\
SVXIIe chip                                  &  0.381  \\
Wire-bonds (above chip)                      &  0.500  \\ \hline
Total                                        &  4.761  \\ \hline
\end{tabular}
\end{center}
\vspace{0.6cm}
\end{table}

\begin{table}[ht]
\begin{center}
\caption{Radial heights above the bulkhead mounting surface for an installed
 ladder.}\label{mech_T5}
\vspace{0.6cm}
    \begin{tabular}{|l|c|}
\hline
\bf {Element} &  \bf {Radial} \\
              &  \bf {height (mm)}  \\
\hline
Adhesive                                    &  0.051  \\
Beryllium                                   &  0.406  \\
Adhesive                                    &  0.075  \\
Silicon sensor                              &  0.300  \\
Adhesive                                    &  0.075  \\
Beryllium                                   &  0.406  \\
Adhesive                                    &  0.075  \\
HDI under chips                             &  0.117  \\
Adhesive                                    &  0.075  \\
SVXIIe chip                                 &  0.381  \\
Wire-bonds above chips                      &  0.500  \\ \hline
Total                                       &  2.461  \\ \hline
\end{tabular}
\end{center}
\vspace{0.6cm}
\end{table}

\item The space needed for a 3.175\,mm\,$\times\,$7.366\,mm cooling  
channel placed a lower bound on the sub-layer radial spacing. 

\item A minimum space of 5.516\,mm was left between inner and outer sub-layer
ladder mounting surfaces for cables. Components which extend from the ladder
surfaces reduce the clear space for cables to about 2.7\,mm. 

\item Space between adjacent ladders of a sub-layer was left for 
a 9.4\,mm wide layer-to-layer coolant connection.  This limits 
the minimum radius of a layer, given ladder physical widths and 
the number of ladders in the layer. 

\end{enumerate}

Final machining of the active and passive bulkheads of a barrel was done 
with the bulkheads clamped together as a unit.  This ensured that bulkhead 
ladder mounting surfaces and mounting features match on each pair of 
bulkheads and helped ensure that installed ladders would be parallel to the 
beam line.  Mounting surfaces were machined to a flatness of approximately 
25\,$\mu$m, and the mounting surfaces on the 9.525\,mm thick active 
bulkhead were made perpendicular to the plane of the bulkhead to 25\,$\mu$m.  
This establishes a maximum ladder slope at the active bulkhead of 
$25/9525=2.6$\,mrad. Assuming this slope, perfect alignment of the active 
and passive bulkheads, and the appropriate elastic properties of a ladder, 
implies about 53\,$\mu$m maximum radial deflection of the ladder from a 
straight line.

Ladders were positioned laterally with the aid of posts on the active 
and passive bulkheads which extend from the bulkhead mounting surfaces 
and engage notches in the ladder beryllium.  Only the beryllium piece 
closest to the bulkhead mounting surface and the edges of the posts 
closest to the ladder centerline were used for this purpose. Clearances 
between notches in the beryllium piece of the ladder and the posts were 
chosen so that ladders could be placed on each bulkhead with a transverse 
accuracy of $\pm$6\,$\mu$m. Malleable pins through holes in the posts 
hold the ladder in place. In addition, the outermost layer ladders were 
glued in place to provide a rigid coupling between the active and passive 
bulkheads. The ladders play a very important structural role in the barrels.  
Each barrel is a complete, internally aligned unit. All barrel positioning 
was done with respect to the active bulkhead. The passive bulkhead is held 
and accurately positioned by the ladders and sets only the relative positions 
of ladders.

The barrel structure formed by the ladders and bulkheads must be sufficiently
stiff that internal alignment criteria are satisfied under forces from the
cables, the coolant connections, and thermal contraction.  The required 
stiffness is provided without the use of additional structural members, 
by fastening ladders to the bulkheads with epoxy.  The total barrel 
structure has a torsional compliance of 0.11\,$\mu$m/N$\cdot$m assuming 
that the passive bulkhead positions ladders relative to each other only 
in $r$ and $\phi$. Uncontrolled forces on the barrels are approximately 
1\,N.  This leads to a negligible shift in ladder $\phi$ positions. Shear 
stresses in the most highly stressed epoxy joints with a 1\,N force applied 
to the passive bulkhead at a radius of 100\,mm are 0.0050\,MPa, a factor of 
more than 1300 below the shear strength of the electrically and thermally 
conductive adhesive used ({\it e.g.,} TraCon 2902 silver-filled epoxy). 

\begin{figure}
\begin{center}
\centerline{\psfig{figure=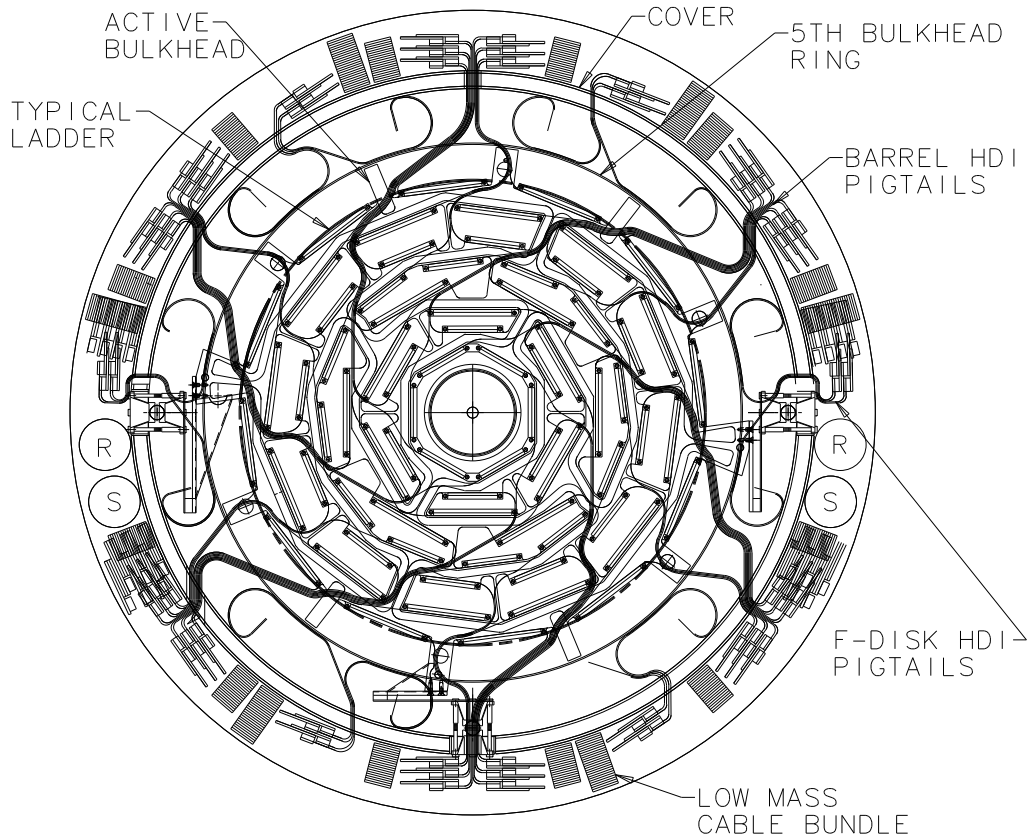,height=6.0in,angle=270.0}}
\caption{Barrel cable paths.}\label{mech_F6}
\end{center}
\end{figure}

The pigtails that emerge at the HDI edge follow paths between and around 
ladders to the outer radius of the barrel, penetrate the support 
half-cylinder or its covers, and connect to low mass cables mounted on 
the outside.  An extra bulkhead ring, made of carbon fiber composite, is 
provided at a radius of about 130\,mm to anchor the cables. The cables 
are grouped at this ring into bundles which are brought through the 
half-cylinder and its covers 
at 2:00, 4:00, 6:00, 8:00, 10:00, and 12:00 o'clock 
as shown in Fig.~\ref{mech_F6}. 

\subsection{F-Disk and F-Wedge Structure}
\begin{figure}
\begin{center}
\centerline{\psfig{figure=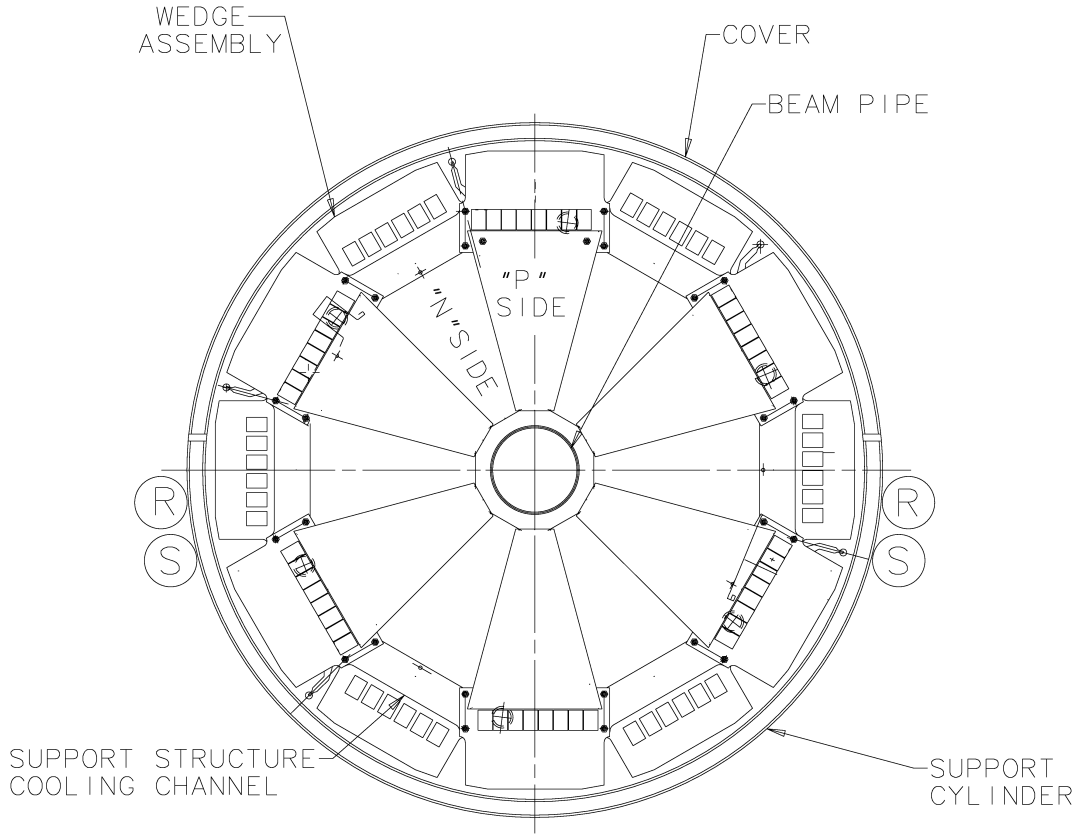,height=6.0in,
angle=270.0}}
   \caption{F-disk detectors.}\label{mech_aF1}
\end{center}
\end{figure}

As mentioned in Sec.~2, each barrel is accompanied by an F-disk, and 
three more F-disks each are installed in the forward and backward regions. 
The schematic for an F-disk is shown in Fig.~\ref{mech_aF1}. It contains 
12 double-sided wedges which are installed with alternating p- and n-sides. 
The dodecagon support structure is made of beryllium and has a 
cross-section of 15.7\,mm$\,\times\,$2.5\,mm. A 7.5\,mm$\,\times\,$1.0\,mm 
channel in the beryllium provides a path for coolant flow to cool the wedges. 
Disks are oriented so that the wedges match the barrel towers. 

The disk thickness contributes to the $z$-gap between adjacent barrels. To
minimize this thickness, the HDI was placed at the outer periphery 
of the wedge. The thickness of the structure is 4.850\,mm at the 
detector sensor and 6.322\,mm at the SVXIIe chips, including 1\,mm 
for two sets of wire-bonds. With additional clearance to avoid 
potential interference and accommodate thickness variations, the 
interleaved disks fit within a barrel-to-barrel gap of about 8\,mm. 

A disk was attached to the active bulkhead of each barrel to form a 
disk/barrel module.  This permitted the disk and barrel to be accurately 
matched and to be installed as a unit.  Each of the interleaved
disks was attached to a barrel at three points spaced evenly in $\phi$. 
The posts are permanently attached to the beryllium disk support structure.  

\subsection{Module Support}
The disk/barrel modules and end disks were mounted from the support
half-cylinders in a way which adequately resists the known gravitational forces
and the somewhat variable forces from cabling and coolant connections.  
Provided that temperature variations are sufficiently independent of 
$\phi$, thermal contraction effects are predictable.  The thermal 
expansion coefficient of beryllium, 11.6\,ppm/$^\circ$C, leads to 
a radial contraction of 21\,$\mu$m for a 100\,mm radius structure 
which is cooled from $23^\circ$C to $5^\circ$C.  Since the connections 
from disks and disk/barrel modules to the support half-cylinder are all 
nearly identical, the vertical center-lines of these structures move 
together and good relative alignment is maintained.

The support half-cylinder is described in Sec.~3.7. 
Carbon fiber composite mounts attached to the active bulkheads connect 
disk/barrel modules to the half-cylinder at 3:00, 6:00, and 9:00 o'clock 
as shown in Fig.~\ref{mech_F3}. The 3:00 and 9:00 o'clock mounts control 
$y$ and $z$ positions, while the 6:00 o'clock mount controls $x$ and $z$ 
positions. Pairs of leaf springs allow motion in one  direction while 
remaining stiff in the other two directions.  A sapphire ball between the 
half-cylinder portion of the coupling and the barrel portion of the 
coupling permitted angular orientations to be matched during assembly; 
four screws around the ball allowed for adjustment and locking of the angular 
orientation. 

\subsection{H-Disk and H-Wedge Structure}
\begin{figure}
\begin{center}
\centerline{\psfig{figure=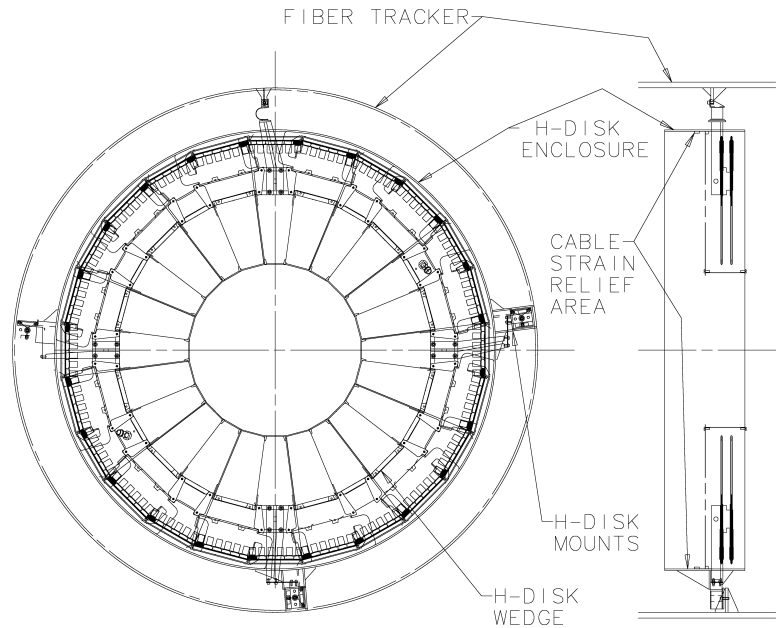,height=4.5in,angle=270.0}}
\caption{H-disk assembly.}\label{mech_hF1}
\end{center}
\end{figure}

An H-disk is shown in Fig.~\ref{mech_hF1}. It  contains 24 wedges which
alternate on each surface of a beryllium cooling pipe. Four disks were 
deployed for Run\,IIa, at $z\approx \pm 100$ and $\pm$ 121\,cm. In 
Run\,IIb the two outermost disks were removed. Within a disk, the 
alternating wedges are separated by 9.1\,mm in $z$ and the sensors are 
separated by $\approx 1.2 $\,mm within a wedge. The disks are supported 
by a ball-and-cone mount from the CFT carbon fiber support cylinder. 
The disk supports were installed and aligned on the CFT support cylinder 
using a coordinate measuring machine during the assembly of the CFT to 
insure precise alignment during installation of the disk assemblies in \D0.

\begin{figure}
\begin{center}
\centerline{\psfig{figure=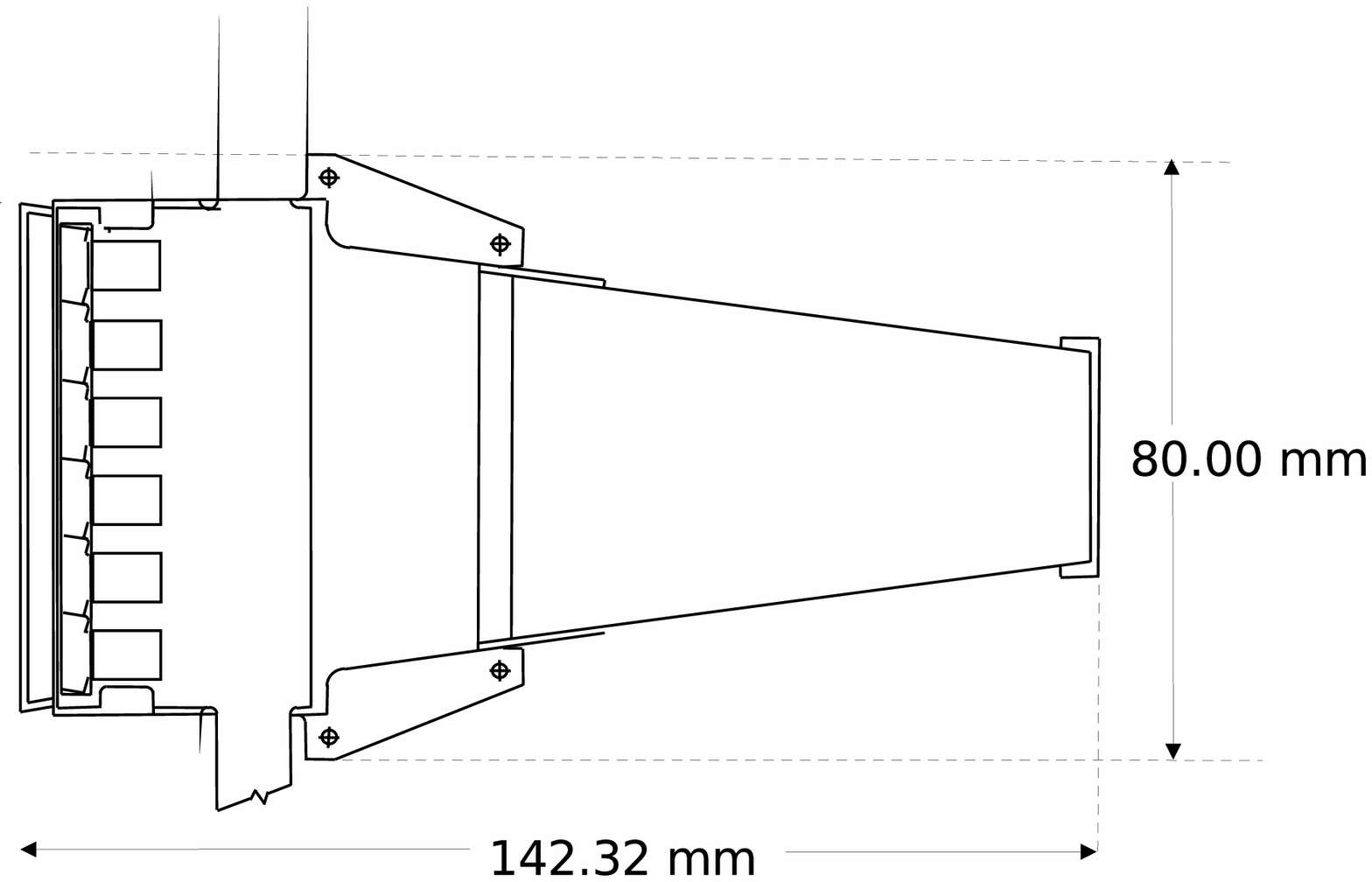,height=3.0in}}
\caption{H-wedge assembly.}
\label{Hwedge}
\end{center}
\end{figure}

An H-wedge was fabricated from two back-to-back, single-sided, detector 
assemblies (Fig.~\ref{Hwedge}). The inner and outer radii at the wedge 
centerline are 9.61\,cm and 23.6\,cm, respectively. The detector pair is 
supported by a beryllium plate, which also forms the wedge support for 
attachment to the beryllium cooling ring. Front and back detectors on a 
wedge were aligned using a specially fabricated dual microscope alignment 
system.

\subsection{Cooling}
The heat to be removed from detector-mounted readout electronics is summarized
in Table~\ref{mech_bT1}.  The combined power dissipation for the SVXIIe chips
and other HDI components is about 3\,mW per channel. Heat loss in the cabling 
brings the total power dissipation to about 2400\,W.  

\begin{table}[ht]
\begin{center}
\caption{Detector power dissipation (watts).}\label{mech_bT1}
\vspace{0.6cm}
    \begin{tabular}{|l|c|c|c|}
\hline
\bf {Elements} & \bf {Dissipation} & \bf {Number of} & \bf {Dissipation} \\
       &         \bf {(watts)}    & \bf {Modules} &   \bf {(watts)} \\ \hline 
Barrel         &                  &         &                      \\ 
3 chip       & 1.15 per ladder & 36 per outer barrel & 41 per outer barrel \\
6 chip       & 2.30 per ladder & 36 per inner barrel  & 83 per inner barrel \\
9 chip       & 3.46.per ladder & 36 per barrel  &  124 per barrel\\
\hline  
F-disks      & 5.38 per wedges &  12 per disk   &  65 per disk \\
H-disks      & 2.30 per wedges &  48 per disk   &  110 per disk \\
\hline  
Outer barrels  & 165 per barrel  &  2 barrels   &   331 \\
Inner barrels  & 207 per barrel  &  4 barrels   &   826  \\
F-Disks        & 65 per disk     & 12 disks     &   778 \\ 
H-Disks        & 110 per disk    &  4 disks     &   442  \\ 
\hline
Detector total     &                     &              &   2377  \\
\hline
\end{tabular}
\end{center}
\vspace{0.6cm}
\end{table}

The ethylene glycol/water cooling was chosen because of the favorable 
heat capacity of water and the simplicity of water systems. The cooling 
system consists of a reservoir, pump, flow control valve, de-ionizer, 
vacuum system, supply lines, return lines, and instrumentation. The pump 
circulates approximately 60\,l/min of coolant. At this flow rate, the 
barrels, disks, and manifolds produce an aggregate pressure drop of 
55\,kPa.  To prevent coolant from leaking onto the detector components if 
leaks occur, the coolant pressure of the supply manifold at the detector 
is maintained at about 90\,kPa. This pressure is achieved with an open 
bath on the suction side of the system from which fluid is drawn through the 
detector; the pressure drop in the piping to the detector and 
elevation difference between the open bath and the detector determine 
the pressure at the inlet to the detector. The pump is located at the floor 
of the collision hall to take advantage of the available heat to ensure 
adequate pressure on the suction side of the pump to avoid cavitation in 
the pump.

With 30\% ethylene glycol/water mixture the system is operating at 
$-8^\circ$C. The temperature within sensors varies with location 
from approximately $-5^\circ$C to $+2^\circ$C.  The cooling system 
was designed to allow temperature to be lowered another $7^\circ$C 
if compensation for the effects of radiation damage proves to be necessary.  

Four coolant manifolds, two supply and two return, provide coolant flow to the
barrels and disks.  Each barrel requires two supply and two return connections,
as shown in Fig.~\ref{mech_F3}. The cooling channel of each layer of a barrel
bulkhead is divided into equal left and right portions.  The left portions of
all layers are connected in series between the left supply and return
manifolds.  Similarly, the right portions are connected in series between the
right supply and return manifolds. Layer-to-layer connections were machined as
part of the bulkhead beryllium structure.  

Disk coolant flow is divided between equal length upper and lower paths.
The two paths share coolant  connections at 3:00 and 9:00 o'clock. To 
equalize flow rates in the supply and return manifolds, the flow direction 
is reversed from one disk to the next. A set of supply and return manifolds 
is located on the outer surface of the half-cylinder near 3:30 o'clock; a 
second set is located near 8:30 o'clock. The supply manifolds enter the 
half-cylinder region as 15.875\,mm inside diameter (ID) lines. Approximately 
700\,mm along the half-cylinder, after supplying the end disks and the 
first barrel/disk module, both lines reduce to 12.7\,mm. The return 
manifolds have the same configuration, starting as 12.7\,mm ID pipe 
and then increasing to 15.875\,mm ID pipe between the second and third 
barrel/disk modules. This configuration of reducing supply manifold size 
and increasing return manifold size produces evenly divided flow to the 
detectors. The manifolds are fabricated from thin-walled PVC pipe. Coolant 
connections from the disks and disk/barrel modules to the manifolds were 
made with flexible tubing. 

\subsection{Half-Cylinder and Cover}\label{mech_strough}
\begin{figure} 
\centerline{\psfig{figure=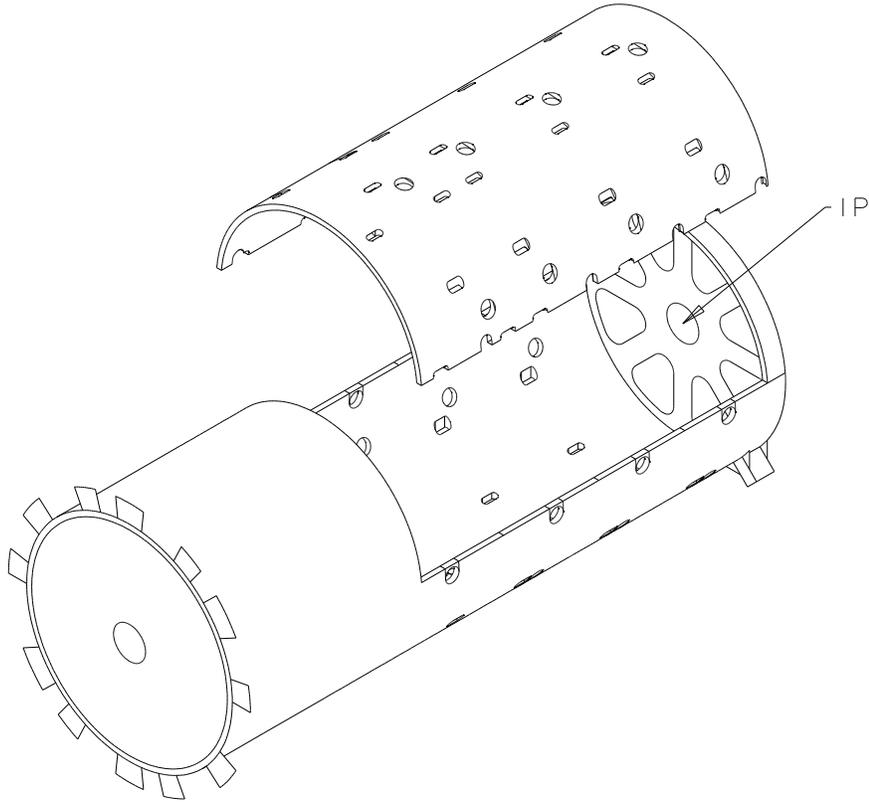,height=5.0in, angle=270.}}
\caption{Support half-cylinder and cover. IP denotes the nominal 
central interaction point.
The holes seen on the barrel are where the cables and cooling lines pass 
through it. The structure on the left disk is support for the 
cables.}\label{mech_dF2} 
\end{figure}

Carbon fiber half-cylinders (Fig.~\ref{mech_dF2}) provide accurate and
stable support of detector elements.  The half-cylinders are 1.1\,m long, 
have an inside radius of 145\,mm, and outside radius of 153\,mm.  The
half-cylinders consist of 0.9\,mm thick inner and outer shells joined 
by 0.4\,mm thick webs. The webs are spaced every 25.65\,mm in $z$. 
The outer ends of the half-cylinders are closed by 2\,mm thick full 
circular membranes which have appropriate openings for the beam pipe, 
cables, and coolant manifolds. The inner ends, near $z$=0, have 0.5\,mm 
disks with large wedge-shaped cutouts to reduce mass in this critical 
region.  A third circular membrane is present just beyond the outermost 
F-disk assembly.

The support half-cylinders were made from high modulus (900\,MPa) carbon fiber 
prepreg plies. The high modulus carbon fiber has a low thermal expansion 
coefficient, high stiffness to mass ratio, reasonable cost, and acceptable 
fabrication properties. A low coefficient of thermal expansion ensures that 
temperature gradients do not cause significant displacement of detector 
elements in the face of the large thermal gradients from the coexistence of 
cooled surfaces and hot electronic chip modules.  The high modulus minimizes
deflections caused by gravitational, cabling, and coolant connection forces. 
Multiple high modulus carbon fiber plies are combined at specific angles to
achieve a thermal expansion coefficient of approximately 0.9\,ppm/$^\circ$C.
The elastic modulus of the laminate is approximately 124\,GPa 
($18\times 10^6\,$psi). This provides a stiffness to radiation length ratio 
which is five times that of steel and three times that of aluminum.  
The high modulus carbon fiber is an excellent electrical conductor for 
the frequencies of concern for electronic noise (up to 1\,MHz). The outer 
surface of the half-cylinder is wrapped with 2" wide aluminum foil tape 
near the outer end to provide a connection from the carbon fiber to the 
electronics ground.

Finite element analysis, confirmed by analytical calculations, give a 
maximum half-cylinder deflection of 100\,$\mu$m and rotations about the 
x-axis of 70\,$\mu$rad at the locations of the first and last barrels. 
Only a fraction of this maximum deflection occurs over the length of the 
silicon tracker itself. Final alignment of each barrel/disk and end-disk 
assembly were done after the half-cylinders were fully loaded 
(including cables) with the cylinders supported off their final mounting 
points so errors due to cylinder deflection were effectively removed.

The shells and webs were made as separate components from several layers 
of unidirectional carbon fiber prepreg tape. Each component was formed 
and cured at an elevated temperature. The components were assembled to 
form the half-cylinder structures using room temperature curing epoxy.  

The half-cylinders are supported from the CFT near $z$=0 and at $z$=830\,mm.  
At the inner end the cylinders have two feet located at $\pm 45^{\circ}$ 
from the bottom with spherical ceramic ends resting in reinforced mounting 
points attached to the inner wall of the CFT.

Covers for the half-cylinder provide a thermal barrier, mechanical and light
protection, electrostatic shielding, and a mounting surface for the low mass 
data cables.   The covers have the same mechanical configuration as the 
half-cylinders of an inner and an outer skin coupled with thin radial webs.    
Holes are provided to allow alignment of the completed assembly.  

\subsection{Run\,IIa Beam Pipe}
The part of the Tevatron Run\,IIa beam pipe that passes through the SMT has 
a 38.1\,mm outside diameter and a 0.508\,mm thick wall. Bellows and 
knife-edge flanges are provided at each end of the 2.578\,m long beam pipe
assembly. The Run\,IIb beam pipe is described in~\cite{layor0}.

The beryllium beam pipe was fabricated by rolling a
beryllium sheet into a round cylindrical shape and joining the edges 
with a vertical strip, providing four fillet locations for brazing.  All 
beryllium to beryllium, and beryllium to stainless steel joints are 
brazed with an aluminum filler material. 

\newpage
\pagebreak

\section{Electronics}
\subsection{Introduction}

\begin{figure}
\vskip 1.0in
\centerline{\psfig{figure=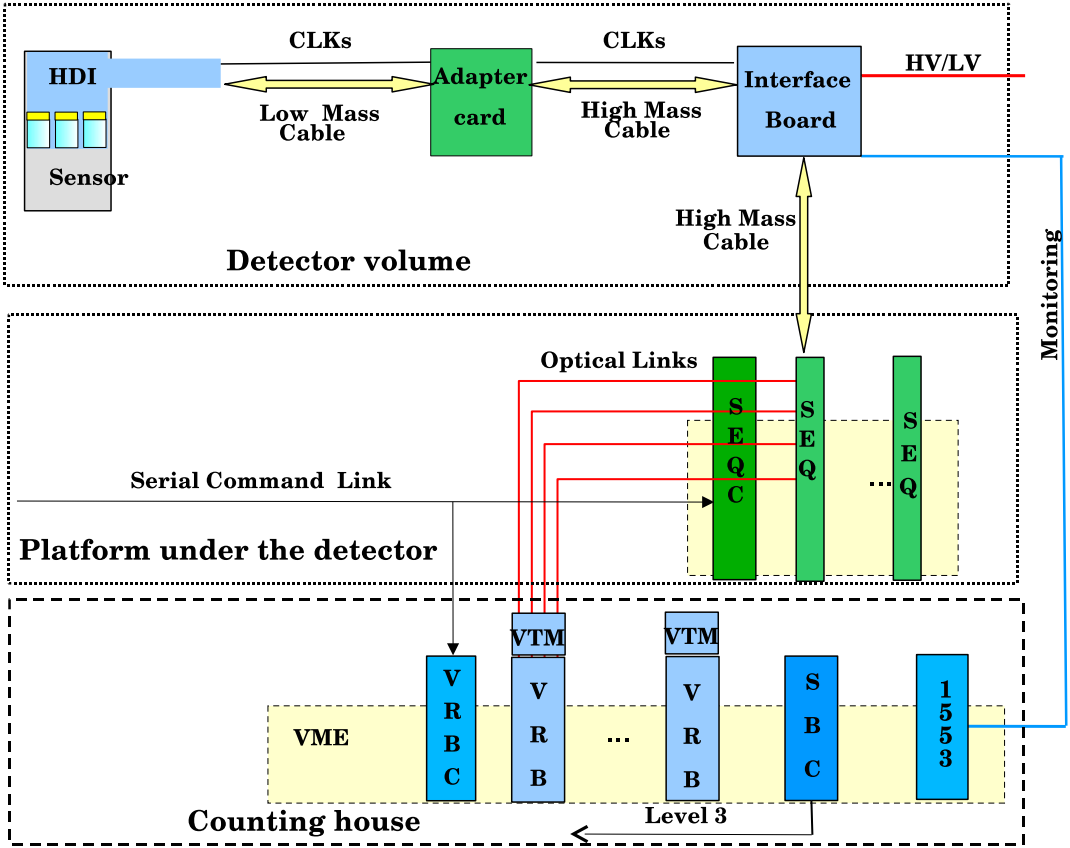,height=5.0in}}
\caption{\D0 SMT readout system.}
\label{f:readout}
\end{figure}

The data acquisition and readout system for the SMT is responsible for 
gathering the charge from the silicon sensors, digitizing these data, 
and sending it to the \D0 data acquisition system.  The MIL-STD-1553 
connection~\cite{1553} to the D0 Control System provides the ability to 
download and monitor the readout electronics. The readout system is 
depicted in the block diagram shown in Fig.~\ref{f:readout}. 
The locations of the components are also indicated in the figure.

The electronics design for the SMT was dictated by the physics needs of the 
experiment, the engineering resources that were available, and the chip 
technology. The SVXIIe chip has 128 channels and a 32 deep pipeline delay 
to allow time for trigger formation; it can operate with 132\,ns bunch 
spacing.  The chips are mounted on the HDI, a polyimide hybrid. 
One HDI supports between three and nine chips depending on the sensor 
arrangement. The pigtail on the HDI  allows the chip signals to be routed 
to the outer edge of the detector before installing a connector.

From the connector at the outer edge, a special polyimide cable, called the 
low mass cable, carries the signals for a distance of about 3\,m to a 
small passive circuit board referred to as the adapter card. The adapter 
card interfaces the low mass cable to an 80-conductor 3M pleated foil cable. 
Two low mass cables are connected to each 80-conductor cable. Signal 
dispersion in this two-cable combination limits the combined length 
to about 15\,m.  The 15\,m point is at the base of the liquid argon 
calorimeters. Since there is not enough space in this area to house the 
readout, a repeater board, called the interface board, was installed to 
retransmit the signals to the electronics area under the detector. The 
power supplies for the SMT are also located on the side of the central 
calorimeter cryostat near the interface board.

The repeated signals are sent to a sequencer module (SEQ) located under 
the detector on the \D0 platform.  The sequencer generates the sequences to 
download parameters into the SVXIIe chips, to put the chips into data 
acquisition mode, and to have the chips digitize and read out data 
in response to a trigger. During readout, the sequencer sends serialized 
data over a fiber optic link to a VME module, called a VME Read Out 
Buffer (VRB)~\cite{vrb}. The optical data are converted to 20-bit parallel 
data words by the VRB Transition Module (VTM) before reaching the VRB.  
The VRB buffers the data from each Level~1 triggers and waits for 
a Level~2 trigger accept. If one occurs, data are sent over the back plane 
to a VME readout controller and then sent over ethernet to the Level~3 
processor farm.  

The following sections provide more detailed  descriptions of the various 
components and the reasons for the design choices.  
  
\subsection{Low Noise Design}

Designing a precision, small-signal-level detector involves a complex  
interplay of mechanical and electrical design constraints. Many of the 
mechanical components are electrical conductors so they also have an 
affect on the electrical performance of the detector. The main goals were 
to minimize induced noise in the calorimeter, properly ground the 
beryllium mounting plates, and eliminate ground loops and other unwanted 
current paths through the detector.  

The SMT is situated in the center of the calorimeter and many components are 
mounted on the calorimeter cryostat so preventing noise pick up required 
careful design.  It was decided not to take additional triggers while the 
SMT was digitizing and reading out data,  which means that when a Level~1 
trigger occurs, all further Level~1 triggers are inhibited until all the 
data is read out from the current trigger. This results in Level~1 dead 
time for each Level~1 accept. The design of the SMT was simplified by this 
decision since the calorimeter signals are already stored in the 
calorimeter readout electronics when the SVXIIe digitization starts.   
In fact, many of the design choices such as single-ended output from 
the SVXIIe and unshielded low mass cables would not have worked with 
simultaneous acquisition and digitization. Tests have shown that SMT 
digitization and readout generates noise signals in the calorimeter.

The only SMT signal that is active during calorimeter data taking is 
the clock, which is carefully shielded.  The signal is sent 
differentially on a small, high quality coaxial cable using RF connectors.  
The pair of coaxial lines that are in the calorimeter bore are housed 
in a single braided shield, and both lines are terminated on the HDI.  
There is no observable noise from the clock system.

The second area of concern was the grounding of the beryllium plates.  
There are two important grounds for these plates.  The first is the 
connection to the HDI.  If this ground connection is not properly made, 
the beryllium will act as a floating capacitor, and any voltage induced 
on the plate will spread over the entire plate and then capacitively 
couple into the readout channels.  This ground is achieved by bonding 
the ground plane on the back of the HDI directly to the beryllium with 
conductive epoxy. The other concern was the double sided detectors. 
These modules have beryllium plates on each side of the sensor, so they 
form a parallel plate capacitor with the sensor in the center. Any ground 
loop through this capacitor will induce noise into the sensor. The original 
solution was to use silver epoxy to form a conducting bridge between the 
two pieces of beryllium.  There is no direct contact in this case, so the 
connection is dependent on the silver epoxy. Some of these bonds became 
resistive so all of the modules were repaired by wiring the pieces together. 
The wire attachments to the beryllium were made with indium solder. 

The final issue was ground loops through the SMT. The SMT is read out from 
both ends, and each end is locally grounded. A potential difference might 
occur between the two ends and current will then flow in any path connecting 
one end of the detector to the other. If this current flows through the 
signal path, the detector will see this current as noise. Therefore an 
electrical break was made at $z=0$ for both the detector structure and the 
cooling path. The cooling pipes were also electrically isolated by plastic 
tubes at both entrances to the SMT detector. 

The common ground point for each half of the SMT was achieved by
an aluminum ring which was constructed on the face of the calorimeter 
with all the adapter cards mounted on it. The ring was electrically 
isolated from the calorimeter and the adapter cards were all grounded 
to the ring.  The ground connection to the SMT is made through the parallel 
combination of all the low mass cables.   

\subsection{SVXIIe Chip}

The SVXIIe chip~\cite{svx2} is a 128 channel, full custom, mixed 
analog and digital integrated circuit. The main  parameters are 
listed in Table~\ref{svx}.  

\begin{table}[h]
\begin{center}
\caption{SVXIIe properties.}
\label{svx}
\vspace{0.6cm}
\begin{tabular}{|l|l|}
\hline
\bf {SVXIIe Property} & \bf {Value} \\
\hline 
Process technology & 1.2 micron CMOS \\ \hline
Chip size & 6.3 by 8.7 mm \\ \hline
Number of channels & 128\\ \hline
Maximum interaction rate & 132 ns between interactions \\ \hline
Detector input  capacitance & 10 to 35 pF  \\  \hline
Number of bits in ADC & variable up to 8 bits \\ \hline
Number of pipeline stages & 32 \\ \hline
Preamp dynamic range & 240 pC \\ \hline
Wilkenson ADC clock frequency & Maximum of 53 MHz \\ \hline
Readout clock frequency & Maximum of 26.5 MHz \\  \hline
On chip zero suppression & Yes \\ \hline
On chip test pulser & Yes \\  \hline
Daisy chained operation & Yes \\  \hline
Simultaneous trigger and readout & No  \\ \hline
\end{tabular}
\end{center}
\vspace{0.6cm}
\end{table}

\begin{figure}
\centerline{\psfig{figure=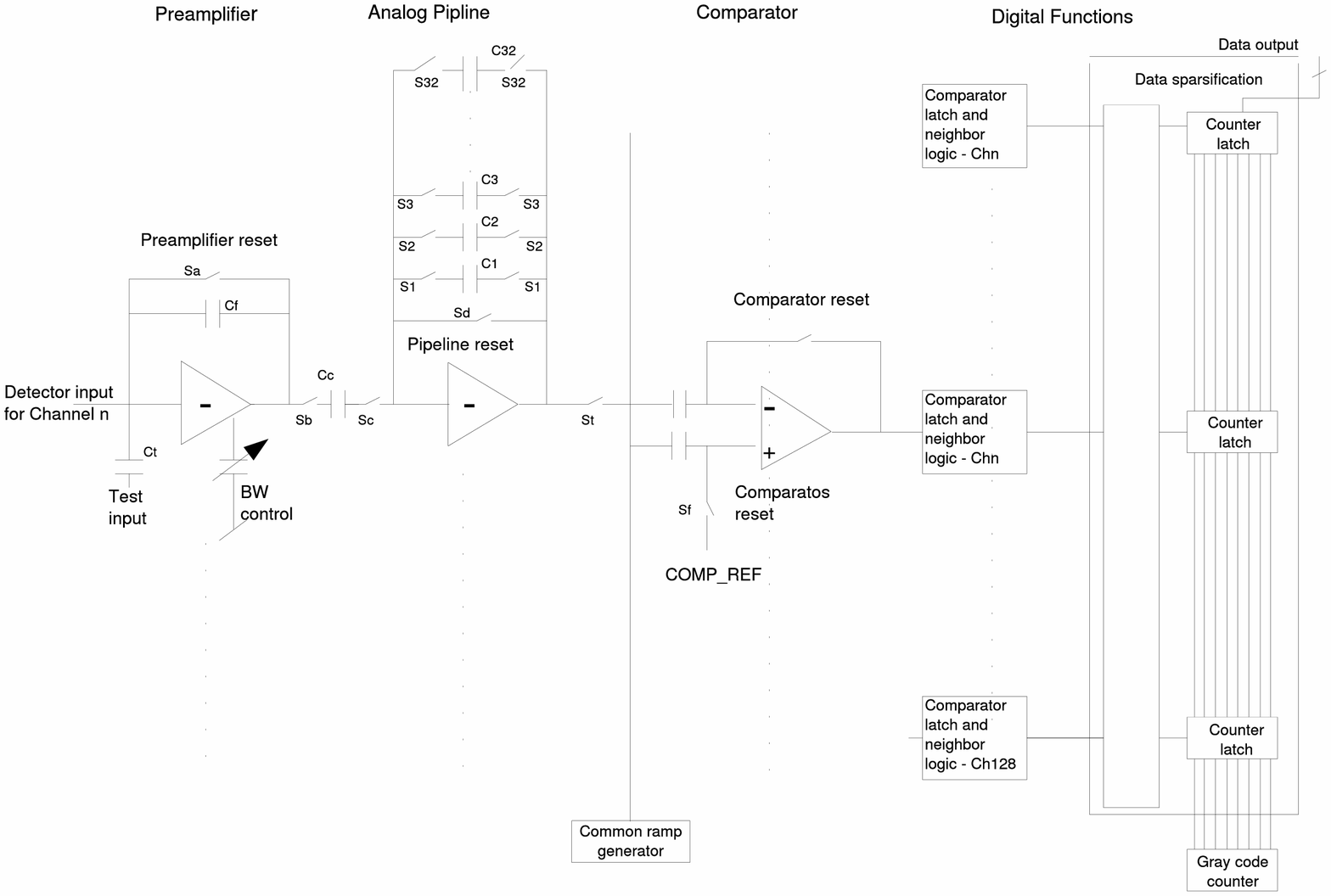,height=5.0in}}
\caption{A block diagram of a single channel of the SVXIIe chip.}
\label{f:block-chip}
\end{figure}

The chip was designed by a collaboration of engineers at Fermilab and 
Lawrence Berkeley Laboratory, and it was fabricated in the UTMC radiation 
hard 1.2\,$\mu$m CMOS technology. Each channel consists of an integrating 
preamplifier which converts charge to voltage, a 32 deep pipeline which 
is implemented as a string of 32 capacitors, and an 8 bit Wilkensen type 
ADC. A simplified diagram of one of the channels is shown in 
Fig.~\ref{f:block-chip}. All 128 channels are connected to a data readout 
section which can either read out all channels, just those above a digitally 
set threshold, or those above threshold plus one channel on either side of 
those above threshold. The latter mode can cross chip boundaries to 
adjacent chips mounted on the same HDI. The integrator integrates charge 
continuously until it is reset, but the charge into each pipeline capacitor 
is only the increment in charge between beam crossings. This means that 
there is a correlated double sampler between the integrator and the 
pipeline, so that only the change in charge between the previous beam 
crossing and the current beam crossing is put into a pipeline capacitor.  
The pipeline is controlled by two counters which are configured as
ring counters which means that the pipeline is a ring of capacitors. There 
is a write counter which points to the next available capacitor in the ring. 
When a new beam crossing occurs, the value of the input signal is written 
into the capacitor pointed to by the write counter. The read counter is 
identical to the write counter, but it is delayed by a down-loadable 
pipeline delay. The beam crossing clock that samples the charge also steps 
the write pointer and the read pointer ring counters. If there is a trigger,
the counters are stopped and the capacitor pointed to by the read counter is 
digitized and read out. 

The SVXIIe was designed to minimize both power consumption and detector 
mass.  It consumes approximately 3\,mW per channel and has only 19 
connections to the outside world. This small number of connections was 
achieved by making the function of many of the lines mode dependent.  
For example, the 8 bit readout bus is used for control functions in all 
modes but readout. 

\begin{figure}
\vskip 1.0in
\centerline{\psfig{figure=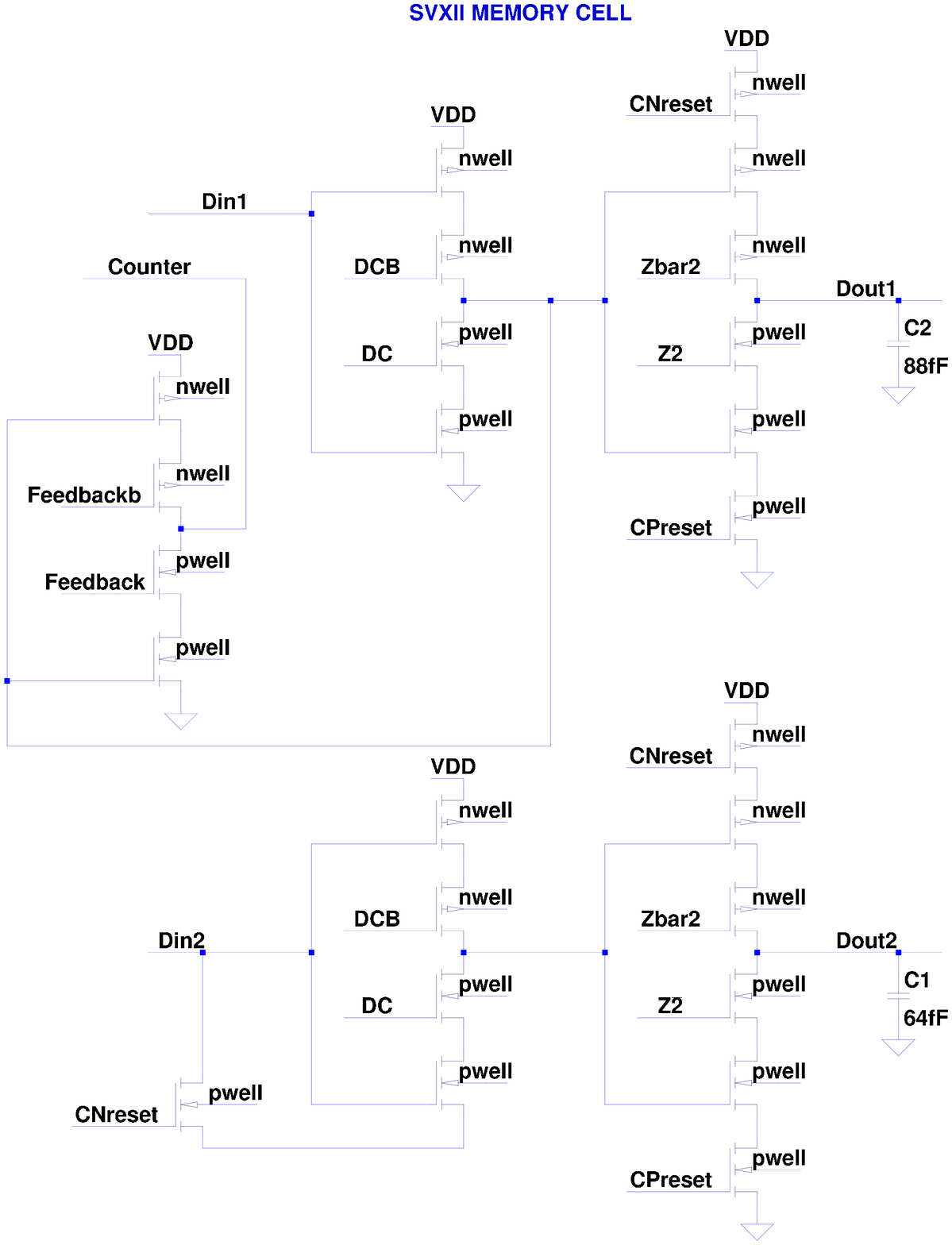,height=6.0in}}
\caption{A portion of the SVXIIe schematic showing the dynamic 
memory cell.}
\label{f:channel}
\end{figure}

A portion of the SVXIIe schematic containing the dynamic 
memory cell is shown in Fig.~\ref{f:channel}. 

The chip has four modes of operation with only one mode active at a time.  
This means that it is not possible for the chip to acquire data at the 
same time it is digitizing or reading out other data.  

The first mode is the download mode where settable parameters are loaded 
into the chip.  There are 190 bits in the download register for each chip.  
The download is implemented as a long serial shift register using only two 
chip pads called top neighbor and bottom neighbor and the clock.  Data are 
shifted into the top neighbor pad and out of the bottom neighbor pad.  
The output from the last chip's bottom neighbor is read back into the 
sequencer.   Thus, a three chip HDI requires sending 570 bits over the 
MIL-STD-1553 link~\cite{1553}. The chip does not have a read command to 
read the data back. The data shift through the chip string two or more 
times and the sequencer compares the read-back data to the input.

The second mode is acquire. In this mode, every time a clock pulse arrives, 
the chip samples the change in charge on the input capacitor and stores 
the data into a pipeline capacitor.  The ring counter pointers are incremented 
and the process is repeated.  There is no specific trigger line.  
When a trigger occurs, the sequencer stops the beam crossing clock 
and changes the mode to digitize.  The correct capacitor to digitize 
is pointed to by the read counter.

Digitization is accomplished by transferring the charge from the 
selected capacitor onto a capacitor connected to a comparator. The 
pipeline capacitor is then reset and the offset voltage is subtracted 
from the charge on the comparator capacitor.  This minimizes any offsets 
that may be present in the pipeline.  Actual digitization is accomplished 
by putting a linear voltage ramp on the other input to the comparator 
and starting a counter at the same time. The 53\,MHz digitization clock 
increments this counter on both edges of the clock pulse giving an 
effective clock rate of 106\,MHz.When the ramp voltage reaches the input 
value, the counter is stopped. 

The final mode is readout.  The counter values are shifted into a 
pipeline and the pipeline is compressed if zero suppression is enabled.  
Since the chip is designed to be used in a multi-chip configuration, the 
readout bus is configured as a tri-state bus during readout. Control of 
the bus is passed between chips through the top neighbor and the bottom 
neighbor connections.  A signal on the top neighbor during readout gives 
the chip control of the output bus.  When it is finished it signals 
the bottom neighbor to give control to the next chip or, if it is the last 
chip, to tell the sequencer that readout is done.  

Once the pipeline has been compressed and the chip has control of the 
output bus, the data is read out, one byte at a time. A 7 bit channel 
ID and an 8 bit data word are read out on each cycle. The clock is run 
at one-half of the normal 53\,MHz, and data is read out on both edges of 
the clock. This reduces the bandwidth requirement on the readout cable 
but it places severe restrictions on the clock duty cycle. Small changes 
in the clock duty cycle can cause one edge of the data strobe to be missed 
which then interchanges the channel ID and data for the rest of the 
readout. In sparsification mode, when only channels over 
a threshold are read out, this error is not easily detected.

The SVXIIe has a sophisticated method for testing itself and the 
entire downstream readout electronics.  The test circuit, called 
cal\_inject, consists of a capacitor connected by a switch to each 
of the front end channels at the input to the preamplifier. The switch 
is activated through a transition on bus line 7 in acquire mode. The 128 
bit cal\_inject mask (1 bit per channel) is a down-loadable parameter 
so all possible patterns can be generated. The pulse amplitude can either 
be set by an external source or from an internal one that is controlled 
by a 3 bit DAC.  The DAC setting is also a down-loadable parameter.

All of the power connections are at the back of the chip, including the power 
for the front end amplifiers.  The chip's metal layers have too-high  
resistance for the front end amplifiers to get enough power. A row of double 
wire bonds was made along exposed sections of the upper metal layer 
connections for front end amplifier power to reduce this resistence by 
creating parallel paths for current flow. The spacing of the exposed
sections kept the wire bond distances reliably short.

Even though the average chip power consumption is low, the current varies
by nearly a factor of two between an idle state and readout, so if there 
is no local monitoring of the voltage by the power supply, often referred 
to as remote sensing, there could be a substantial change in chip voltage.  
The layout of the system makes remote sensing quite difficult. One supply 
feeds a large number of HDIs. Two HDIs are fed through the 80-conductor 
cables. These cables have many different lengths since they had to be as 
short as possible in order to fit in the space between the central and 
end cryostats. Finally, the number of chips varies between HDIs. The best 
option was to locate the power supplies as close to the detector as 
possible. The place with adequate space was at the base of the central 
calorimeter near the interface boards.  The local magnetic field is about 
200\,G at this location, so the power supplies had to have magnetic 
shielding.  In  addition, safe practice required fusing each power line 
for each of the three voltages for groups of four strings of SVXIIe chips.  
This required a fuse panel with about 40 fuses at four locations at the 
base of the calorimeter.

\subsection{High Density Interconnect}

The HDI is a two-layer flexible printed circuit with 50\,$\mu$m 
polyimide with 125$\mu $m line spacing and 50\,$\mu$m vias. The circuits 
are laminated to beryllium heat spreaders. One HDI and its beryllium 
substrate are epoxied directly to one silicon sensor. Precision notches 
machined into the heat spreaders provided the alignment of the silicon 
ladders on the bulkhead. The HDI  consists of SVXIIe chips, resistors, 
capacitors, and an interface circuit, which are mounted in the area which 
are in direct contact with the silicon sensor. The SVXIIe chips are wire 
bonded directly to the sensors. The second part of the HDI, the pigtail, 
is the extension of the signal and power traces to the outer radius of 
the silicon detector region.

\begin{figure}
\vskip 1.0in
\centerline{\psfig{figure=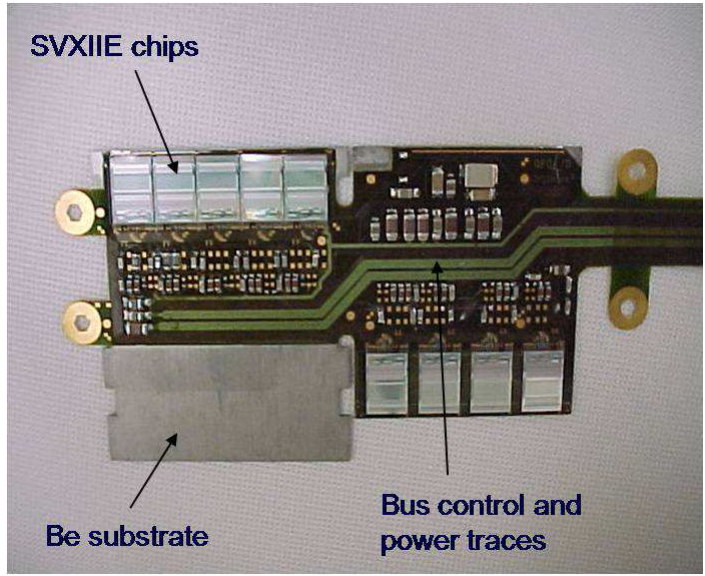,height=3.0in}}
\caption{An HDI for a 9-chip detector (prior to folding and lamination).}
\label{f:hdi}
\end{figure}

A picture of an HDI for a 9-chip detector is shown in Fig.~\ref{f:hdi}.  

Multiple layer polyimide circuits with very fine spacings are difficult 
to make because the polyimide stretches easily. A ceramic HDI would require 
several layers to provide the required number of signal, power, and ground 
planes.  The small number of connections on the SVXIIe and clever use of 
wire bonds made it possible to make a polyimide HDI with only two layers.  
The output bus and control lines were laid out in a row of parallel lines in 
front of the string of chips. Connections from each chip to the different 
lines were made by running the wire bond over the intervening traces. 
Bypass capacitors were mounted outside the bus lines and longer wire bonds 
were used to connect to the capacitors.  Thus, the equivalent of a 
multilayer circuit was made with wire bonds.  The longest wire bonds 
approach the maximum allowable length so if there were many more lines 
to connect, this method would not have worked.
 
Six different types of HDIs were needed to accommodate the various detector 
and readout geometries. In the case of double-sided ladder detectors, a 
single circuit contained the readout for both the p- and n-sides with the 
polyimide folded over to sandwich the sensor. The double-sided F-wedges 
have one separate HDI on each side. A single HDI would have implied 
daisy-chaining 14 readout chips, which was considered too risky in case of 
a chip failure and too expensive in terms of readout time. The barrel 
detectors employ 9-chip HDIs for the double-sided ladders, 6-chip HDIs for 
the 90-degree stereo ladders, and 3-chip HDIs for the single sided ladders. 
The F-wedges are read out by an 8-chip HDI for the p-side, and a 6-chip HDI 
for the n-side. Each single H-wedge is read out with one 6-chip HDI. 
 
All HDIs have the same functionality. They provide one 8-bit data 
bus connecting all of the SVXIIe chips to the data path interface 
circuit, distribute the differential input clock and single-ended 
mode control signals to all the SVXIIe chips, distribute the various 
analog voltages required by the SVXIIe chips, and silicon bias,
and provide monitoring of the temperature of the beryllium heat spreader. 

\subsection{Low Mass Cables and Adapter Card}

The HDI readout cables are routed between the ladders in the barrel to a 
larger radius outside of the barrel vertex volume. There the pigtails 
are trimmed to length and attached to the low-mass cables. The low-mass 
cables are flex-circuit striplines, to minimize the amount of material 
in the sensitive volume. They were fabricated by Honeywell~\cite{honey}  
and carry the digital signals and power between the hybrids and the adapter 
cards. To minimize any deterioration of the clock signals, the clock signals 
are brought to the hybrids through mini-coaxial cables. Signal traces and 
broad power and ground traces are located on both sides of a polyimide 
dielectric about 100\,$\mu$m thick. The low-mass cables are routed along the 
half-cylinder and coupled to the 80-conductor pleated foil cables at a 
ring of adapter cards located between the calorimeter cryostats. Each 
adapter card takes the signals from two HDIs and, without further signal 
processing, launches them on the high quality 80-conductor pleated foil 
cable to the interface board. The clock signals are carried by separate 
coaxial cables attached to the 80-conductor cable.     

\subsection{Interface Board}

The interface board serves as a repeater for the signals between the 
SVXIIe and the sequencer. Each board handles up to eight detector modules.
It also provides the distribution of the bias voltage to the sensors, allows 
remote control (on/off) of the power for the  SVXIIe chips, and provides 
monitoring  of the chip voltages, and currents.

The interface boards are housed in eight 9U$\times$280\,mm crates with custom 
backplanes, using 16--18 slots per crate. Two crates are placed at each of 
four locations at the base of the central calorimeter.

Temperatures of the beryllium heat spreaders are monitored on the interface 
board. A jumper on the interface board provides a selectable temperature trip 
at $40^\circ$C or $65^\circ$C. The circuit is designed not to trip due to an 
open or short circuit. Calibration is adjusted by a trim resistor and is 
accurate to about $1^\circ$C. 

The current limiters circuits for the HDI power connections were designed to 
trip at 0.7\,A in order to protect the wire bonds. Response time is a few ms. 
The time constant for integrated signals (output of the current monitor) 
is of order 100\,ms to avoid trips on glitches. Each HDI power line is 
monitored, in total 24 current limiter circuits per board.

The high voltage (HV) control for the detector bias is provided by the 
HDI enable circuit and photovoltaic relays (no moving parts) on the 
interface board. There are two HV lines provided for each HDI to allow 
operation of double-sided sensors in split mode. HV is supplied via a 
34-conductor twist-and-flat strip cable to a connector on the backplane.
For single-sided detectors, the primary bias voltage is grounded and 
the secondary bias has a positive voltage. For double-sided detectors 
the primary bias is at +HV/2 or less and the secondary bias at $-$HV/2 
or greater. This allows for asymmetric breakdown characteristics.
The ground reference is the HDI ground. Bias current from groups of four 
HDIs can be measured via the MIL-STD-1553 interface at any time.

\subsection{Sequencer Controller}
\label{seqc}

The sequencer controller (SEQC) is a 9U$\times$340\,mm Eurostyle module 
residing in each sequencer crate, used to give coarse timing commands to 
the SVXIIe sequencers which control acquisition and readout of the SVXIIe 
chips.  It is designed to be plugged into slot 1 of the custom sequencer 
backplanes.

The controller receives beam crossing timing and triggering 
information via a single coaxial cable called the Serial Command Link 
which originates at the Trigger Framework\footnote{The Trigger Framework 
gathers digital information from each of the specific Level~1 trigger 
devices and chooses whether a particular event is to be accepted for further 
examination.}.  Information from this link is interpreted and transformed 
into general SVXIIe data acquisition commands that go to each sequencer 
via the backplane.  These commands are encoded onto a serial link called 
the NRZ (non-return-to-zero) link which is a 7-bit repetitive code sent on 
dedicated backplane lines to each sequencer slot. These lines are used in 
conjunction with a dedicated differential 53\,MHz clock sent to each 
sequencer slot. The seven bits are the framing bit, the crossing bit, 
four bits for an encoded data command, and the parity bit. The seven bit 
structure was chosen to remain synchronous with the Tevatron beam structure.  
A new code packet is sent every 132\,ns. The framing bit is always set high 
and is monitored by a state machine to ensure that the link is synchronized. 
The crossing bit is only set if there is beam for that particular 132\,ns 
period.  The sixteen possible encoded data commands include commands such 
as Acquire, Preamp Reset, Level~1 Accept, Digitize, Readout, and Readout 
Abort.   

The NRZ link has an adjustable delay system to synchronize the crossing 
command with actual beam crossings so that SVXIIe charge acquisition 
happens at the optimal time with respect to beam.  There are three 
delays, 132\,ns, 18.8\,ns, and 2\,ns steps, writable from a MIL-STD-1553 
serial communication link. This MIL-STD-1553 link is also used to set various 
diagnostic modes.

To prevent missed triggers, the SEQC sends a Busy signal back to the Trigger 
Framework based on a logical OR of busy signals from the sequencers. This is 
necessary because the SVXIIe chips may operate in sparsification mode, 
reading out only the channels over threshold, and readout may end at various 
times after a trigger.  Since there is only one buffer on the SVXIIe, a new 
trigger may be accepted only after readout has finished.

A Diagnostic mode is provided to exercise the data acquisition system if 
the Trigger Framework signals are not operational.  This mode is exercised 
via MIL-STD-1553 link.

\subsection{SVXIIe Sequencer}
\label{seq}

The basic task of the SVXIIe sequencers is to coordinate data acquisition in 
the SVXIIe chips and serialize the resulting data onto high speed fibers 
to be sent to data storage.

The sequencers are 9U$\times$340\,mm circuit boards that reside in slots 2 
through 21 in six Eurocard crates on the detector platform. Geographic 
addressing is designed into the backplane for each slot for MIL-STD-1553 
Remote Terminal identification. Each sequencer is connected to an interface 
board via four 50-conductor 3M pleated-foil cables, a VRB via four optical 
fibers for data readout, the SEQC via the backplane, and the control system 
via the MIL-STD-1553 link which is also plugged into the backplane.

In initialize mode, the sequencers interpret data from the MIL-STD-1553
data bus and then clock the appropriate download data pattern into the chips.  
This pattern is readable from the chips only by shifting new data into
the chips so a simultaneous download of all chips on an HDI must occur.  
In acquire mode, the sequencers advance the pipeline clock with each beam 
crossing, and the SVXIIe chips' preamplifiers are reset during beam gaps.  
When a trigger occurs, a specific complex manipulation of the control 
signals occurs which extracts charge out of the correct pipeline cell, 
and then the sequencer sends a 53\,MHz clock which the chip uses to 
digitize this charge for each channel. In readout mode, clocks are sent 
to the SVXIIe chip at 26.5\,MHz, and the chip then sends alternately 
channel ID and data back to the sequencer.  The sequencer serializes this 
data into a 1.062\,Gb/s data stream, adds header and trailer information, 
and sends it via optical fiber to the VRBs in the counting house.

Diagnostic features are interfaced to the MIL-STD-1553 bus and include a 
snapshot register to read the current state of important SVXIIe control 
lines and a built-in logic analyzer that records the same control 
lines for about 75\,$\mu$s after a selectable trigger. A pattern of 64 
words may be written via the MIL-STD-1553 link and sent to the VRBs 
for testing the gigabit links. The Finisar laser ~\cite{finisar} 
drivers' diagnostic links may be read via MIL-STD-1553, monitoring power 
output, temperature, and other parameters. Other MIL-STD-1553 link registers 
include SVXIIe chip power on/off control, module serial number, and a remote 
programming register.

An optional Readout Abort feature is used to guard against system hang-ups 
from non-responding chips.  Normally the sequencers use the 
Priority\_Out handshake signal from the last chip in a chain to 
determine when readout is finished.  If this handshake ever fails, 
a 45\,$\mu$s timeout in the SEQC propagates to the sequencers 
and puts the system back in acquire mode, and Busy is released.

\subsection{VME Readout Buffer}

The VME Readout Buffer (VRB) is a 9U$\times$400\,mm multiport memory 
that buffers the data for transfer to the higher level data 
acquisition system. It contains ten independent input ports and a 
common VME output port. For SMT operation, eight channels on each board 
are used.

The VRB acts as a buffer as data wait for the conclusion from
the second decision level (Level~2) of the hardware trigger 
system~\cite{d0nim}. Buffer management is provided by an external System 
Controller (VRBC) through a dedicated control port. A controller serves 
up to 12 VRB modules in one VME crate.

The VRB receives data via serial optical connections from the sequencers 
via the VTM which converts the optical signal to an electrical signal.
Input data is accepted at an aggregate rate of approximately 500\,MBytes/s 
on eight (byte-wide) channels. The output rate is limited by VME transfer 
speeds and by the number of VRB modules sharing the VME bus. For SVXIIe chip 
applications, the VRB inputs data at the accept rate of a few kHz from the 
first decision level (Level~1) of the hardware trigger, and outputs data 
at the Level~2 accept rate of less than 1\,kHz.

The VRB Control Logic performs three basic functions: receive and process 
messages from the system controller, return status signals to the system 
controller, and manage the general flow of data to and from the VRB buffers.
The Control Logic will respond to a message by asserting a status signal, 
typically within 200\,ns.

Upon a Level~1 trigger accept, the VRBC supplies the VRB with a buffer 
number to store the input data. When the VRB Control Logic receives a 
message specifying the next input buffer, it looks up the buffer starting 
address in the shared memory, and broadcasts this information to the 
Receive Logic for all channels. The input data is pushed to the VRB from 
the sequencer. The event data from an SVXIIe chip is logically organized 
by words (2 bytes). The first bytes of each data stream contain a header 
inserted by the sequencer to identify the data source. This is followed by 
a block of data from each SVXIIe chip containing the chip ID, status, and 
up to 128 pairs of channel number and data. The VRB input FIFO (First In 
First Out memeory configuration) will accept data until it recognizes an 
End Of Record word. When all event data is received, the VRB Control Logic 
will read the individual byte counts from each Receive Logic block and 
generate a global byte count for the event. The global byte count is 
available to a VME Scan Processor so that it can perform a single block 
read operation to obtain all data for the event. When all the channels 
transmitting data to a VRB are done, the VRB will inform the VRBC.
This is accomplished by releasing a busy signal on a dedicated line of 
control bus. The busy line is an open collector signal that can be driven 
low by any of the VRBs in the crate. The transfer of the data from the 
SVXIIe to the Level~2 buffer is finished when all VRBs have released 
signal on this line.

Following a Level~2 Accept, the VRBC supplies the VRB with a buffer number 
for data output. Events that are accepted by the Level~2 trigger are 
copied to the VME output data FIFO. For events rejected by the Level~2 
trigger, the VRBC re-uses the buffer number, causing the previous event 
data to be overwritten.
  
\newpage
\pagebreak

\section{Silicon Sensors}\label{s:sensors}  
The SMT detector uses a combination of single-sided, double-sided, and 
double-sided double-metal silicon sensors. 

\begin{table}[b]
\caption{Characteristics and deployment of various sensor types. 
The length of the inner H-disk sensor is indicated with {\it i} and the
outer with {\it o}.}
\vspace{0.6cm}
\begin{center}
\begin{tabular}{ |c|c|c|c|c|c|c|c|} \hline
\bf {Module} & \bf{Type} & \bf {Pitch}        
& \bf {Length} & \bf {Inner}  & \bf {Outer}  & \bf {Manufacturer} \\
       &     &   \bf {p-/n-side}  &  & \bf {Radius} & 
       \bf {Radius} &             \\ 
       &       &  \bf {[$\mu$m]}   &  \bf {[cm]} & \bf {[cm]} 
       & \bf {[cm]} &             \\ 
\hline 

F-disk & DS  &     50/62.5     &   7.93    & 2.57   & 9.96   
& Micron Eurisys \\
\hline 

H-disk  & SS  &   40         & 7.63$^i$,6.39$^o$  & 9.61    
& 23.6   &  Elma    \\   
         &     &   80 readout &                   & & & \\
\hline 

Central & DSDM &  50/153.5     &  12.0    & 2.72   & 7.58    &  Micron  \\
Barrels & DS   &  50/62.5    &  6.0     & 4.55   & 10.51  &   Micron       \\
\hline 

Outer  &  SS   &   50        &  6.0     & 2.715  & 7.582  &  Micron  \\
Barrels & DS   &   50/62.5   &  6.0     & 4.55   & 10.51  &  Micron  \\
\hline 

\end{tabular}      
\end{center}
\label{tab:sensors}
\vspace{0.6cm}
\end{table}

Single-sided and double-sided devices were produced from high resistivity 
4" silicon wafers with crystal orientation $<111>$ and $<100>$.  The 
double-sided double-metal detectors were manufactured using $<100>$ 6" 
wafers. Isolation on the n-side of all double-sided devices is provided 
by p-stop implants. AC coupling aluminization was specified to be 
2\,$\mu$m inside the strip to limit edge fields which might cause 
micro-discharge breakdown.  All traces are biased using poly-silicon 
resistors. 

\begin{figure}
\begin{center}
    \epsfxsize=10.0cm
    \epsfbox{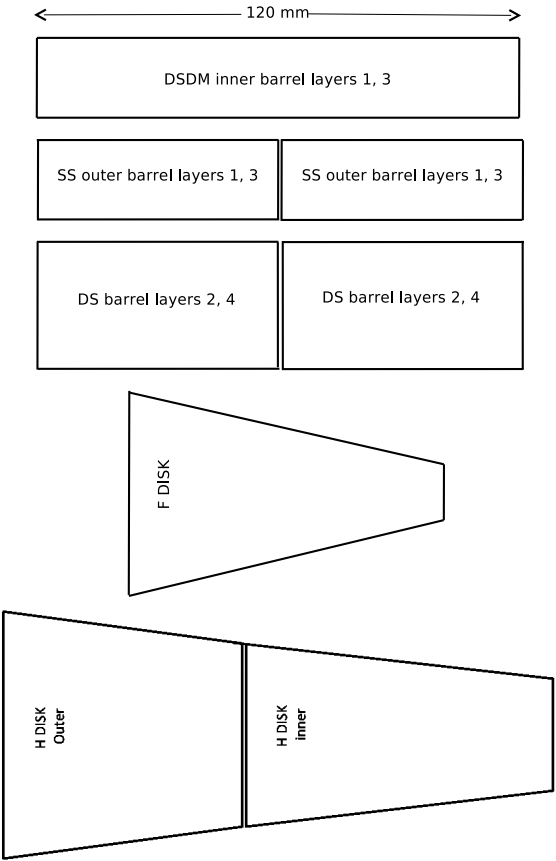}
    \caption{The geometry of all individual detectors showing their
             relative dimensions.}
    \label{fig:sens1}
\end{center}
\end{figure}

Of particular importance for the double-sided devices is the quality and 
robustness of the AC coupling capacitors, since these are required to isolate 
the readout electronics from the detector depletion voltage. To minimize 
the stress on these capacitors, the detectors are in general operated with 
the bias split between the n- and p-sides. The voltage ratio is limited by 
the micro-discharge effect discussed in Sec.~\ref{sec:micro}.
Table~\ref{tab:sensors} summarizes the sensor locations and types, and   
Fig.~\ref{fig:sens1} summarizes the geometry of the individual detectors.

\subsection{Barrel Sensors}
Three types of sensors are used in the central barrels.  The second 
and fourth layers use double-sided stereo detectors with the n-side 
implants at a $2^\circ$ angle with respect to the p-side axial strips.  
Two of these sensors are bonded together to form a 12\,cm long ladder.

On the first and third layers of the outer barrels, single-sided sensors 
with axial strips are used. Also in this case two sensors are bonded 
together to form a 12\,cm long ladder.  The inner four barrels use $90^\circ$ 
stereo sensors.  Ninety degree readout was achieved using a second metal layer 
on the n-side insulated from the first metal layer by 3\,$\mu$m of PECVD 
(Plasma Enhanced Chemical Vapor Deposition) silicon oxide. Two readout strips 
on this side are multiplexed to a single readout channel. Implants on the 
n-side are isolated by individual p-stop frames in addition to a common 
p-stop enclosure. The use of 6'' silicon wafers allowed production of 
12\,cm long DSDM sensors as single pieces.

\subsection{Disk Sensors}
The disk sensors are trapezoids with readout strips arranged parallel to the 
long edge of the devices.
In this arrangement the strip length varies for strips which originate 
past the base of the trapezoid.

The F-disk detectors provide $\pm 30^\circ$ stereo point measurements which 
provide good impact parameter resolution in both $r-\phi$ and $r-z$.

For the H-disks, two  detectors are bonded together to form the readout module.
The strips of the inner and outer sensors are bonded together, leading to 
a maximum strip length of 14.24\,cm.  The single-sided modules are assembled 
back-to-back on a beryllium substrate forming a 15$^\circ$ stereo angle.

In both detector types, a small region in the upper corner of the 
device, where strip lengths would be less than about 1 cm, is left unbonded.

\subsection {Sensor Production and Testing}
The sensors were produced by three different vendors, Micron 
Semiconductor LTD~\cite{micron}, Elma~\cite{elma}, and 
Canberra Eurisys Mesures~\cite{eurisys}. Those produced by Micron were 
tested at the Micron factory by \D0 personnel; Elma sensors were tested 
at Moscow State University; and Eurisys sensors were tested by the company. 
The bulk of the sensors used in the experiment were produced by Micron and 
the testing of these sensors is described in more detail below. 
Table \ref{tab:sen2} summarizes the specifications for the Micron devices.

\begin{table}[h]
\caption{Sensor specifications for the double-sided sensors 
fabricated by Micron.}
\vspace{0.6cm}
\begin{center}
\begin{tabular}{ |c|c|c|} \hline

\bf {Parameter}   &  \bf {F-disks and DS sensors}  &  \bf {DSDM sensors} \\
\hline

$U_\mathrm{depletion}$(V)  &  $ 20 < U_\mathrm{depletion} < 60$ & $20  
< U_\mathrm{depletion} < 60 $ \\
\hline
Total $I_\mathrm{leakage}$ @ $U_\mathrm{depletion}+20$V & $ < 10 \mu$A
    &  $ < 10 \mu$A   \\
\hline
$U_\mathrm{breakdown}$ defined as $I=15 \mu$A & $ > 100$  V  &$  > 100$ V \\
\hline
$C_\mathrm{coupling}$  &  $ > 15 $ pF/cm  & $ > 15 $ pF/cm  \\
\hline
p-side $C_\mathrm{coupling}$ failures & $ < 2 \% $ /side & $  < 2 \% $ 
/side  \\
\hline
n-side $C_\mathrm{coupling}$ failures & $ < 2 \% $ /side & $ < 4 \% $ /side  \\
\hline
$R_\mathrm{bias}$  &  1 M$\Omega < R_\mathrm{bias}$ 
$ < 10$ M$\Omega$  &   1 M$\Omega < R_\mathrm{bias}$ $ < 10$ M$\Omega$\\
\hline
$R_\mathrm{bias}$ uniformity & $ \pm 25\% $  & $ \pm 25\% $ \\
\hline 
$R_\mathrm{interstrip}$  &  $ > 100$ M$\Omega $ & $ > 100$ M$\Omega $ \\
\hline 

\end{tabular}      
\end{center}
\vspace{0.6cm}
\label{tab:sen2}
\end{table}

\begin{figure}
\begin{center}
    \epsfxsize=12.5cm
    \epsfbox{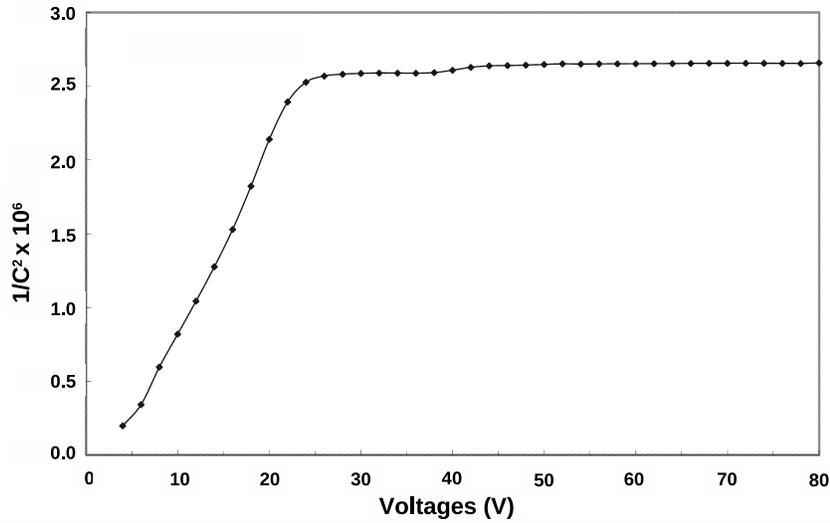}
    \caption{$1/C^{2}$ as a function of the bias voltage for a 9-chip sensor.}
    \label{fig:depl}
\end{center}
\end{figure}

One of the most important electrical characteristics of the silicon sensors 
is the depletion voltage. It was estimated by studying the capacitance-voltage
relationship measured at a dedicated station at Micron. A typical 
measurement is shown in Fig.~\ref{fig:depl} where the inverse of the 
capacitance squared is plotted as a function of bias voltage.The depletion 
voltage is given by the intersection of the plateau, seen in figure, with 
the $1/C^2$ slope.  

Other basic detector tests included:
\begin{itemize}
\item{measurement of IV characteristics and breakdown voltage}
\item{ measurement of AC coupling capacitor value}
\item{measurement of AC coupling capacitor leakage and 
pinholes\footnote{Broken AC coupling capacitors caused by 
flaws in the lithography could cause holes in the insulator.}  to 100\,V.}
\end{itemize}

\begin{figure}
\begin{center}
    \epsfxsize=10.0cm
    \epsfbox{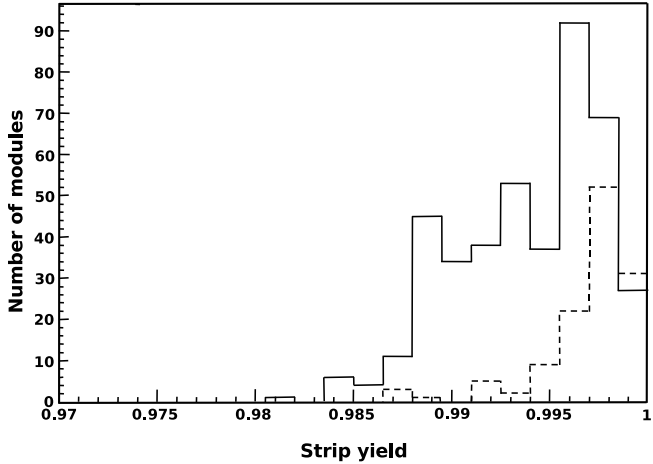}
    \caption{The yield of good strips for accepted detectors for 
     DS ladders (solid line) and F-wedges (dashed line).}
    \label{fig:sens2}
\end{center}
\end{figure}

Figure~\ref{fig:sens2} shows the yield of good strips for accepted 
sensors for DS ladders and F-wedges.  

Several problems were encountered during production which necessitated more 
detailed testing.  Several devices showed regions of low interstrip resistance.
This seemed to depend not only on device processing but also post-processing 
handling. These areas could be detected by current and resistance measurements
on the DC contacts of the individual strips, and these tests were instituted 
for all detectors. Several detectors exhibited high leakage currents after the 
AC coupling capacitor tests.  These detectors were cleaned and retested and 
accepted if the currents returned to previous levels. 

Early in the production phase of the double-metal detectors some sensors were 
found with regions which exhibited large noise and high strip currents. By 
visual inspection, it was found that these regions were associated with 
flaws in the island p-stops on the n-side. These were traced to areas where 
the p-stop isolation implants were contacting the n$^+$ strip-defining 
implants with no intervening bulk material. In-process testing of n-side 
strips at Micron was introduced to identify these problems at an early stage 
of the processing.  A strip current limit of 1\,$\mu$A at 80\,V was used to 
identify p-stop problems and no detectors were accepted with more than one 
such flaw. 

\subsection {Micro-discharge}\label{sec:micro}  
\begin{figure}
\begin{center}
    \epsfxsize=10.0cm
    \epsfbox{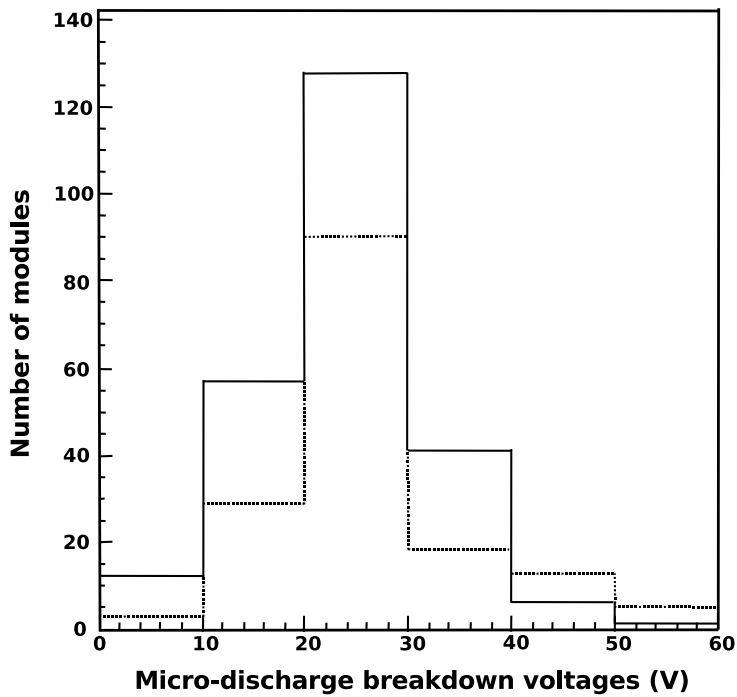}
    \caption{N-side micro-discharge breakdown threshold for barrel layer 
             DSDM detectors (solid line) and F-disk detectors 
             (dashed line).}
    \label{fig:sens3}
\end{center}
\end{figure}

During the initial production testing, a number of devices were found 
that exhibited breakdown which depended on the voltage applied to the 
p-side of the device.  The breakdown was due to ``micro-discharge''~\cite{md}:
avalanche breakdown of the p-n junction when the potential between the 
negatively biased p-implant and grounded AC pad increases the junction field. 
This was confirmed by measuring the temperature dependence of the associated 
current, which increases with decreasing temperature due to the 
increase in carrier mobility. The threshold for this breakdown 
varied from detector to detector and was worst for those devices which had 
obviously misaligned implants and AC coupling strips.  The threshold for the 
breakdown was affected by the amount of fixed positive oxide charge which 
increases the effective field near the p-n junction. The existence of this 
breakdown made it necessary to characterize each device by the maximum 
p-side voltage which could be sustained without breakdown, as illustrated 
in Fig.~\ref{fig:sens3}. The detectors were then biased 
asymmetrically depending on this voltage.

\begin{figure}
\begin{center}
    \epsfxsize=13.0cm
    \epsfbox{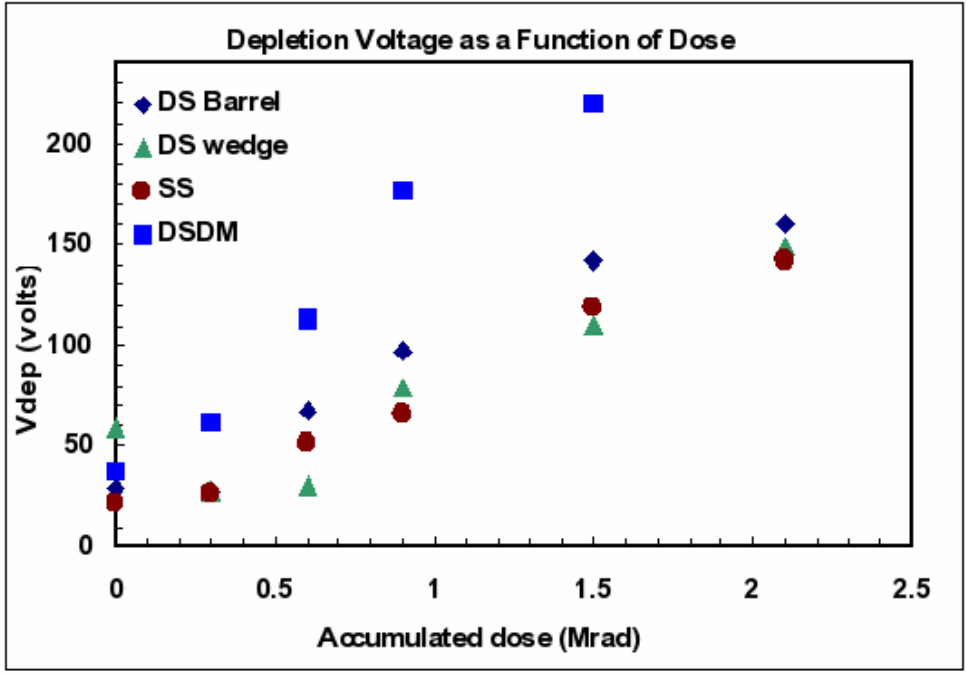}
    \caption{Depletion voltage as a function of dose.}
    \label{fig:sens4}
\end{center}
\end{figure}

Bare sensors were tested at the Fermilab booster irradiation facility 
(8\,GeV protons)  and the neutron irradiation facility at Lowell, MA, 
USA (one MeV neutrons). In addition, ladders were tested to 2\,Mrad in 
the Fermilab booster. In general the sensors behaved as expected from 
the Hamburg model~\cite{moll} of reverse annealing and with a current damage 
constant of  $\alpha = 3.3 \times 10^{17}$\,A/cm$^2$. The one exception to 
this behavior was the DSDM detectors, which exhibit anomalously rapid 
increase in depletion voltage with respect to the $2^\circ$ and disk 
detectors as can be seen in Fig.~\ref{fig:sens4}. Diode test structures 
from the same 6'' wafers were also irradiated in the Fermilab booster.
The behavior of these test structures was consistent with that of the 
detectors without double metalization, indicating that the anomalous 
behavior is likely associated with the PECVD layer utilized in the 
double metal process.

\begin{figure}
\begin{center}
    \epsfxsize=13.0cm
    \epsfbox{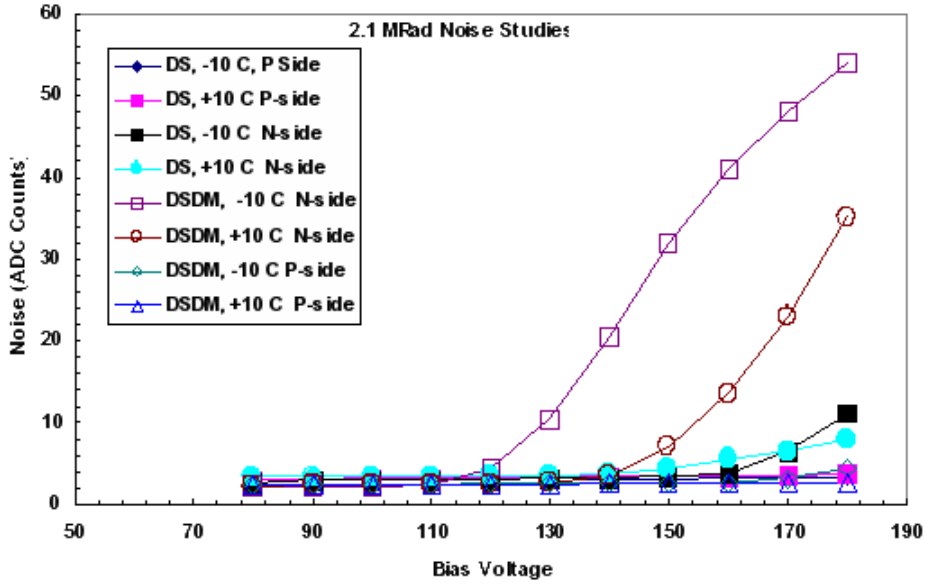}
    \caption{Noise as a function of bias voltage at different temperatures.}
    \label{fig:sens5}
\end{center}
\end{figure}

The onset of micro-discharge will likely limit the lifetime of the SMT. 
After type inversion, the breakdown moves with the junction from 
the p- to the n-side of the sensors. For a junction on the n-side, 
trapped positive charge in the oxide tends to reduce the field at the 
junction and the onset voltage of micro-discharge is consequently higher 
on the n-side of an irradiated device than on the p-side of the same 
detector before inversion. Figure~\ref{fig:sens5} shows the noise 
associated with the onset of micro-discharge after irradiation.
Micro-discharge will limit the practical operating voltage of the 
DSDM detectors to about 75 to 100\,V/side.

An assessment of the radiation damage with an estimate of the 
expected lifetime is described in Sec.10.
 
\newpage
\pagebreak

\section{Production of Ladders and Wedges}

The large scale production of silicon detectors was a challenge 
for the \D0 collaboration. A total of 1392 assemblies, which came in five 
different types, were required for building the SMT. In addition, the need 
for left and right- handed species of all the ladders also increased the 
complexity of the production work. This number excludes the up to 20\% spare 
modules which were built. The large number turned the module assembly into 
a mass production which only functioned smoothly because of well 
defined procedures for the assembly processes.

All ladder and wedge modules were produced at the Silicon Detector 
Facility (SiDet) at Fermilab. This facility provided large clean room 
areas equipped with optical Coordinate Measurement Machines (CMM)
and automatic bonding machines. 

\subsection{Production of 3-, 6-, and 9-chip Ladders}

Since the basic layouts of the barrel ladders are very similar, these 
various ladder assemblies exploited almost identical techniques. The 3-chip 
ladders consist of two single sided sensors with an on-board mounted 
HDI for three readout chips. Their production was a one-stage process 
and therefore rather straightforward, so this assembly will not be described  
here. The 6- and 9-chip ladders however, employed double-sided silicon which 
made the use of a two-stage production process with special fixturing for 
fold- and flip-over necessary. In the following sections, the successful 
two-stage production assembly is described. 

\subsubsection{The Generic Double-sided Ladder Layout of the SMT 
Barrel Detectors}

\begin{figure}
\begin{center}
\includegraphics[width=11cm]{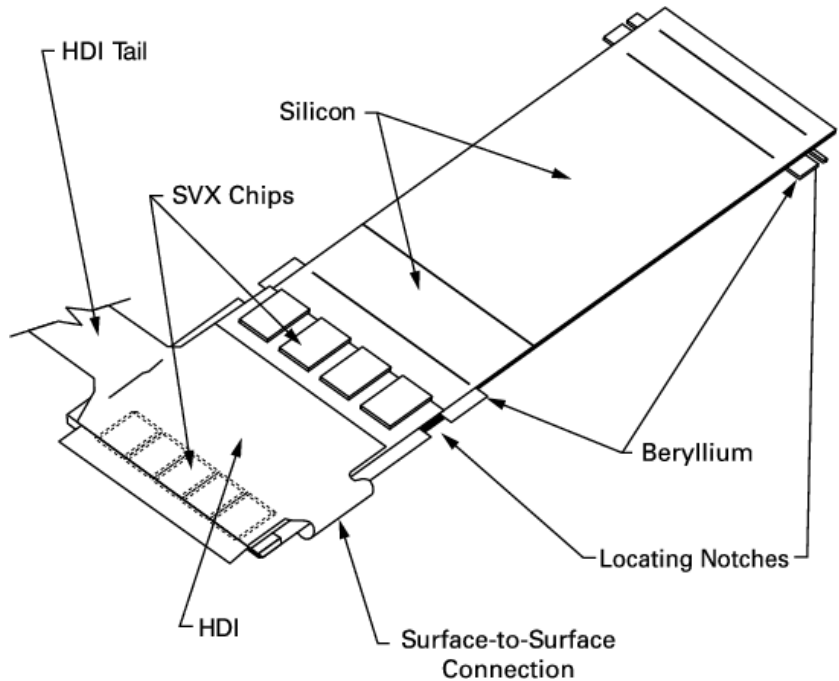}
\caption{Schematic design of a double-sided silicon ladder, isometric view.}
\label{ds_ladder_iso}
\end{center}
\end{figure}

\begin{figure}
\begin{center}
\includegraphics[width=11cm]{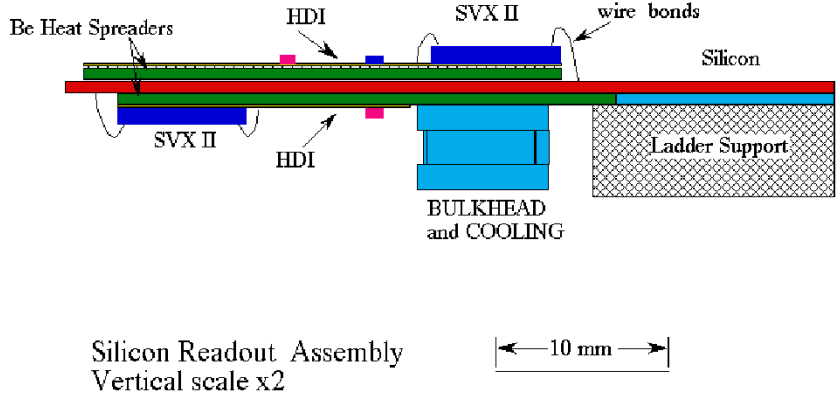}
\caption{Schematic design of a double-sided silicon ladder, elevation view.
         Note that the vertical size is not drawn to scale.}
\label{ds_ladder_elev}
\end{center}
\end{figure}

The basic design of the 9-chip double-sided ladders is shown in
Fig.~\ref{ds_ladder_iso}. The isometric drawing shows a 9-chip ladder with 
5 and 4 readout chips on the p- and n-side respectively. The two silicon 
sensors are ganged by wire-bonds and supported by carbon fiber rails. 
The HDI, laminated on the beryllium pieces prior to ladder assembly, was
wrapped around the two surfaces of the silicon and connected to the silicon
strips on both sides of the sensors. Figure~\ref{ds_ladder_elev} shows an 
elevation view of the on-board mounted readout assembly. 

\subsubsection{The Assembly Components and Assembly Preparations}

Sensors used for ladder production were selected according to the results 
from the measurements described in Sec.~5. In the case of 3- and 9-chip 
ladders, which consist of two sensors, the depletion voltage of the 
individual sensors was matched. 

The selected sensors were visually inspected to ensure a clean surface on 
the bonding pads prior to the assembly process. Spots of organic residue 
on sensors, in particular on the AC-bonding pads, caused a few unreliable 
bonds. 

Beryllium was chosen as the heat spreader material due to its excellent 
thermal and electrical properties, its high specific stiffness, and its 
very long radiation length. The beryllium pieces used in the ladder design 
were machined into their final shapes by the supplier~\cite{brush}, so that 
no further machining was necessary at Fermilab.

Three different beryllium substrates were used in the assembly process of 
the ladders. The ``active cooled'' and ``passive'' beryllium pieces were 
positioned on the p-side of the sensor, whereas the beryllium ``top plate'' 
was located on the n-side. The typical thickness of the beryllium 
substrates is 400\,$\mu$m for the ladders. The beryllium pieces aid in 
stiffening the ladder, provide an improved thermal path, and serve 
handling and reference purposes. The active and the passive beryllium 
pieces were used for the alignment. They have precisely machined notches at 
the side (machining accuracy of notch surface $\pm$5\,$\mu$m) in order to 
locate a ladder into a barrel by press fitting one of the notches of each 
beryllium piece into machined posts at the bulkhead surface. All beryllium 
pieces were visually inspected prior to lamination and the specified 
mechanical dimensions measured. The thickness was determined at five 
different points. The 50\,$\mu$m flatness specification of the substrates 
was verified, and the upper tolerance of the notch-notch distance was checked 
with each beryllium piece on a small fixture.
     
The HDIs were laminated to the active and top plate beryllium pieces by using 
a 25\,$\mu$m thick film and a 75\,$\mu$m thick adhesive film~\cite{ablefilm}. 
An additional thermally and electrically conductive path through silver 
epoxy loaded holes in the HDI was provided so that the beryllium pieces could 
remove the heat load of the SVXIIe chips and be kept on a common system 
ground.  

The beryllium substrates also help to maintain the flatness of the HDI. The 
two beryllium (active and passive) pieces with the locating notches were 
connected by two support rails, each consisting of a carbon/boron fiber, 
Rohacell foam, carbon/boron fiber sandwich. These rails were glued to the 
silicon ladder parallel to the long side in order to make the ladder 
stable and stiff. The carbon/boron fiber composite was designed to match 
the thermal expansion coefficient of the silicon. The carbon fiber support 
rails also serve as low impedance connections in order to ground the passive 
beryllium piece at the ladder's end. 

For each 3-, 6-, or 9-chip ladder, there were two basic ladder types produced: 
left-handed (LH) and right-handed (RH) ladders. Both types were produced in 
equal number. They differ by the relative orientation of the beryllium 
notches in the bulkhead posts with respect to the orientation of the ladder 
axis\footnote{In addition, the LH and RH ladders have two different HDI tail 
lengths.}. Two different fixtures were employed for RH and LH ladders. These 
assembly fixtures were made out of either aluminum or steel and were coated 
with a layer of Teflon to allow a gentle sliding of the silicon sensors for 
accurate alignment.

The fixtures had pushers at the side to exactly place the silicon sensors 
with respect to the glass scale targets on the fixture. Additional steel 
posts had been inserted into the fixture in order to press fit the notches 
of the active and passive beryllium pieces against these posts. Spring 
loaded clamps on the fixture supported the press fit by pushing the 
beryllium notches against the posts. All the alignment and assembly of 
the silicon ladders was done under control of a CMM.

Before ladder production started, the final position of the glass scales 
and the steel inserts on the fixtures was determined in an iterative 
procedure. The qualification process included the assembly and measurements 
of test ladders as well as the fixture targets until the resulting 
distances between the notch edges of the beryllium and the silicon center 
axis defined by the silicon targets converged to within the desired 
specifications of 5\,$\mu$m. 

\subsubsection{Ladder Production} 

Before the assembly started the components were tested. The testing
procedure is discussed in Sec.~7.
 
The first stage of the assembly, Stage-I, of a 6- or 9-chip ladder was the 
alignment of the silicon sensor piece(s) on the assembly fixture under the 
control of an optical CMM. The axial strip side (p-side) of the silicon  
faced up. Once the silicon was aligned, the vacuum on the fixture was turned 
on to keep the sensors in position. The alignment was then verified with 
the CMM, and if needed, the sensors were readjusted in order to keep their 
position uncertainty below 5\,$\mu$m.

\begin{figure}
\begin{center}
\includegraphics[width=0.7\textwidth]{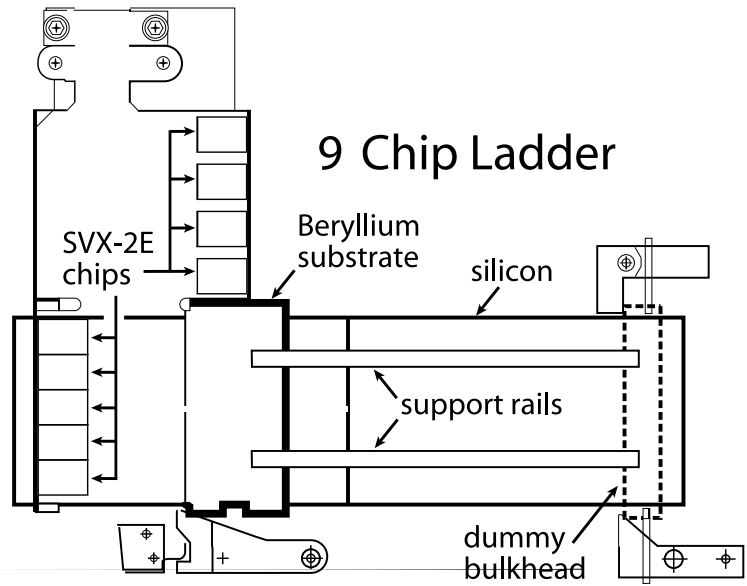} 
\caption{Schematic drawing of a 9-chip ladder resting on the Stage-I 
assembly fixture.}
\label{9chip_ladder_schema}
\end{center}
\end{figure}

\begin{figure}
\begin{center}
\includegraphics[width=0.5\textwidth,angle=-90]{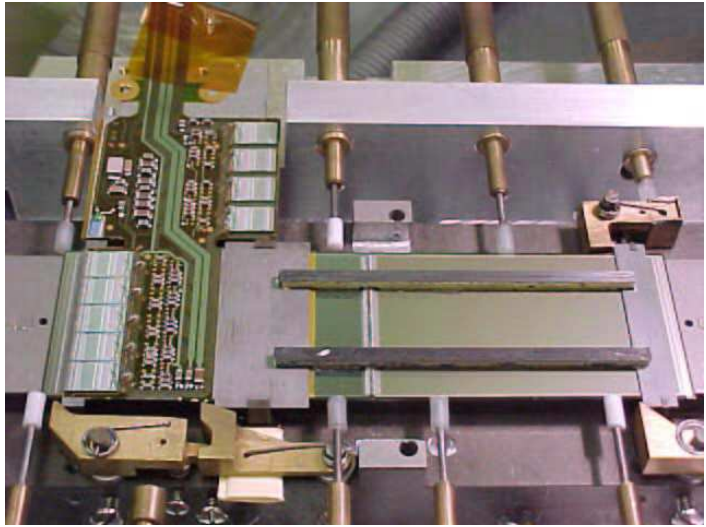} 
\caption{A photograph of a 9-chip ladder on the Stage-I fixture.}
\label{9chip_ladder_photo}
\end{center}
\end{figure}

HEXCEL epoxy glue was applied onto one film laminated beryllium 
carrier piece of the HDI and onto the smaller passive beryllium 
piece at the other end of the ladder. An exact amount of epoxy controlled by 
a glue dispenser was used and the glue was distributed in a fish-bone-like 
pattern on the beryllium pieces to ensure an evenly spread glue joint. 
The HDI with the laminated beryllium pieces was then glued onto the 
silicon surface. It was checked that the spring loaded pushers were pushing 
the notches of the active and passive beryllium pieces tightly against the 
steel posts in the fixture. Figure~\ref{9chip_ladder_schema} shows a 
schematic drawing of a 9-chip ladder resting on the Stage-I fixture.
Figure~\ref{9chip_ladder_photo} is a photograph of a 9-chip ladder on that 
fixture. 

After setting special gluing weights above the adhesive joints between HDI 
and silicon, the ladder alignment was rechecked with the CMM. The gluing 
weights were applied for at least twelve hours. The thickness of the glue 
joint after the epoxy is about 75\,$\mu$m. 

After curing of the glue, the sensor was bonded to the readout chips 
and -- in case of the 3- and 9-chip ladders -- sensor to sensor bonds were 
made using the same assembly fixture as a bonding jig. A visual inspection 
of the wire bonds followed after each bonding step. The carbon/boron fiber 
rails were then glued on the silicon ladder connecting the active and 
passive beryllium piece with a low impedance path. The rails were glued 
using the same HEXCEL epoxy onto the silicon sensors, whereas the adhesive 
joint to the beryllium plates was made by small amounts of conductive 
silver epoxy.

In the second assembly stage, Stage-II,  the ladder was flipped to the 
n-side using a special pickup fixture. The HDI was folded over and centered 
onto the n-side of the silicon sensor with the help of spring loaded pushers 
at the Stage-II fixture. The second part of the HDI serving the n-side was 
then glued with HEXCEL epoxy to the sensor, and another gluing weight was 
applied during the curing cycle. After this, the sensor to sensor bonds on 
the n-side and the bonds to the readout chips were done. 

\begin{figure}
\begin{center}
\includegraphics[width=0.95\textwidth]{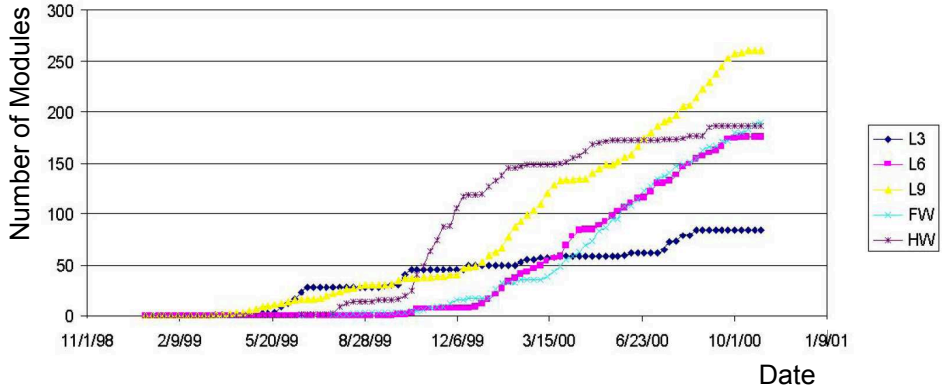}
\caption{Production of ladders and wedges as function of time. The overall 
production pace was mainly determined by the availability of the parts.}
\label{production_pace}
\end{center}
\end{figure}

The production times of the ladders depended on the ladder type. The single 
sided 3-chip ladders could be produced within one day. The double-sided 
6-chip ladders, having only one sensor, could be fabricated within two days,
while almost three days were necessary for the production of  double-sided 
9-chip ladders due to the additional sensor to sensor bonds. While the 
individual production time of a ladder was mainly determined by the curing 
time of the glue, the overall pace of the ladder fabrication was dominated 
by the availability of the parts and components. Duplicate sets of assembly 
fixtures and up to four CMM machines with mounted fixtures were used to keep 
up with the tight production schedule.  The number of assembled ladder 
modules versus time is shown in Fig.~\ref{production_pace}.

\subsubsection{Mechanical Accuracy of the Ladders}

\begin{figure}
\begin{center}
\epsfig{file=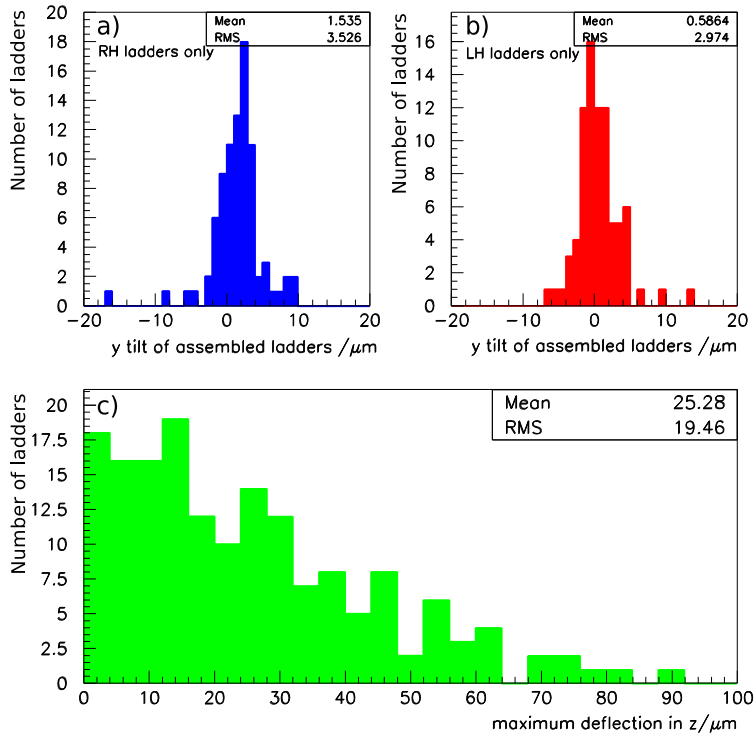,height=10cm,width=13cm}
\caption{Mechanical accuracy of 6-chip ladders for (a) RH ladders, (b) LH 
ladders, and (c) all ladders.}
\label{pic912}
\end{center}
\end{figure}

Since the ladders were inserted into the bulkheads by pressing the  
notches of the active and passive beryllium pieces against the bulkhead 
posts, the final location of the ladder in the bulkhead and hence the 
axial alignment of the silicon strips are determined by these two notches.
Any tilt between the two beryllium notches with respect to the silicon 
strips leads to a misalignment. The achieved mechanical accuracy in the 
ladder assembly is shown in Fig.~\ref{pic912}. Figure 28a and b show the 
maximum tilt between the active and passive ends of the beryllium for all 
production 6-chip ladders on the LH and RH fixtures. The distributions are 
well centered and have an rms value of about 3.5\,$\mu$m. Figure~\ref{pic912}c 
shows the absolute value of the maximum deflection normal to the sensor plane 
of all produced 6-chip ladders. The flatness measurements were performed on 
an optical metrology machine. Almost all ladders are well below 50\,$\mu$m, 
which is the maximum radial displacement permitted for the inner layer.

\subsection{F-Disk Wedge Production}

\subsubsection{Gluing Process}
Qualified sensors and HDIs were selected to be glued together. The HDI  was 
mounted on the same carrier bar that was used for testing and shipping. This 
bar was also used to help handle the HDI for the gluing procedure. There were 
two important considerations for the gluing process: a) relative alignment of 
the sensor to the HDI, and b) the thickness of the glue joint such that good 
contact was insured. The process was performed on a specially machined 
fixture that was mounted on a CMM.

To control the thickness of the glue, the thicknesses of the HDI substrate 
as well as the silicon sensor had to be known. The  silicon sensor 
thickness as measured by the manufacturer was used but the HDI thickness 
was measured directly. Once the thicknesses were known, the correct thickness 
gauge blocks were chosen to mount on the fixture. The HDI was mounted on top 
of these gauge blocks in the fixture on the CMM. A coordinate system was 
defined from the HDIs fixed on the gluing fixture, with the $x$ ($y$) 
coordinate being in the direction of the base (length) of the wedge, and the 
$z$ coordinate perpendicular to the sensor surface.

The sensor was mounted on a flat pedestal and secured via a vacuum connection.
This pedestal could be translated or rotated relative to the HDI using two 
micrometer adjustable screws. The plane height in the $z$ direction was fixed 
relative to the HDI. The alignment of the silicon relative to the HDI 
coordinate system was performed using two fiducial marks on the sensor. The 
first fiducial was in the center of the sensor right next to the HDI, the 
second was at the center-most part of the wedge as it would be placed on a 
disk. An iterative procedure was performed using the micrometer screws to 
allow both fiducials to line up at the correct places. This alignment was 
done to better than 5\,$\mu$m in the $x-y$ plane.

The same epoxy that was used for ladders was used to glue the wedges.
The HDI was held down onto the silicon using spring loaded screws and 
left to cure overnight. After curing, the alignment was verified before 
the vacuum was released so that the new F-wedge could be dismounted from 
the gluing fixture and stored in its handling box.  Because there are bias 
bonding pads on the very corner parts of the wedge, an extra step was taken 
to inject glue at these points. A trained technician using a small bore 
syringe performed this procedure. 

A total of four gluing fixtures were made so that four F-wedges could be glued 
per day.  The alignment and gluing procedure for these wedges could be 
completed in one hour during the afternoon.  The subsequent corner gluing 
and visual inspection took another hour the following morning. Once the 
procedures were developed, there were very few problems during gluing.  
The main problem was assuring that the sensor did not come into contact 
with the alignment pin on the fixture during dismounting.  The yield for 
this procedure over the entire production was over 90\%.
  
\subsubsection{Channel Bonding}

The F-wedges have a double-sided silicon sensor with one HDI with eight chips 
on one side and another HDI with six chips on the other side. On the side 
with six chips there was also a jumper, which is a pitch adapter to adapt 
from the chip spacing to the silicon spacing. Four production wire bonding 
operations were needed for each wedge. First the bonding from the chips to 
the sensors was done on the 8-chip side. Then the detector was turned over 
to the 6-chip side and the bondings from the chips to the jumper and from 
the jumper to the sensor were done.  Finally the bias bonds to the detectors 
on both sides  were attached. The total number of wire bonds for all wedges 
is about 370,000.

The 8-chip channel bonds were bonded using automatic wire bonding machines.
This procedure took approximately 20 minutes per wedge. Only minor problems 
were encountered with this operation during development when the automatic 
bonder operating parameters were being determined. 

The wedge-to-wedge height differences and non-similar jumper surfaces on 
the 6-chip side made the wire bonding on the jumper side more problematic.
Because of the awkwardness in bonding to the jumper, both of the bonding 
steps on the 6-chip side of the wedge were performed on  semi-automatic 
bonding machines. The wedge was placed into a vacuum fixture for these steps.
Bonding parameters occasionally had to be readjusted during this process, 
but the bonding was accomplished with very few unbondable channels. The 
two wire bonding steps on the 6-chip side took on average 45 minutes to 
one hour to complete.

The bias bonds are located on the corners of the sensor where the amount 
of glue underneath the sensor made the bonding difficult and therefore 
the bias bonds were made using a manual bonding machine.  

Aside from the wire bonding technician, another technician helped to 
mount and dismount wedges from the fixtures and perform the quality 
assurance procedures.  The majority of problems encountered were due to 
mishandling of the wedges in the fixtures.  Only one misalignment on the 
wire bonding machine broke a wedge.  Only very minor problems were 
encountered with quality assurance of the bonding itself. These included 
tool marks and some wire bonds that broke after bonding.  

\subsection{H-Disk Wedge Production}
\begin{figure}
\centerline{
\psfig{file=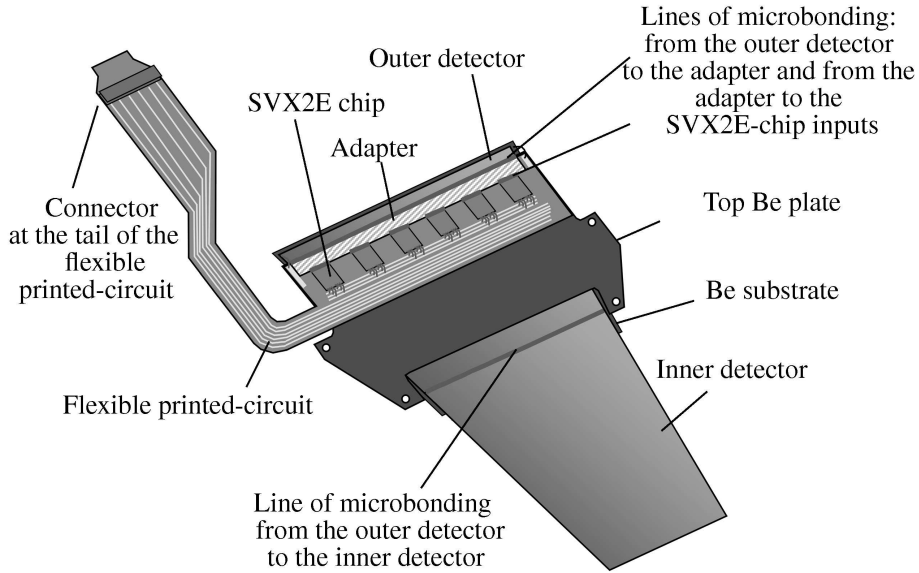,height=8cm}}
 \caption{Simplified configuration of the single-sided half wedge of an 
H-disk.}
 \label{half_wedge}
 \end{figure}

The H-disk wedges are two-layered and each wedge is composed of two 
single-sided half
wedges which were glued back-to-back.
A simplified configuration of the half wedge is shown in 
Fig.~\ref{half_wedge}.
Each half wedge consists of inner and outer silicon sensors mounted on 
a beryllium plate.

The single sided silicon sensors are shaped as trapezoids; their strips 
are parallel to one of the trapezoid's sides. The strips of the inner 
and outer sensors were bonded together, leading to a maximum strip length 
of 14.24\,cm. A beryllium plate was attached on top of the outer sensor 
providing a mounting surface for front end readout electronics installed 
on the HDI. The HDI was laminated in a similar way to the barrel ladders 
on top of the beryllium plate. The outer sensor strips were connected with 
the input pads of the chips with wire bonds. A special silicon based pitch 
adapter, fabricated from the same wafer as the wedge sensor, was used.
A part of the top beryllium plate surface of one of the half-wedges was used 
for mounting of the wedge on the cooling ring of the disk. A full wedge 
module was produced by joining together two half wedges back-to-back.

\subsubsection{Wedge Production for the H-Disks}
Before assembly started, silicon sensors, beryllium substrates, and HDI flex 
circuits were thoroughly tested. Every step of the assembly process was 
followed by an electrical test of the subassembly. The half wedge assembly 
process started with lamination of the flex HDI circuits on the beryllium 
substrates. To adhere a large area of the outer silicon sensor to the 
beryllium plate, a thermal conductive tape based on an acrylic adhesive 
(3M9882) was used. This allowed stress relief on the silicon during 
assembly and cooling of the detector during operation. The substrates were 
aligned and bonded onto an outer silicon sensor using special assembly 
fixtures. To bond the inner silicon sensor to beryllium, the epoxy 
adhesive HEXCEL was used, the same type as was used in the ladder 
assembly process. The sensor alignment was done under CMM control. Wire 
bonding of the wedges was done in two steps. First, the SVXIIe chips were 
bonded to the pads on the HDI circuits. Second, the input pads of the chips 
were wire bonded to the strips from the outer sensor via a pitch adapter on 
the HDI, and the strips from the outer sensors were wire bonded to the 
corresponding strips of the inner sensor. All wire bonds were encapsulated 
on each of the half wedges to prevent damage during their assembly and 
installation on the mounting ring. A special bi-facial machine consisting 
of two co-axial microscopes was developed to align and bond together two 
half wedges back-to-back with precision better than 15\,$\mu$m. The 
assembled wedges were installed on the beryllium cooling channel ring using 
the rotary table of a CMM with a precision of 25\,$\mu$m. 
 
\newpage
\pagebreak

\section{Testing of Ladders and Wedges}
Thorough testing of each ladder and wedge at various stages of the production 
was required to avoid costly or even impossible back-tracking during the 
detector assembly. This section describes the procedures by which the 
modules were electrically tested during production and the criteria used 
in assigning an electrical grade to the finished detector modules.
A more detailed description can be found in Ref.~\cite{d0note_3841}.

\subsection{Testing Sequence}
\label{s:sequence}
The essential building blocks of a detector module are the silicon sensors,
the SVXIIe chips, and the HDIs. 
Each component was tested as described in Sec.~\ref{s:sensors}, 
\ref{s:chip}, and Sec.~\ref{s:hdi}.
The production testing sequence is described below, and can be 
summarized as follows:
\begin{enumerate}
\item Bare HDIs that passed the functionality test at the manufacturer 
were sent to Fermilab for long-term testing. 
\item Sensors and burned-in HDIs were assembled into a detector module.
\item Detector modules underwent initial functionality test. 
\item Detector modules that passed the initial functionality test were
burned-in. 
\item Burned-in detector modules were laser tested. 
\item Detector modules were assigned an electrical grade. 
\item At each stage of the production sequence, malfunctioning modules 
were sent to be repaired. 
\end{enumerate}

Debugging and repair was done by expert physicists; the burn-in and laser 
tests were operated by non-expert physicists on shift with experts, who 
coordinated the activities and helped with the set-up of the tests.
Detailed instructions of the daily activities and the procedures to be 
followed proved to be crucial to the operation, as was the meticulous 
bookkeeping of all activities. In the following sections, each step of 
the production testing sequence is described in detail. The overall yield 
of detector production was about 85\%. The lost detectors failed to pass 
either electrical or mechanical tests.

\subsection{Testing of the SVXIIe Chips}
\label{s:chip}
The SVXIIe chips were produced on 12.7 cm wafers, with 148 usable dice on each
wafer. A total of 206 wafers were produced and 155 of those were tested using 
an automatic probe station. One-hundred-twenty wafers were tested in the 
first half of 1997 (first batch), and 35 wafers were tested in the spring of 
2000 (second batch). The testing and analysis procedures were identical 
for both groups of wafers. 

The testing procedure consisted of driving all chip inputs with a 
computer-specified pattern, sampling the output lines, and checking that the 
output states were correct. The tests were separated into three categories:  
digital, analog, and miscellaneous. The digital tests exercised the readout 
and pipeline sections of the chip and checked that the chip current draw was 
within specifications for all supply voltages. The analog tests exercised 
the front-end preamplifiers and pipeline amplifiers, in both positive and 
negative modes. Pedestals and gains were measured for all 32 pipeline buckets 
for each of the 128 channels. Finally, the miscellaneous tests exercised all 
of the auxiliary registers, which control the calibration voltage, the ramp 
rate used for digitization, the input bandwidth, the pedestal level, and 
finally the chip bias current. 

A probe station was used to physically position the wafer so that the 
probe card made contact with a specific die. One die was processed at a time;  
all the tests were run on that die, then the probe station moved the next die 
into contact with the probe card. Once the tests were completed, the wafer 
was diced and chips visually inspected for mechanical damage, such as 
scratched surfaces and missing corners. Two wafers were broken during the 
dicing process. 17593 chips (77\%) passed the digital tests, 14701 chips 
(64\%) passed the analog and miscellaneous tests, and 13988 chips (61\%) 
passed the visual inspection after dicing and were distributed.
The yields from the first and second wafer batches were the same.

\subsection{Testing of Bare HDIs}
\label{s:hdi}
The HDIs are a two layer, 0.127\,mm thick flex circuit of polyimide
film with gold/nickel plated copper pads. The requirements of 0.127\,mm 
wide pads and 0.051\,mm wide via feedthroughs made the HDI a technical 
challenge for production. The tests checked for continuity and shorts 
between each of the pads on the front end and the 28 lines of the tail.
This was accomplished through the use of a Ruker \& Kolls semi-automated wafer 
probe station, controlled through GPIB interface with a PC running LabView.
Given a coordinate map of the HDI circuit, the probe station stage positioned 
each pad under a probe tip. Connection was made through a GPIB multiplexing 
box,  and a GPIB controlled multimeter checked for continuity or shorts 
between the appropriate line on the tail and the probe.Approximately 60 pads 
were checked per SVXIIe chip mount point on the circuit.

The yields of good bare HDIs varied significantly between vendors and 
production batches. Litchfield Precision Components~\cite{lpc} had a very low 
yield of $\approx 30\% $ after ``fixing'' (zapping the shorts with a high 
current). Dyconex~\cite{dyconex} only delivered HDIs after internal testing, 
thus delivering good parts with excellent yield, but at a significantly 
higher price. A good compromise between cost and yield was finally reached 
with two additional vendors: Speedy~\cite{speedy} and 
Compunetics~\cite{compunetics}. Obtaining working bare HDIs was one of the 
major issues of the early stages of the SMT production.

\subsection {PC Based Test Stands}
\label{s:hardware}

All the electrical tests performed on readout HDIs and detector modules 
used the same type of test stand, based on the Stand Alone Sequencer 
Board (SASeq), developed at Fermilab. The SASeq-based test stands
were developed independently from the full \D0 readout system. They were 
replicated and distributed at various locations at Fermilab and at remote 
institutes long before the final version of the full readout system was 
available. Using the PC-based version of the readout system proved crucial 
to the success of the production testing effort. The test stations used for 
the burn-in could operate 16 HDIs or detector modules at once. Separate 
one-SASeq test stands were also built which could read out two devices at 
once. The one-SASeq test stands were used for initial functionality tests 
of HDIs, debugging of detector modules, laser tests, repair, and 
for barrel/disk assembly. Nine stations were set up at the Fermilab Silicon 
Detector Facility, and three more stations were available at remote 
institutes. 

\begin{figure}
\includegraphics[width=\textwidth]{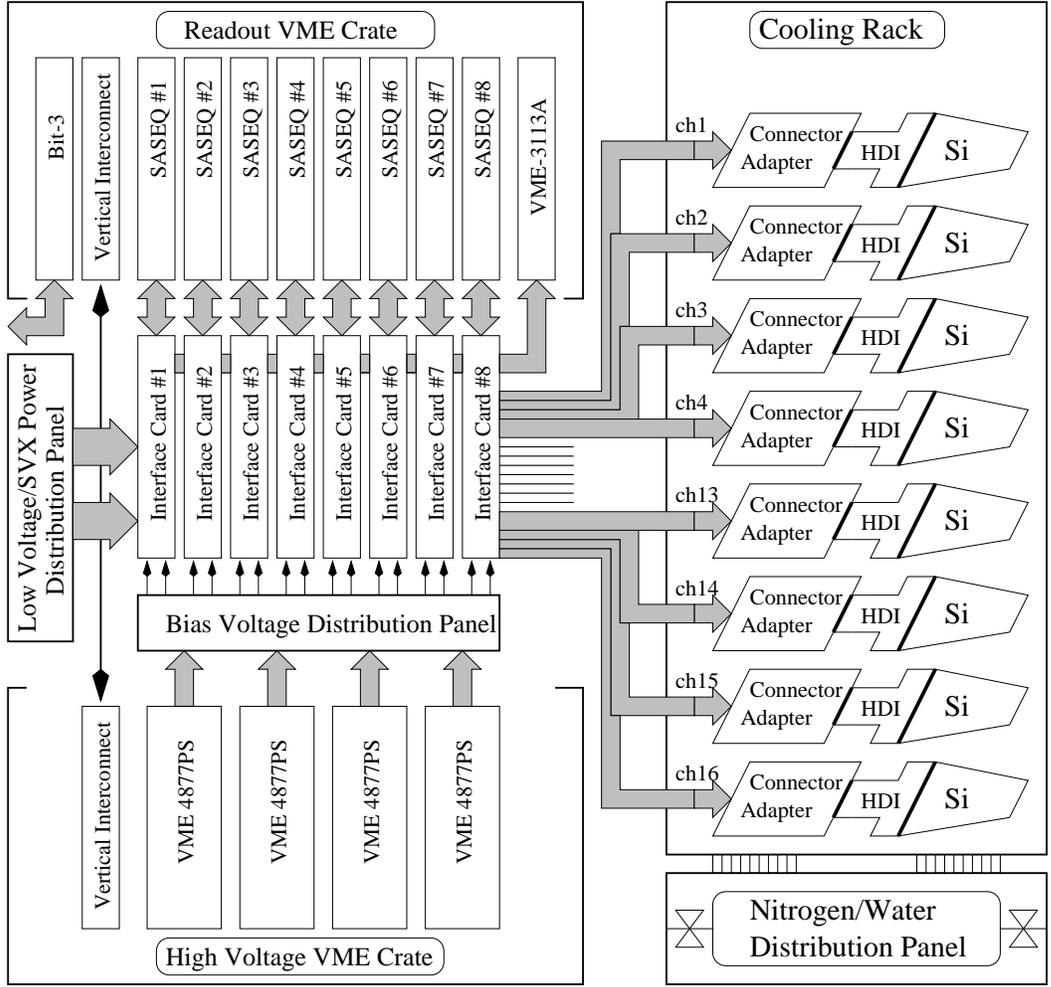}
\caption{Burn-in test setup. For debugging/repair/laser stations 
only one SASeq/IC pair was used,  and no cooling was provided.}
\label{hardware}
\end{figure} 

The hardware configuration for the burn-in stations is shown in 
Fig.~\ref{hardware}.
It consisted of a VME crate that contained a Bit-3 VME controller 
card, eight SASeqs, a scanning 12-bit 64 channel analog-to-digital converter 
board (VME-3113A) for current and temperature measurement, and a master 
vertical interconnect board for high voltage crate control. The two-channel 
SASeq board was a self-contained data acquisition card designed to interface 
to the SVXIIe chips. Its basic functions were to control the SVXIIe chip 
for data acquisition, collect the data when a data cycle was requested, and to 
relay the data to the processor in the crate. The crate also housed eight 
interface cards (IC). Each IC had two independent channels and was used as a 
bi-directional interconnect between the SASeq and the HDIs. The IC provided 
the control of the SVXIIe ``power on'' and ``power off'' sequence and prepared 
the monitoring of currents corresponding to the three SVXIIe operating 
voltages and to the temperature measurement. 

Each SASeq was connected to an IC by a 3\,m long, 50-conductor cable with 
an impedance of 82\,$\Omega$. Three low voltage power supplies were used to 
supply the three operating voltages needed by the SVXIIe chips. An SVXIIe 
voltage power distribution panel was located between the power sources and 
the IC crate and was used for the distribution of the SVXIIe power and the 
interface board power for each of the eight IC. 

Bias voltage was provided independently for each detector module via a bias 
voltage distribution panel. The panel was located between the high voltage 
sources and the IC crate. A set of switches on the panel allowed for three 
different schemes for the biasing of double-sided detectors: positive bias 
applied to the n-side, negative bias applied to the p-side, and 
``split bias,'' in which both positive and negative bias was applied at 
the same time to the n- and p-side, respectively. The burn-in stations
needed a separate VME crate to house the high voltage sources to bias the
detectors. For the one-SASeq test stations, the HV module was connected to 
the same VME crate as the SASeq. 

The HV VME crate for the burn-in stations contained four 
VME 4877PS~\cite{vme4877ps} motherboards and a slave vertical interconnect 
board for high voltage crate control. Every motherboard carried eight HV 
pods. Thus the HV crate provided 16 positive and 16 negative independent 
voltages for silicon detector biasing and supported the current monitoring. 
The 4877PS Motherboard allowed the voltage to be set from 0 to 5000\,V, 
with a maximum current of 2\,mA per channel. To ensure the safe operation 
of the burn-in stations, the over-voltage hardware protection of the HV 
supply was set to 120\,V.

During testing, the HDI tails were inserted into Hirose connectors on the 
connector adapter boards. These boards contained the standard 3M 50 pin 
connectors and the Hirose connectors and were used as connector adapters 
between the signal cables coming from the ICs and the HDI tails. The 
ICs were connected to the adapter board by a 3\,m long, 50-conductor 
coaxial ribbon cable with an impedance of 75\,$\Omega$.

The burn-in stations were outfitted with a cooling system to operate the 
detectors at low temperature. Up to 16 detectors were placed on shelves 
inside a regular rack that had been thermally isolated. The chiller 
temperature was set to 3\degC, and the detectors ran at temperatures 
between $5^\circ$C and $15^\circ$C, depending on the number of chips on 
the HDI. Each detector module was placed on a custom made, 
178$\times$229\,mm aluminum plate, designed to accept all types of HDIs or 
detectors. Every aluminum plate was equipped with a pipe for cooling water 
and special holes to provide nitrogen flow through the box that enclosed 
the device under test. Two aluminum plates together with the connector 
adapter boards and signal cables were placed on a plywood board. This 
board was equipped with sliders to simplify loading and unloading of the 
devices under test. Plywood was chosen because it was a cheap, low 
thermo-conductive material that helped reduce the condensation on surfaces 
inside the rack. Every board had its own water and nitrogen pipe. The 
control over the water and gas flow was provided by a control panel outside 
the rack. A software based interlock system monitored the temperature 
on each device and shut down the power in the event that the temperature
exceeded  $50^\circ$C. The aim was to stay below the glass transition 
temperature of the epoxy to avoid damage to the detector assembly.

\subsection{Testing of Detector Modules}
\label{s:debug}

The testing of detector modules was the first attempt to read out and 
bias a completed ladder or wedge detector after module assembly.
The testing also included some remediation of issues and is thus also 
referred to as ``debugging.'' The likelihood for damage to occur during 
assembly, mainly during wire bonding, was substantial. About 25\% of the 
assembled detectors failed to download or to read out correctly. In 
addition, the leakage current was unacceptably high for about 95\% of the 
double-sided assembled detectors. The functionality of the detectors had 
to be restored before any further tests were done.

First the modules were visually inspected to ensure that no mistakes had
been made during wire bonding and no mechanical damage had occurred. The 
electrical resistances between the active beryllium pieces and the HDI 
ground were measured, and if a resistance was greater than 10\,$\Omega$, 
the grounding was improved either with a small wire or directly with a 
silver epoxy trace. 

The functionality test of the readout was done without applying bias to 
the detector. Download and readout failures were mainly caused by 
damaged SVXIIe chips, misplaced or missing wire bonds, or damaged HDIs.
About 10\% of the SVXIIe chips had to be replaced.

Once a detector module had been successfully read out, it was biased.
The biasing scheme depended on the type of detector under consideration.
For the single-sided detectors, positive bias was applied to the n-side, 
with no segmentation. Because single-sided detectors do not have the bias 
applied on the coupling capacitor, the leakage current was low. No repairs 
of the single-sided detectors were needed, since they would remain 
operational even if a capacitor was broken during the wire bonding.
Double-sided detectors were ``split biased'' by applying  positive bias to 
the the n-side and negative bias to the p-side, corresponding to the axial 
strips for the ladders. Broken AC coupling capacitors caused by flaws in 
the lithography caused holes in the insulator, called pinholes. If one of 
the AC coupling capacitors failed for a channel, leakage current traveled 
directly through the readout electronics connected to the broken AC capacitor.
This caused high leakage current for the detector, even at low bias voltages. 
Typically, the leakage current was as high as 100\,$\mu$A at a bias 
voltage of about 10\,V applied on the n-side during testing.To remedy this, 
channels with broken AC coupling capacitors were identified and disconnected 
from the readout electronics by pulling the wire bond between the silicon 
sensor AC bonding pad and the SVXIIe preamplifier.

\begin{figure}
\includegraphics[width=\textwidth]{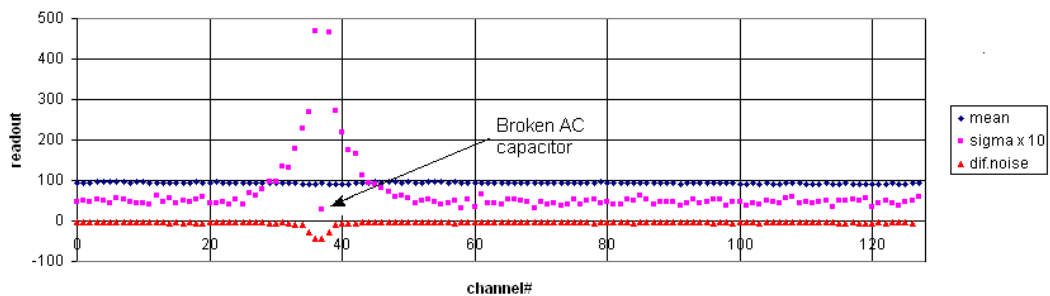}
\caption{Distributions of pedestal(diamonds), noise (squares), and 
differential noise (triangles)
for a chip containing a broken AC capacitor. The noise (sigma $\times$ 10)
is low for broken channels and high for neighboring channels. The differential
noise is plotted multiplied by $-1$ for better visual clarity.}
\label{testfig2}
\end{figure}

To identify the broken AC coupling capacitors at one side of the detector, 
bias voltage was applied on that side and the other side was kept grounded.
Typical plots of pedestal, noise (defined as pedestal fluctuations for each 
channel), and differential noise (defined as fluctuation of the noise 
difference between two consecutive channels) are shown in Fig.~\ref{testfig2} 
for a chip containing one broken AC capacitor. Broken AC coupling capacitors 
caused the input of the preamplifier to be dominated by the DC component, 
resulting in an almost constant output of the preamplifier. Consequently, 
the level of noise was low for broken channels and high for the neighbors 
as shown in Fig.~\ref{testfig2}.This phenomena is further discussed in 
Ref.~\cite{marvin} 

The bonds for broken capacitors were pulled, and  after that the positive 
bias voltage applied to the n-side could be raised up to 
$V_{TS} + 30\;\rm V$. $V_{TS}$ is defined as the depletion voltage measured 
on the test structure, and it  was  determined from the sensor probing 
discussed in Sec.~5. The leakage currents at the test voltage 
$V_{\rm{test}}$, defined as $V_{TS} + 20\;\rm V$, or 90\,V, whichever was 
lower, were measured to be in the range from 5\,$\mu$A to 25\,$\mu$A.

A few detectors with high leakage current did not have any broken AC coupling 
capacitors. It was concluded that the  leakage current was due to defects in 
the bulk silicon of the sensors, which could not be repaired.

The leakage current at depletion voltage was one of the considerations 
in the final electrical grade of the detector. Detectors with very high 
leakage current were not installed in the SMT.

One additional consideration for the detectors from Micron was that the 
negative bias voltage applied to the p-side was severely limited by 
micro-discharges  \cite{md}. For a module with no broken AC coupling 
capacitors, $V^{-}_{\rm{max}}$ was defined as the highest voltage that 
could be applied on the p-side before the current started to grow 
exponentially. Above this value the leakage current, and also the noise, 
increased rapidly. In the absence of micro-discharges, $V^{-}_{\rm{max}}$  
was set to $\frac{1}{2} V_{\rm{test}}$. 

For the F-wedges, three additional problems were observed. The sensors had 
a tendency to develop pinholes during the first half hour of a burn-in test, 
resulting in high currents. In order to limit the time spent on burn-in, 
the wedges first were biased at the test station at $V_{\rm{test}}$ and 
left at that voltage for 30 minutes. A second problem which developed for 
the Micron F-wedges was that there were regions in which pinholes seemingly 
propagated to several adjacent strips. When examining these wedges under 
a microscope, p-stop implantation faults were seen in the affected areas.
However, this correlation was not complete, as some wedges affected did 
not exhibit visible faults. Probably this problem was linked to faults 
in the masking procedure of the silicon. The third problem that affected 
Micron F-wedge modules was related to the stability of the bias currents.
The Eurysis F-wedge sensors generally exhibited fewer problems than the 
Micron sensors. While the Micron sensors in most cases had problems on 
the n-side, the Eurysis sensors exhibited a larger number of pinholes on 
the p-side. In addition pads that only should be used for bonding had been 
used to probe the sensors, thus the chips were bonded to pads with probe marks.
For these sensors, the I-V characteristics showed a clear junction breakdown 
when applying negative bias.

One problem of concern was the general appearance of micro-discharge 
at around 70\,V in positive bias, which indicated that the high voltage 
would be limited to less than the design value for the wedges. This could 
become an issue after type inversion if the depletion voltage needs to be 
increased past the micro-discharge voltage.

For double-sided detectors, 0.8\% of the channels were disconnected from the 
readout electronics during the debugging process. The bare silicon sensors 
were tested before assembly, and only working channels were bonded. It was 
thus concluded that the capacitors broke during wire bonding.

The debugging was the most time consuming step of the production testing 
effort,  and therefore its bottleneck.

\subsection{Burn-in Tests}
\label{s:burnin}

The burn-in test was a long-term functionality test performed during 
the production testing. First, the burn-in test was done on the bare HDI 
after its initial functionality test.  The second burn-in test was carried 
out after ladder or wedge assembly. 

The goal of the burn-in test was to run each HDI or detector module 
for 72 hours, monitoring its performance and measuring pedestals, total 
noise, random noise, gain, and occupancy in sparsification mode. Additional 
parameters were monitored, including temperature, SVXIIe chip current for 
the three SVXIIe chip voltages, and for ladders and wedges, also detector 
bias voltage and dark current. Typical problems revealed by the burn-in test 
were SVXIIe chip failures, broken and shorted bonds, grounding problems, 
noisy strips, and coupling capacitor failures. The burn-in setups were able 
to accommodate 32 devices. Two burn-in cycles were run per week, which 
allowed keeping up with the peak production rate of $\approx 20$ detector 
modules per week. Several different steps were incorporated into the 
burn-in tests, and they are briefly described below.

\subsubsection {Temperature Sensor Test}

The goal of the temperature sensor test was to check the operation of the 
temperature sensor mounted on each HDI. The temperature sensor is a thermal 
resistor, and its  voltage is digitized and converted into temperature.
The temperature was  measured five times with an interval of 20 seconds 
between measurements. The test  was performed with all readout electronics 
off and with the sensor at room temperature. The most common failure of the 
temperature test was due to a bad contact of the thermo-resistor to the HDI. 
These HDIs were easily repaired.

\subsubsection{Data Integrity Check}
 
The goal of the data integrity test was to check the SVXIIe chip downloading.
The SVXIIe chips were downloaded about 100 times. For each download cycle, 
the program compared the initial download file with the one read back from 
the chip and counted the errors. The parameters that were compared included 
chip ID, channel number and sequence, and pedestal values. Only those HDIs 
that had no download errors were accepted for future use. 

\subsubsection {Long Burn-in Test}

During the long burn-in tests, the devices were left powered and biased 
for 72 hours. Bare HDIs were run on an aluminum plate that acted as a 
heat sink, and effectively operated at approximately $30 \degC$. Higher 
temperatures could prevent the SVXIIe chips from downloading reliably. 
Modules were run on a cooled plate, and operated at about 15 to $20 \degC$, 
depending on the number of chips. During the burn-in process, each device 
was sequentially tested in so-called runs. Overall, ten runs were taken.
Each individual test took approximately three minutes per chip. i.e. 10 
minutes for each 3-chip ladder, 20 minutes for each 6-chip ladder, etc. 
Once the last device had been tested for the first time, the system 
would wait for a user determined delay. During that time the units were 
left powered and biased, and monitoring information was collected.
Several tests were performed during each run. They are summarized below.

The run started by taking one hundred pedestal events in data modes 
to obtain the average pedestal and random noise of the device under test.
The pedestal mean and its width were also recorded for each readout channel 
separately and average values determined. For every event, the common line 
shift (average pedestal over 128 SVXIIe chip channels) was also calculated.
This was used to record common mode subtracted pedestals for every channel, 
from which the random noise was determined. This sequence was repeated with 
the chip in cal\_inject mode, to obtain pedestal values to be used during 
the chip calibration.

The next step was the chip calibration. Each channel of the SVXIIe chip was 
connected to a built-in test input capacitor. The test capacitor was used 
for gain studies, during which a known charge was injected into each channel.
Every eighth channel was pulsed simultaneously in order to not to get cross 
talk from adjacent channels and at the same time have a reasonable run time
for the test. Four calibration pulses with different charges were applied 
one after another to the chip. One hundred events were recorded for each 
value of the calibration voltage. The same analysis cycle was repeated for 
eight channel patterns so that the charge was injected in every channel of 
the SVXIIe chip. 

The chips were also tested in sparsification mode. In this mode, only the 
channels with a response that exceeds the preset threshold and their 
immediate neighbors were read out and the frequency of false readouts was 
studied. The chip performance in sparsification mode was evaluated by counting
the frequency of false readouts for every channel. The readout was called 
false if the channel appeared in the data more often than expected (noisy 
channels) or less often than expected (dead or low gain channels). For 
double-sided devices calibration and sparsification readout testing were 
performed separately for the p- and n-side.

Monitoring information was collected once per minute for the duration of the 
burn-in test. Bias voltage and current were measured for ladders and wedges 
directly through the high voltage power supply. 

The burn-in test proved to be an efficient and reliable tool to test the 
long term reliability of the detector modules before installation 
into the SMT. Bare HDIs were selected for detector construction based on 
the results of the burn-in test. Together with the laser test, the 
burn-in test was used to characterize the detector modules and select 
those devices to be installed in the tracker. Over a thousand HDIs 
and detector modules were tested during the SMT production testing project. 
Taking into account the average load of 75\%, this corresponds to 
approximately 6,000 hours of stable running by each of the two burn-in 
stations. 

\subsection{Laser Test}
\label{s:laser}

The laser test was performed on every detector module that passed the burn-in 
test. The modules were characterized by measuring the depletion voltage and 
by determining the numbers of dead and noisy channels. This  information had 
already been obtained by other tests: the burn-in was able to identify dead 
and noisy channels, and the probe tests on the sensors determined their 
depletion voltage. However, early tests indicated discrepancies between the 
depletion voltage measured during sensor probing and the one obtained on the 
assembled module. This led to the implementation of the laser test, and  
although it provided redundant information, it was considered a useful 
tool in the characterization of modules. This was particularly true in the 
characterization of irradiated detectors, to study noisy strips, and to 
investigate irregular charge collection observed on some ladders. 
The test stand was set up to run completely automatically and was very fast.

\begin{figure}[htb]
\begin{center}
\begin{tabular}{c}
\mbox{\epsfig{file=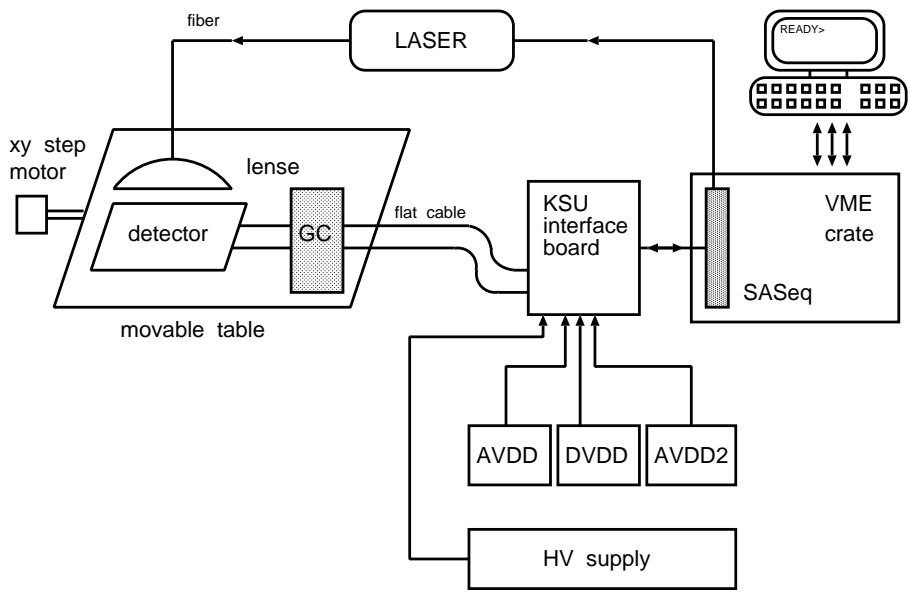,width=12cm}}  
\end{tabular}
\end{center}
\caption{Simplified scheme of the laser test stand setup.}
\label{laser}
\end{figure}

A simplified diagram of the laser test stand is shown in Fig.~\ref{laser}.
It was based on the one-SASeq test stand, with the addition of an
infrared laser and a movable table. The solid state laser had a wavelength 
of 1064\,nm. Given the attenuation length of the silicon of 206\,$\mu$m, 
the laser penetrated the entire depth of the 300\,$\mu$m thick detector 
and not just a surface layer. The laser was connected to an optical fiber. 
At the end of the fiber, a focusing lens collimated the laser spot. 
Completed modules were placed on an {\it x-y} movable table and shuttled 
underneath the laser head. Two laser test stands were used during production.

The depletion voltage was measured by monitoring the charge collected on 
the readout strips that detected the laser, as a function of the bias voltage. 
The procedure for the single-sided detectors was the following: first, the 
range of the voltage variation was set from $0\;\rm V$ to $V_{\rm{test}}$. 
Then, the bias voltage was increased automatically in steps of 5\,V, and after 
each step the amplitude of the laser signal in ADC counts was measured. The 
leakage current of the detector at each voltage setting was also recorded.
We found an almost linear increase of the amplitude $A$ of the signal 
with the bias voltage $V$ up to a value $V_{\rm{min}}$, after which the $A(V)$ 
curve reaches a plateau. As a first approximation, the $A(V)$ curve can be 
described by a slope and a plateau that fit the data linearly. The 
depletion voltage was defined as the voltage corresponding to the point 
where the two linear fits intersect. The results using this method 
for determining the depletion voltage $V_{\rm{depl}}$ was in most cases in 
good agreement with the results obtained with the C--V method of the 
depletion voltage measurements used on silicon sensors. The minimal 
operating voltage was defined as the voltage at the beginning of the plateau. 
The operating voltage $V_{\rm{oper}}$ of the detector was obtained by 
adding 5\,V to the minimal operating voltage. The procedure was very similar 
for double-sided detectors. 

The 3-chip ladders, the 9-chip ladders, and the H-wedges which consist of 
two silicon sensors had their strips wire-bonded by pairs to each other. 
For these devices, the depletion voltage was measured separately for 
the two sensors. Because sensors were paired into detectors based on their 
depletion voltage as determined during sensor probing, these two measurements 
were in general very close. The higher of the two operating voltages was 
taken as the detector operating voltage.

The laser scan was also used to identify dead and noisy channels. This was 
performed with the detectors biased at $V_{\rm{oper}}$, as determined from 
the laser test. Every channel was read out ten times resulting in a set of 
raw amplitudes. The average amplitude and the standard deviation were 
calculated over the ten measurements. 

Each detector module was tested with the laser at least once during the 
production testing effort. The information obtained was used as input for 
the grading of the modules.

\subsection{Diagnosis and Repair of Defective Modules}
\label{s:repair}

HDI and detector modules with download or readout problems had to be 
diagnosed and repaired. A total of 339 problematic HDIs and detectors were 
found, and most of them were also repaired. Although some of the repairs 
were performed by the physicists working on diagnosing the problems, most 
of the repairs required work done by a specially trained technician. 
Malfunction of HDIs was usually due to problems during stuffing at the 
commercial vendors, or bad packaging before shipping. Detectors were also 
damaged during the production and testing process, or during installation 
on the rings or on the barrels.  

The stations used for the diagnosis and repair effort were based on 
the one-SASeq test stations with the addition of a microscope, a probe, 
an oscilloscope, and a logic analyzer. This additional equipment made it 
possible to probe the signals on the SVXIIe chips directly. One very 
important feature of the diagnosis stations was that on the IC, the top 
neighbor and bottom neighbor lines were disconnected, and the signal was sent 
to the SVXIIe chips via the probe. This made it possible to download the 
chips one by one, which turned out to be a very effective tool to find bad 
chips on a device that was showing download errors.

The procedure to diagnose modules with problems was the following:
\begin{enumerate}
\item Visual inspection to look for broken or shorted bonds. On HDIs that 
failed to work properly at the initial functionality test, the visual 
inspection was extended to check that all capacitors and resistors were 
mounted properly. 
\item Check the HDI tail for shorts between the three voltage power lines 
and ground. Shorts tend to develop at the trimmed end of the tail of the 
HDI and could be fixed by sanding the tail with sandpaper.
\item Shorts would also develop due to solder on the tail or under a surface 
mounted component making contact with other pads. Several H-wedge HDIs 
developed shorts between one trace and ground due to sparks from a capacitor 
between the bias voltage line and ground to this trace.
\item Broken traces on the HDI could in most cases be repaired by adding an 
extra wire. This kind of problem was very common on the H-wedges, for which 
traces often broke close to the Hirose connector that was mounted on the 
HDI tail. 
\item Shorts on the SVXIIe chips themselves were identified by first isolating
the faulty chip by pulling the relevant bonds, and then  measuring the 
resistance between the shorted channel and ground. If a short was found on a 
chip, the chip was replaced.
\item If no short was found between the power lines and ground, the next
step was to try to download the chips. Usually at least one of the chips 
would give download errors. To identify the problem, the lines were checked 
with the logic analyzer.
\item To further investigate which chip had download errors, chips were 
downloaded individually. Chips were isolated by pulling the top neighbor and 
the bottom neighbor bonds. When the chip with the download errors was 
identified, it was checked once more for bad bonds and if none were found, 
the chip was replaced. 
\item If a chip was downloading correctly but had a readout problem such as
no gain, it was replaced.
\item Occasionally, detectors would stop working after encapsulation
of the wire bonds. In those circumstances, the encapsulant was removed, which
caused the wire-bonds to come off too, and the detector re-bonded.
\item On some of the detectors, neighboring chips were damaged during 
replacing of a bad chip. This was especially common for the F-wedges. 
\end{enumerate}

Of the 339 devices (181 bare HDIs and 158 detectors) that were diagnosed 
by the repair group, 264 were fully recovered. From these, 74 had only minor 
problems, such as bad grounding or missing or damaged bonds. Thirty-eight 
detectors could not be repaired, and for 37, the repair was not attempted, 
because enough spares of the same type were available. A total of 288 SVXIIe 
chips was replaced: 145 because of readout problems, 83 did not download, 
and 60 had a short. A total of 14 components was found missing on the HDIs. 
Problems with the tail or the traces on the HDIs occurred  for 51 devices.
 
The types of devices with the most problems were F-wedges and 9-chip 
ladders: 113 and 114, respectively. Seventy-eight H-wedges, 32 6-chip 
ladders, and seven 3-chip ladders were sent to the repair group. F-wedges 
often took a long time to repair since they were frequently damaged during 
repair. H-wedges were difficult to diagnose since they often had many broken 
traces close to the connector on the tail.

The effort of the repair group was essential to completing of the SMT 
because there were not enough spares to replace broken detector modules or 
non-functioning HDIs.

\subsection{Electrical Grading and Characteristics of Detector Modules}
\label{s:grading}

The results collected during detector burn-in and laser test were used to 
assign an electrical grade to each detector module. We defined two types 
of ``exceptional'' channels:
\begin{itemize} 
\item A channel was called {\bf DEAD} if its response to the laser in
the laser test was less than 40 counts (the average response was 100 counts).
For detectors with two sensors, the one with the larger count of dead 
channels was considered in the electrical grading.
\item A channel was called {\bf NOISY} if its random noise in data mode
during burn-in was larger than 6 counts (average noise was 1.5 counts)
\end{itemize}

A detector was assigned an ``A'' grade if the sum of the number of dead and 
noisy channels was less than 2.6\% of the total number of channels. This 
corresponds to 10 channels for a 3-chip ladder, 20 channels for a 6-chip 
ladder, etc. No distinction between n-side and p-side bad channels was made 
in the grading. A detector was assigned a ``B'' grade if the sum of the 
number of dead and noisy channels was more than 2.6\% but less than 5.2\%. 
Detectors with more than 5.2\%  bad channels were graded as ``C.''  In 
addition, we required that the leakage current be less than 
10\,$\mu$A at operating voltage and that the burn-in results were stable 
over time. Mostly ``A'' grade detectors were installed in the SMT, 
no ``C'' grade detectors were installed.

The performance of detector modules installed in barrels and F-disks is 
summarized in Table~\ref{tab:barrel} and  Table~\ref{tab:disks}, respectively.

\begin{table}[htbp]
\caption{Fractions of dead and noisy channels in barrel detectors given in 
percent. }
\vspace{0.6cm}
\begin{center}
\begin{tabular}{|c|c|c|c|c|c|c|c|c|}
\hline
\bf {Barrel}
 & \multicolumn{2}{c|}{\bf {Layer 1}}
 & \multicolumn{2}{c|}{\bf {Layer 2}}
 & \multicolumn{2}{c|}{\bf {Layer 3}}
 & \multicolumn{2}{c|}{\bf {Layer 4}} \\
\cline{2-9}
      & \bf {Dead}   & \bf {Noisy}  & \bf {Dead}  & \bf {Noisy}  
      & \bf {Dead}   & \bf {Noisy}  & \bf {Dead}  & \bf {Noisy}\\
\hline
  1   & 0.85   & 0.4     & 1.99   & 0.8    & 0.93   & 0.4    & 1.77   & 0.6 \\
  2   & 0.95   & 0.56    & 2.13   & 0.43   & 1.34   & 0.22   & 2.01   & 0.44\\
  3   & 0.70   & 0.30    & 1.4    & 0.4    & 0.77   & 0.12   & 1.27   & 0.3 \\
  4   & 0.56   & 0.18    & 2.1    & 0.5    & 0.84   & 0.06   & 1.50   & 0.2 \\
  5   & 0.77   & 0.51    & 1.65   & 0.27   & 1.22   & 0.32   & 1.54   & 0.2 \\
  6   & 1.1    & 0.07    & 2.7    & 0.4    & 1.48   & 0.12   & 1.85   & 0.16\\
\hline
\end{tabular}
\end{center}
\vspace{0.6cm}
\label{tab:barrel}
\end{table}

\begin{table}[htbp]
\caption{Fractions of dead and noisy channels in F-disks given in percent. }
\vspace{0.6cm}
\begin{center}
\begin{tabular}{|c|c|c|}
\hline
\bf {Disk}       &  \bf {Dead}          & \bf {Noisy} \\ \hline
    1      &         0.92    &      0.23\\
    2      &         1.0     &      0.21\\
    3      &         1.75    &      0.05\\
    4      &         1.0     &      0.32\\
    5      &         1.65    &      0.05\\
    6      &         1.02     &     0.25\\
    7      &         1.25      &    0.33\\
    8      &         1.4      &     0.13\\
    9      &         1.22      &    0.45\\
   10      &         1.32      &    0.15\\
   11      &         1.15      &    0.45\\
   12      &         1.25      &    0.35\\
\hline
\end{tabular}
\end{center}
\vspace{0.6cm}
\label{tab:disks}
\end{table}

It can be seen that the characteristics of the devices are very similar 
throughout the different barrels and disks, resulting in a uniform acceptance 
of hits produced by charged particles.  

\newpage
\pagebreak

\section{Assembly of the Central Detector}
The central detector assembly was built in two cylinders. Each cylinder 
contains four sub-assemblies, three barrel-disk modules, and one end-disk 
module.  This section begins with descriptions of the assembly of the 
barrels and the F-disks.  Following these is a description of the 
mating procedure used to assemble these elements into the two module 
types, barrel-disk and end-disk.  The final subsection covers the 
installation and alignment of these modules in the support cylinders.

\subsection{The Barrel Assembly}

The primary mounting of the ladders was to the cooled (active) 
bulkhead, with a secondary mounting point at a thin membrane (passive 
bulkhead). The cooled bulkhead is the primary mechanical object, supporting 
the F-disks mated to the barrel and providing the connections to the support 
cylinder. The ladders themselves form the mechanical connection between the 
two bulkheads so that the passive bulkhead serves as a precise spacer at the 
cantilevered ends of the ladders.  

The barrel detectors are used in the Level~2 track trigger~\cite{STT}, 
which utilizes mainly the $r-\phi$ information to look for tracks with 
large impact parameter. The polar angles of the tracks are not available 
to the trigger. Precise placement of the sensors is therefore essential 
to maintain the trigger hit resolution.  The constraints on assembly 
accuracy are discussed in the next section.

The beryllium plates in the ladders were used to locate the ladders on 
precisely machined beryllium bulkheads. The original intent was to have 
no adjustment in this mounting, but machining precision and ladder 
assembly precision were not sufficient, so that some alignment 
was done during ladder installation, as described below.

The survey data recorded as part of the quality assurance during assembly, 
described in detail below, provided a very good starting point for 
off-line alignment with tracks.

\subsubsection{Requirements on Assembly and Survey Accuracy}
\label{bas_s_1}

Deviations from the ladder nominal position can be characterized by
three shifts and three rotations. If the deviations are small, rotations
effectively commute so all six parameters were considered independently.

\begin{figure}[htbp]
\begin{center}
\epsfig{figure=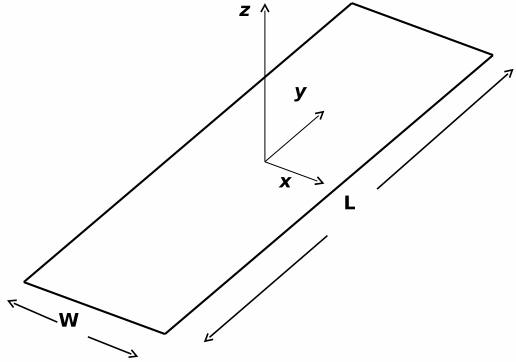,width=100mm,angle=0}
\caption{
\label{fig:bas_f_1} Ladder and its local coordinate system. 
Axial strips run along the ladder ($y$-axis).}
\end{center}
\end{figure}

Shifts $\Delta_x$ and $\Delta_y$, with the local coordinate system given 
in Fig.~\ref{fig:bas_f_1}, result in  mis-measurements that are uniform 
across the ladder and equal to the shifts. A shift $\Delta z$ and all 
rotations result in mis-measurements which are functions of the position 
within the ladder. 

\begin{figure}[htbp]
\begin{center}
\includegraphics[width=0.7\textwidth]{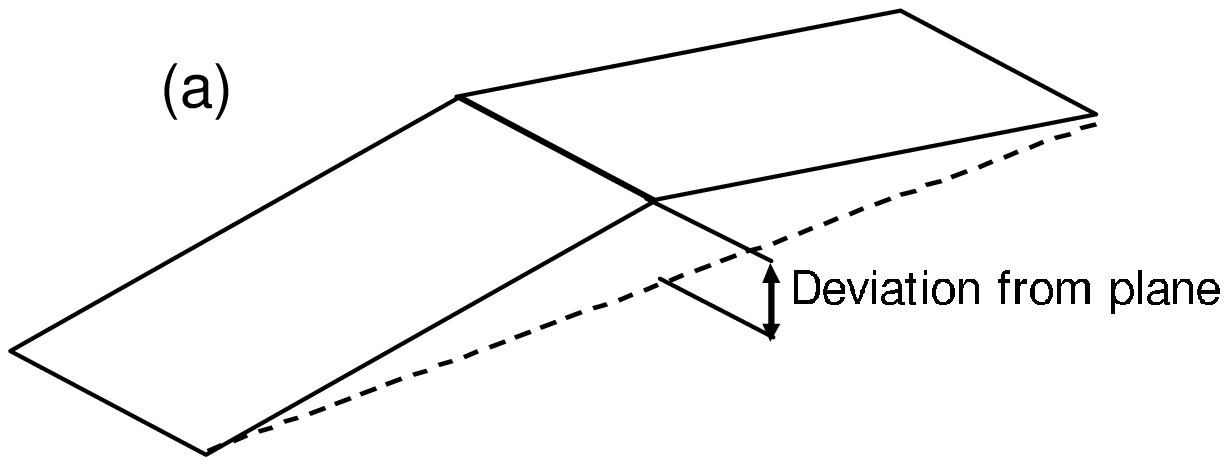}
\includegraphics[width=0.7\textwidth]{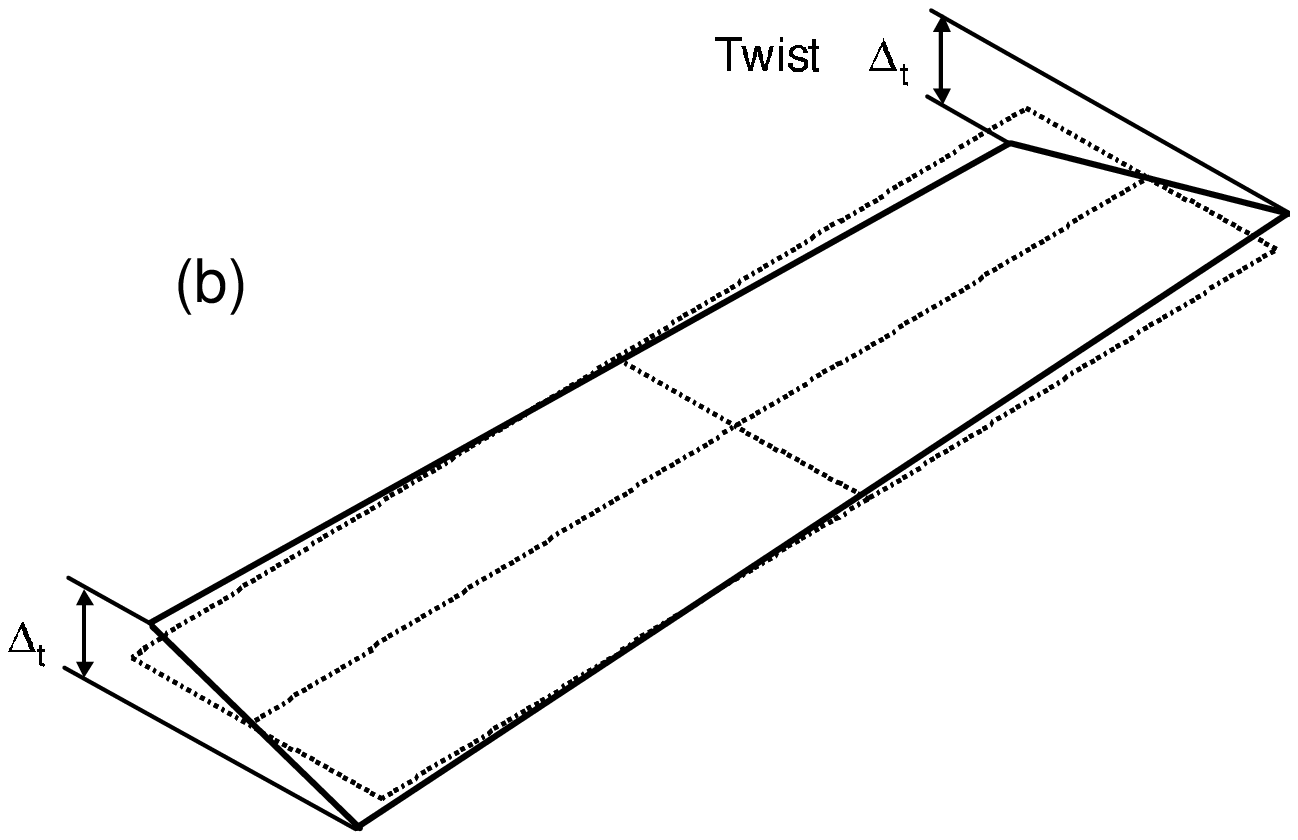}
\caption{Ladder deviations from plan: (a) ``Tent''-like deviations 
and (b) twist.}
\label{bas_2-3}
\end{center}
\end{figure}

The rotations in the $xy$ and $yz$ planes introduce mis-measurements 
which depend on $y$ and therefore can not be corrected at the trigger level.
These two rotations dictated the assembly precision required.  The dominant 
impact parameter uncertainty at the trigger level is the beam size (about 
30\,$\mu$m), and alignment errors should not compromise it. Detailed 
simulations were performed to see how alignment errors in individual layers 
propagate to the error on the impact parameter. The misalignment of the 
geometry was simulated one layer at a time, with the limit of the impact 
parameter error to be less than half the beam spot size. Each layer was 
allowed to contribute equally to the impact parameter error. This led to 
allowed limits for the misalignment of 10, 15, 20 and 15\,$\mu$m for 
layers 1 through 4. Since the simulation assumed 100\% hit efficiency, 
we conservatively took 10\,$\mu$m to be the limit on alignment errors 
in each layer which translates into 
$$ \Delta_{xy} = 25\,\mu \rm{m}$$
$$ \Delta_{yz} = 160\,\mu \rm{m}~ {\rm for ~layers~ 1~ and ~2}$$
$$ \Delta_{yz} = 320\,\mu \rm{m}~ {\rm for ~layers~ 3 ~and~ 4},$$
where $\Delta_{xy}/\Delta_{yz}$ is the rotation in the $xy/yz$ plane 
corresponding to the movement of the edge of the ladder. In the 
calculations done in order to reach these numbers, perfectly planar 
ladders were assumed. Two deviations from planarity were further considered:
``tent''-like shape [Fig.~\ref{bas_2-3}(a)] and a twist 
[Fig.~\ref{bas_2-3}(b)].  

The former contributes to the trigger resolution in the same way 
as $\Delta_{yz}$. The effect of twisting is more than a factor of three 
smaller than that from $\Delta_{xz}$. Offline, however, these shapes are 
considerably more difficult to incorporate into software corrections and 
{\it in situ} alignment algorithms using tracks.  This means that these 
effects may also impact not only the trigger resolution, but also the final 
offline tracking resolution.  Therefore these contributions were kept to 
a minimum during detector assembly.

\subsection{Measurements of Ladders}
\label{bas_s_3}
The main purpose of the ladder survey was to provide a link between the 
optical targets printed on the silicon sensors and the mechanical elements 
of the ladder structure. These elements could be surveyed with a touch probe 
after the ladders were installed in the barrels and the optical targets were 
no longer accessible.  An important byproduct of these measurements was 
quality control of ladder planarity.  These measurements were all done on 
an optical CMM with automated pattern recognition and auto-focus capabilities.
The machine accuracy was less than 2\,$\mu$m in the $xy$ plane and less than
5\,$\mu$m in $z$ when measuring distinctive features such as the metalization 
on the sensors.

\begin{figure}[htb]
\begin{center}
\epsfig{figure=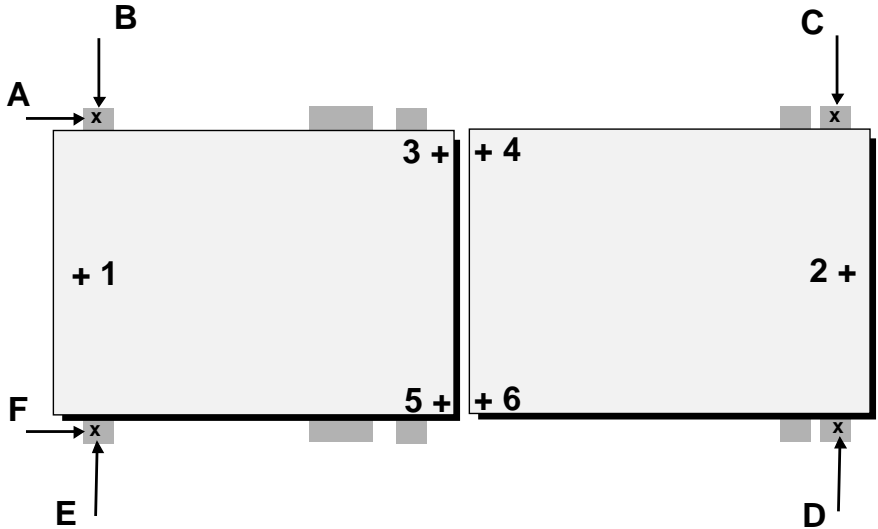,width=140mm}
\caption{
\label{bas_f_3} Ladder measurements. Lighter gray rectangles are silicon 
crystals, darker gray are beryllium support. Silicon fiducials are noted by
+ with a number; measurement on a beryllium surface by x; measurement of 
beryllium edges with
arrows and letters.}
\end{center}
\end{figure}

\begin{figure}[htb]
\begin{center}
\epsfig{figure=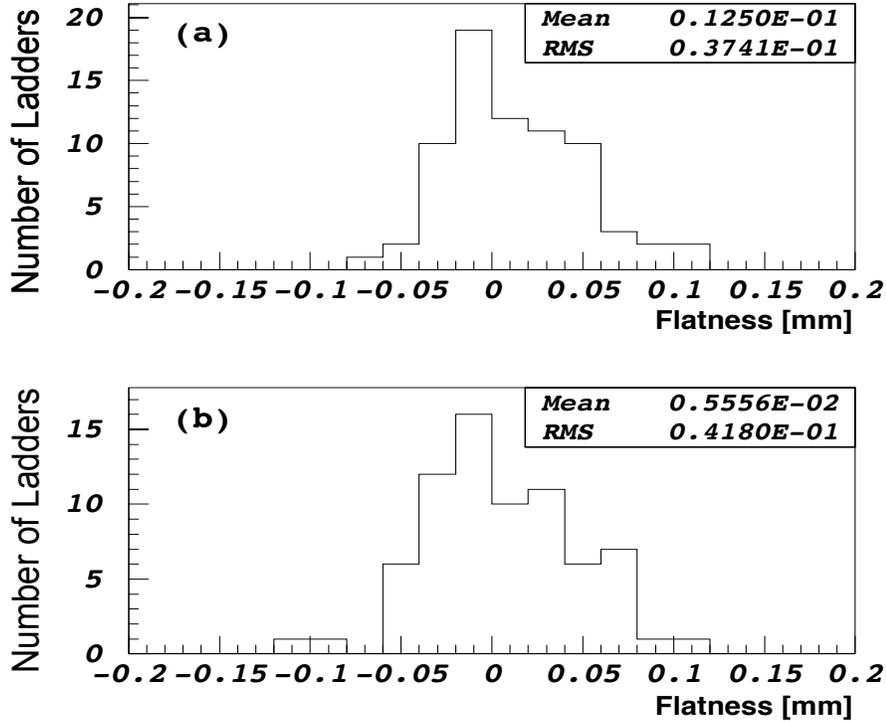,width=120mm,height=100mm}
\caption{
\label{bas_f_4} Ladder flatness for ladders in (a) layer 1 and (b) layer 2. }
\end{center}
\end{figure}

Figure~\ref{bas_f_3} shows the measurement points taken on 9-chip ladders. 
Measurements of 3- and 6-chip ladders were similar.First, six silicon fiducial 
markers, noted by + and a number in the figure, were measured. They were fit 
to a plane, which was defined to be the $xy$ plane of the measurement. The 
origin of the coordinate system coincides with fiducial 1 with the $y$ axis 
passing through fiducial 2. Ladder flatness was defined as the minimum 
separation between two planes parallel to the $xy$ plane which would contain 
all six measured points. It was defined to be positive if the ends of the 
ladder were below the $xy$ plane (as in Fig.~\ref{bas_2-3}a) and negative 
otherwise. Figure~\ref{bas_f_4} shows the flatness distribution for ladders 
installed in the inner two layers of the detector.

\begin{figure}[tbp]
\begin{center}
\epsfig{figure=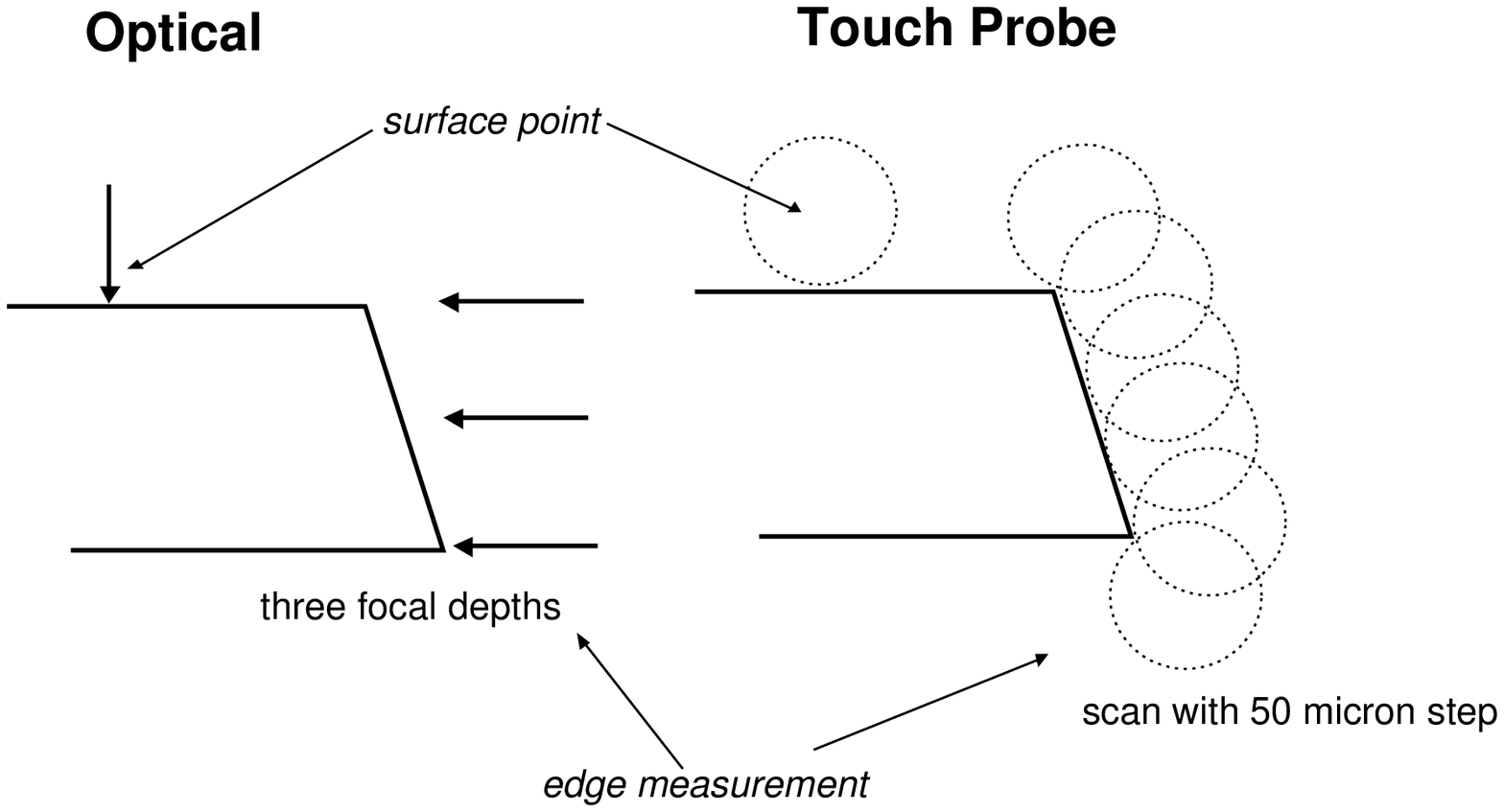,width=140mm}
\caption{
\label{bas_f_6} Schematics of optical and touch probe measurements.}
\end{center}
\end{figure}

After the coordinate system was established, the beryllium surfaces 
(noted by x) and edges (noted by arrows with letters) were measured 
(see Fig.~\ref{bas_f_3}). The edges of the beryllium supports were not 
perfectly machined and generally were not square (see Fig.~\ref{bas_f_6}). 
It was therefore necessary to make a scan across the edge. With the optical 
probe, three measurements were taken at the three different focus points 
indicated in Fig.~\ref{bas_f_6}. The average of the three was taken as 
the final measurement. For touch probe measurements, a scan with a 
50\,$\mu$m step was made, with a probe diameter of 1\,mm, and the largest 
measurement was used.

\begin{figure}[tbp]
\begin{center}
\epsfig{figure=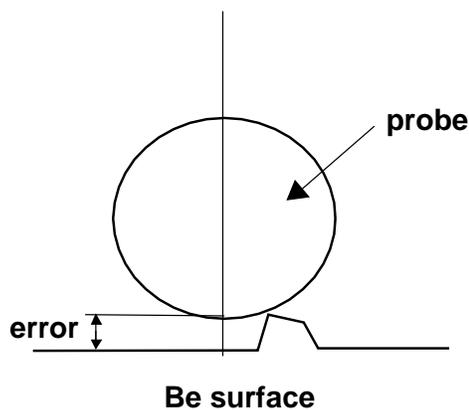,width=80mm}
\caption{
\label{bas_f_5} Effect of edge defects on the match between optical and 
touch probe measurements. The optical measurement may not be affected by the 
bump, while the touch probe was.}
\end{center}
\end{figure}

The edges of the beryllium can have microscopic bumps with a typical
size of 20--30\,$\mu$m. Figure~\ref{bas_f_5} schematically shows the effect
such bumps can have on the measurements. Defects like this were not frequent,
but as a precaution three scans were performed at each location, separated by
100\,$\mu$m. 

\begin{figure}[tbp]
\begin{center}
\epsfig{figure=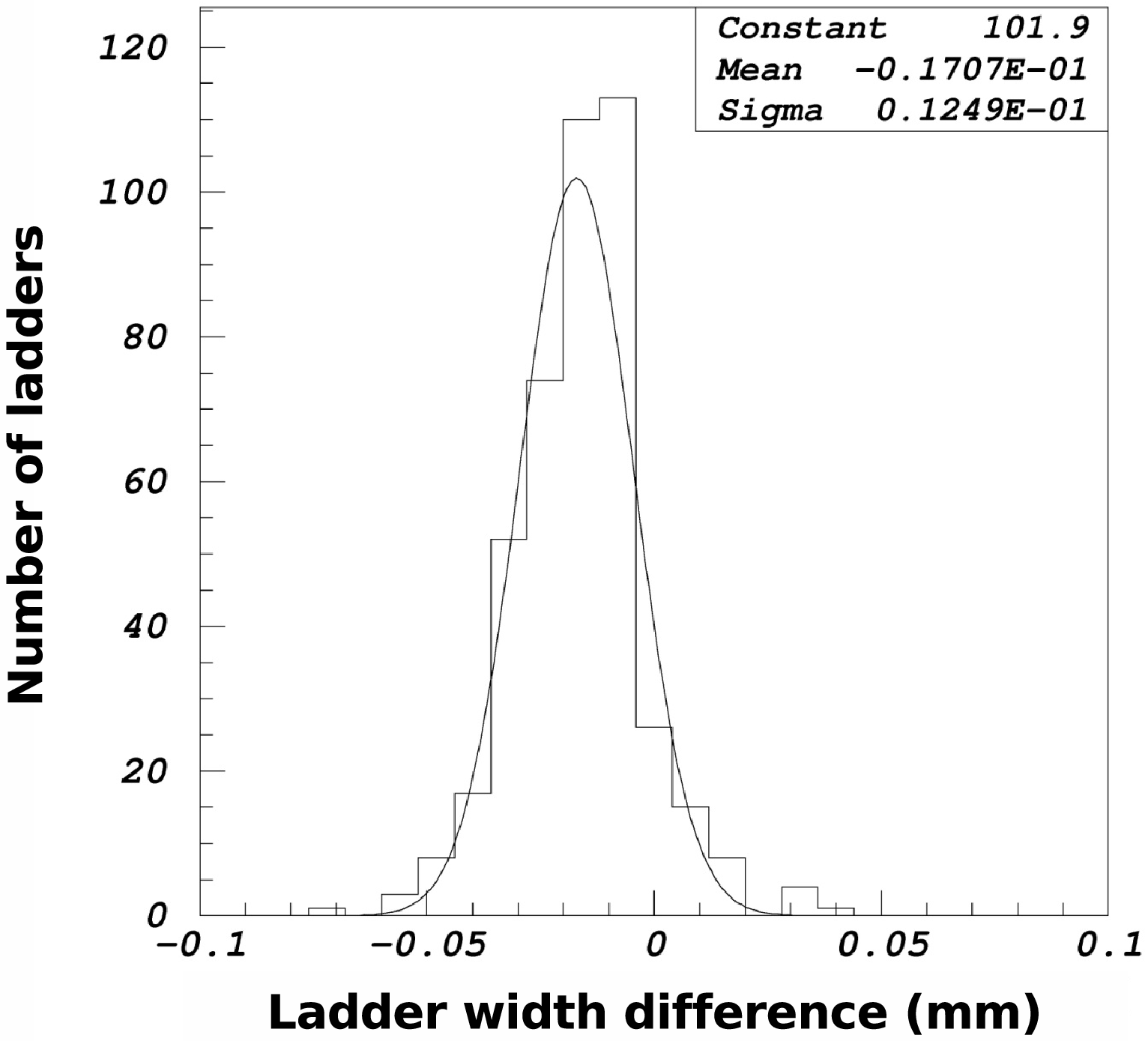,width=100mm}
\caption{
\label{bas_f_7} Difference in ladder width as determined by touch probe and 
optical methods. Negative values mean that touch probe consistently
gave smaller widths.}
\end{center}
\end{figure}

Since the systematic errors associated with the optical and touch probe 
methods were different, the comparison of these measurements provided a 
very powerful consistency check. Figure~\ref{bas_f_7} shows the difference 
between the ladder widths determined by the touch probe and optical methods. 
The absence of a high tail in the distribution confirms that edge 
imperfections are not frequent. The distribution is centered at $-17$\,$\mu$m, 
i.e. on average the touch probe gave smaller widths. This effect was studied 
in great detail and was attributed to the edge finding algorithm, as well as 
the bias introduced by choosing the average of the optical measurements at 
three focal depths versus the largest of the touch probe measurements taken 
through the thickness. However, this systematic bias does not contribute to 
the accuracy of the ladder position which was determined by pairs of 
measurements taken on opposite surfaces. Thus the ladder position accuracy 
was completely determined by the intrinsic accuracy of the CMM machine, 
about 6\,$\mu$m, which was sufficient for quality control. 

\subsection{Assembly and Survey Procedures}
\label{bas_s_4}

The assembly process started with aligning active and passive bulkheads. 
Both were mounted on a rotating barrel support fixture. The cooled bulkhead 
was fixed in place and served as the primary reference system for 
measurements.  The face of this bulkhead and precision holes at 3 
and 9 o'clock were used to establish the reference coordinate system.  
Precision spacers were used to locate the passive bulkhead parallel to 
the cooled bulkhead and to match machined alignment holes at 3 and 9 o'clock.
The passive bulkhead position was adjusted until it was within 10\,$\mu$m 
of nominal and clamped to the fixture.

Next, the ladders were installed using an adjustable mounting fixture.
The assembly sequence was determined by convenience of HDI pigtail routing. 
It was found that it was easiest to assemble the barrels from the inside 
out.  All barrels except the first one were assembled this way. The 
surface where the ladders mate to the cooled bulkheads had thermal 
grease applied to enhance the heat transfer through these joints, with the 
exception of the layer 4 ladders which were glued in place using silver 
epoxy.  These ladders then provided the rigid coupling between the cooled 
and passive bulkheads once the barrel was removed from the assembly fixture. 

After installation of each ladder, an electrical test was performed 
to ensure that it was not damaged. The ladder was powered, nominal bias 
voltage was applied, and pedestal values were recorded. Bias current and 
low voltage currents were checked. The most common damage was to wire bonds 
either from contact with HDI tails from other ladders or during ladder 
position adjustment.  In layer 4, the silver epoxy, which should hold the 
ladders in place, migrated in some cases and produced shorts along the 
exposed edges of the sensors. Damaged ladders were removed for repair and 
the assembly continued with other ladders.

After electrical testing, the ladder was surveyed with the touch probe. 
The touch probe measurements were then fit to the optical measurements 
taken previously. If the measured $\Delta_{xy}$ was more than 25\,$\mu$m, 
the ladder was adjusted and resurveyed. Since it was not possible to 
control the adjustment, sometimes several iterations had to be performed.
In rare cases, it was impossible to reach the desired $\Delta_{xy}$ due 
to ladder or bulkhead imperfections. 

\begin{figure}[tbp]
\begin{center}
\epsfig{figure=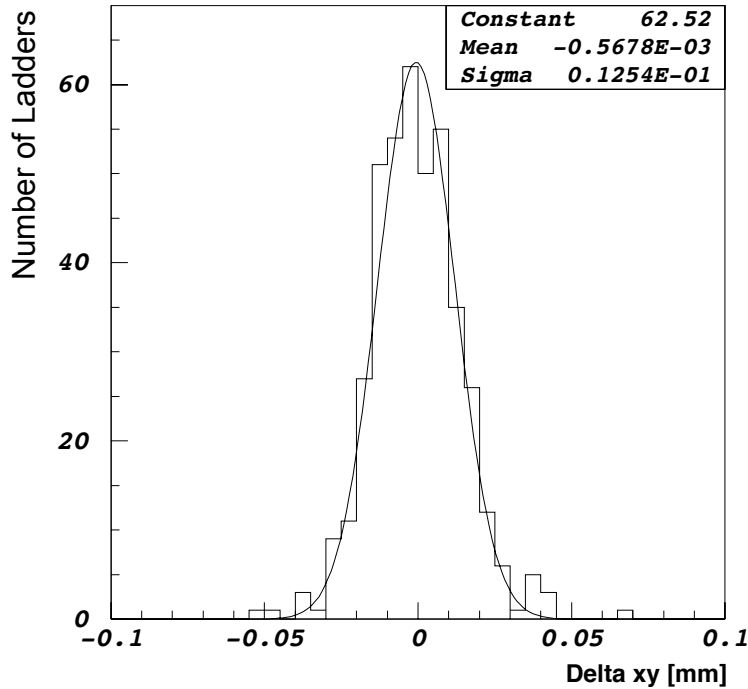,width=100mm}
\caption{
\label{bas_f_8} Barrel survey summary: $\Delta_{xy}$. The width of the 
distribution is well 
below the required 25\,$\mu$m. }
\end{center}
\end{figure}

\begin{figure}[tbp]
\begin{center}
\epsfig{figure=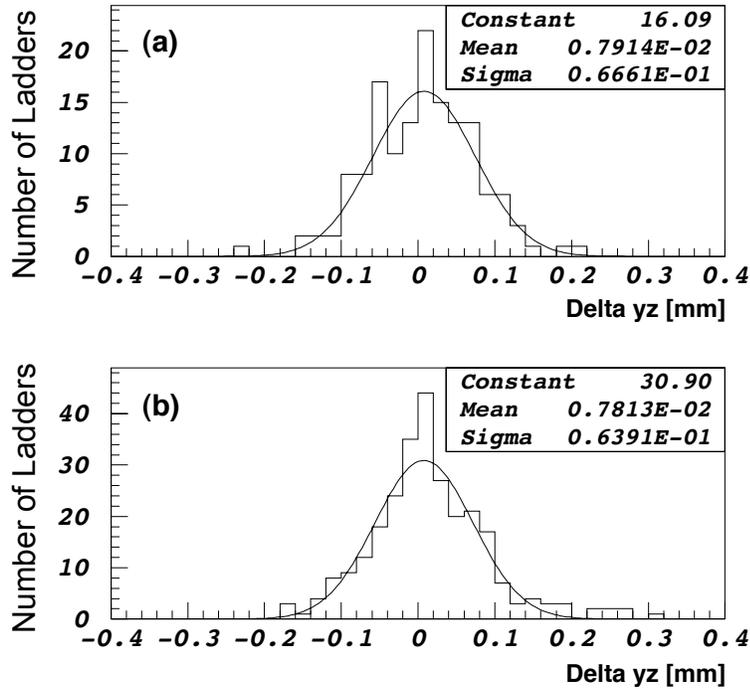,width=100mm}
\caption{
\label{bas_f_9} Barrel survey summary: $\Delta_{yz}$. The width of the 
distribution is well 
below the required 160\,$\mu$m for layers 1 and 2 (a) and 
320\,$\mu$m for layers 3 and 4 (b).}
\end{center}
\end{figure}

When the whole barrel was assembled, dabs of silver epoxy were added to 
the ladders in layers one, two, and three to ensure electrical contact between 
the beryllium bulkheads and the ground of the HDI electronics. This also 
improved the mechanical stability of the barrels.  All ladders were again 
checked electrically, and the final survey of the ladders was performed. 
Prior to the final survey, the HDI tails were permanently strain-relieved 
to a carbon fiber hoop mounted around the outer perimeter of the cooled 
beryllium bulkhead. The full barrel survey was a 2--3 day process for 
each of the six barrels. The distributions of  $\Delta_{xy}$ and  
$\Delta_{yz}$ for all ladders in all barrels are shown in Fig.~\ref{bas_f_8} 
and Fig.~\ref{bas_f_9}, respectively.

After all modules were surveyed, we performed the best fit to find a global 
coordinate system for the barrel, i.e. the system where the ladder rotations 
were minimal. This was done for all six barrels, and it was found that the 
best system was very close to the assembly system. After assembly, testing, 
and final survey of a barrel was complete, the barrel was removed from the 
assembly fixture and stored in a clean, dry air environment until it was 
mated with its F-disk.  

\subsection{The F-disk Assembly}
\begin{figure}[htbp]
\begin{center}
\psfig{file=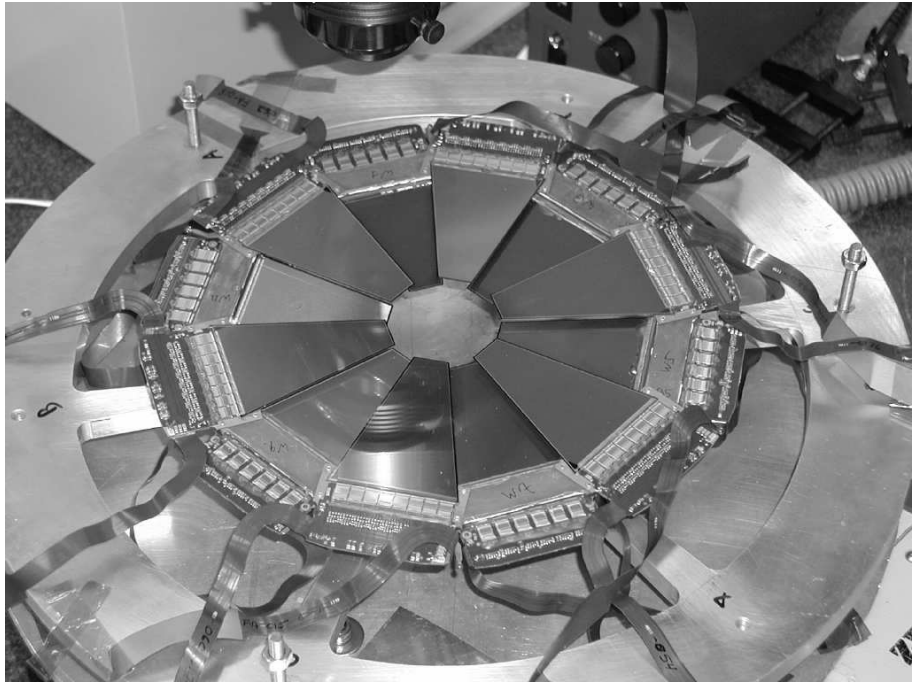,width=12cm}
\caption{First prototype F-disk assembly.}
\label{fig:f_disk}
\end{center}
\end{figure}

This subsection describes the assembly, survey, and quality control 
testing of the F-disks. Figure~\ref{fig:f_disk} shows a photograph of an 
F-disk assembly.  

The F-disks were mounted to the barrels via carbon fiber posts with 
precision machined aluminum inserts which engaged holes in the 
barrels and attached with aluminum nuts on titanium studs.  In the 
case of the end-disk modules, posts from two disks were mated with 
ruby balls and glued together.

\subsubsection{Beryllium Support Rings}
The beryllium support rings provide the infrastructure for the F-disks,
including the mechanical mounts as well as the cooling.  The rings were made 
from two pieces of beryllium laminated together.  The beryllium ring was 
assembled with mounting and aligning hardware as well as plumbing connections.
 
\begin{figure}[htbp]
 \begin{center}
 \psfig{file=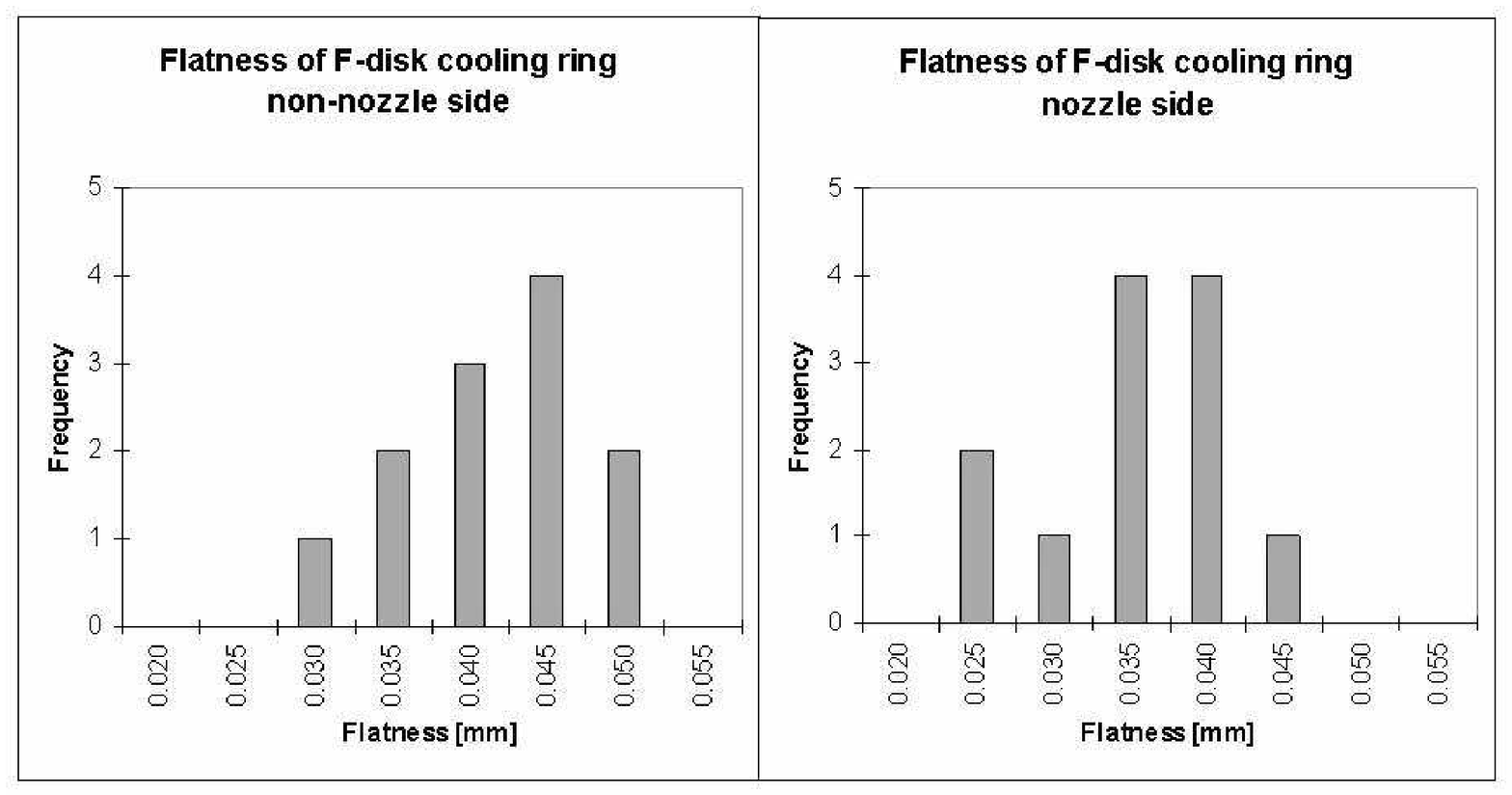,width=12cm}
 \caption{Flatness distributions of beryllium ring halves used in F-disks.}
 \label{f_brflat}
 \end{center}
 \end{figure}

The beryllium was machined by Phoenix Precision~\cite{phoenix}. The most 
important consideration for this process was the flatness of the finished 
piece.  Figure~\ref{f_brflat} shows the distribution of flatness for 
each of the two beryllium halves of the assembly.  They range in flatness from 
25--50\,$\mu$m.  The beryllium pieces contain alignment holes to  
determine the positions of the disks as well as the wedges.  After the 
two sides were laminated together, they were resurveyed for flatness and 
precision of the locating features on the two sides.  

The rest of the assembly was performed on a base plate with a center pin 
as a primary reference.  Threaded studs, on which the wedges would be 
mounted, were inserted and then surveyed with respect to the alignment holes 
on the ring.  The posts that position the disk on either the barrels or 
within the end-ring assembly were then mounted and inspected. Sapphire 
alignment balls, measuring 3\,mm in diameter, were mounted on arms made 
from scrap beryllium and attached to the cooling ring. The balls were 
located in the space between two adjacent wedges near the 2, 4, 8, 10, 
and 12 o'clock positions, 12.5\,mm radially outward from the cooling ring.
After complete assembly, the rings were leak checked and surveyed for 
mechanical precision.

\subsubsection{Installation of Wedges on Rings}
Before the wedges were mounted on rings, the bias bonds on the corners of 
the sensor were encapsulated for protection. This encapsulation was done 
by hand, using a syringe to apply the encapsulant. The encapsulant was 
heat cured for an hour.  

During the assembly of the first F-disk, the procedure was to also 
encapsulate all other bonds on the sensor. Each qualified wedge was 
retested immediately before placement on the ring. Leakage current tests 
indicated that there was some problem with the interaction of the 
encapsulant with the double-sided detector so this step was eliminated  
during assembly of the remaining disks.   

\begin{figure}[htbp]
 \begin{center}
 \psfig{file=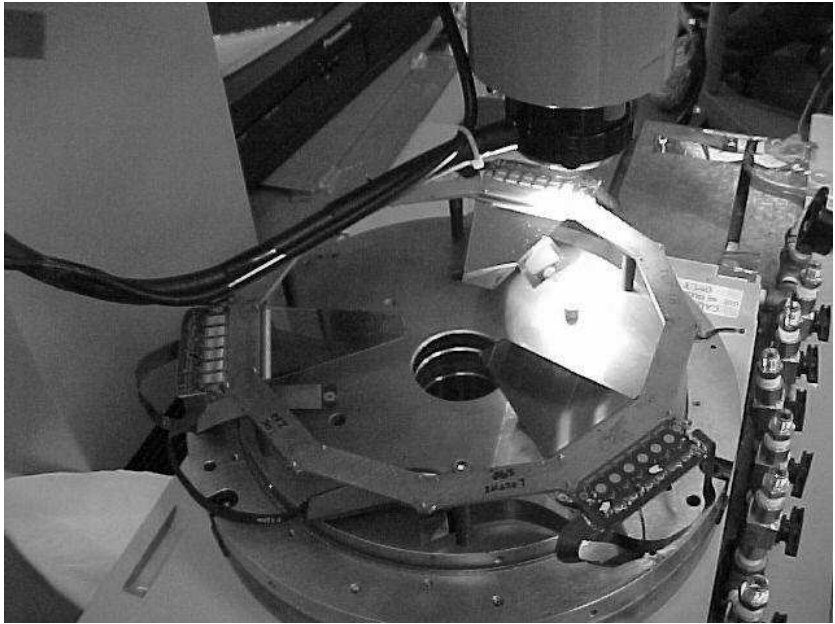,width=11cm}
 \caption{Photograph of the beryllium F-ring with 3 wedges mounted on it.}
 \label{f_ringdwg}
 \end{center}
\end{figure}

The wedges were mounted on the rings using the same fixtures as for the ring 
assembly.  This assembly was mounted on a CMM for precise alignment of the 
wedges.  Each wedge has four mounting holes in the beryllium substructure 
which supports the HDI.  These four holes engage studs mounted on the ring.  
The design clearance between the studs on the rings and the holes in the 
beryllium mounting plates of the wedges was 100\,$\mu$m, allowing for 
adjustment of the wedge positions. The wedges had thermal grease applied 
to the cover layer over the jumper on the HDI where the mechanical contact 
occurs with the ring and good thermal contact needs to be ensured.  
The grease improves the heat flow through this joint by a factor of two.  
All six odd numbered wedges were installed on a ring and then the six were 
aligned relative to the ring mounting holes and center pin of the fixture. 
Figure~\ref{f_ringdwg} shows a partially populated ring assembly on the CMM.

Each wedge was aligned using the same two sensor fiducial marks used for 
gluing the sensor to the HDI.  To move the wedge for alignment, two nuts 
were loosened, and the wedge was carefully pried to change position.  This 
process was repeated until the fiducial positions matched the predetermined 
positions in $x$ and $y$ to within 10\,$\mu$m.  Once a wedge was positioned 
correctly, the nuts were tightened to hold it in place. This process was 
very difficult, and about 7\% of the wedges were damaged and had to be 
replaced. After alignment, a final survey was done to record the wedge 
locations and survey ball locations relative to the ring reference holes.

Once the first six wedges were aligned on the ring, it was tested electrically
to make sure no bonds had been damaged and that the wedges were still working 
correctly.  The disk was carefully flipped over to mount the even numbered 
wedges.  The same mounting and alignment procedure was followed for the even 
numbered wedges.  After all wedges were mounted and aligned on the ring, the 
disk was again electrically tested to ensure proper function.

Working disks were dismounted from the CMM disk assembly fixture and placed 
in individual boxes.  These boxes were made from static free Plexiglas and 
included grounding straps for extra static protection.  The disks were 
stored in a dry box until they were ready for mating with a barrel or with 
another end-disk.  For the disks mated with barrels, a dimpled 0.025\,mm 
polyimide foil was carefully mounted over the posts to try to minimize 
heat transfer from the disks to the barrel detectors mounted close by.  
The HDI tails were carefully dressed into a pinwheel design to hold them 
during mating.

\subsection{Mating of Barrels to Disks and End-disk Assemblies}
The final assembly step consisted of mating F-disks either to barrels 
or to one another to form the end-disk assemblies, followed by final 
dressing of cables and cooling tubes.  

The mating of barrels to disks utilized a fixture consisting of a 
rotating Plexiglas disk from which the F-disk was suspended and a 
lifting table on which the barrel was located.  The lifting table was 
initially used to transfer the F-disk from its storage box onto the 
hangers which suspended it from the Plexiglas disk. Additional locating 
stops limited the movement of the F-disk relative to the Plexiglas 
disk.  The Plexiglas disk was then rotated to achieve the appropriate 
orientation relative to the barrel,  and the barrel lifted into place 
under the F-disk until the posts extending from the F-disk ring engaged 
the mounting holes in the barrel bulkhead.  The relatively unconstrained 
state of the F-disk allowed it settle into place without stress and 
provided visual feedback when it was not seating correctly. The barrel 
was raised until the F-disk assembly was lifted off the hangers before nuts 
were installed to secure the disk to the barrel.  Finally, the F-disk HDI 
tails were dressed to the carbon fiber cable strain relief ring on the 
barrel and the assembly stored in a clean, dry environment until 
installation in the support cylinders.

The end-disk assemblies consist of sets of three F-disks mounted to one 
another.  Because the spacing of these disks is not tight, it was 
possible to put these assemblies together without special tooling.  
The central disks have mounting posts on both sides with 5\,mm ruby 
balls at the ends, while the inner and outer disks have posts only on 
the side facing the center disk with conical receptacles which accept 
the 5\,mm ruby balls.  The central disk was left mounted in its storage box 
while the inner disk was manually installed on it.  Structural epoxy 
was used on the ruby balls to permanently bond the rings together.  
Spring clips were used at each post-ball-post interface to ensure 
proper seating of the balls while the epoxy cured.  The assembly was 
then flipped over and the outer disk added in a similar fashion. 
Cables and cooling tubes were dressed and the assembly stored in 
a clean, dry environment until installation in the support cylinders.

\newpage
\pagebreak

\section{Installation of the Detector}

\begin{figure}[b]
\begin{center}
\includegraphics[width=100mm]{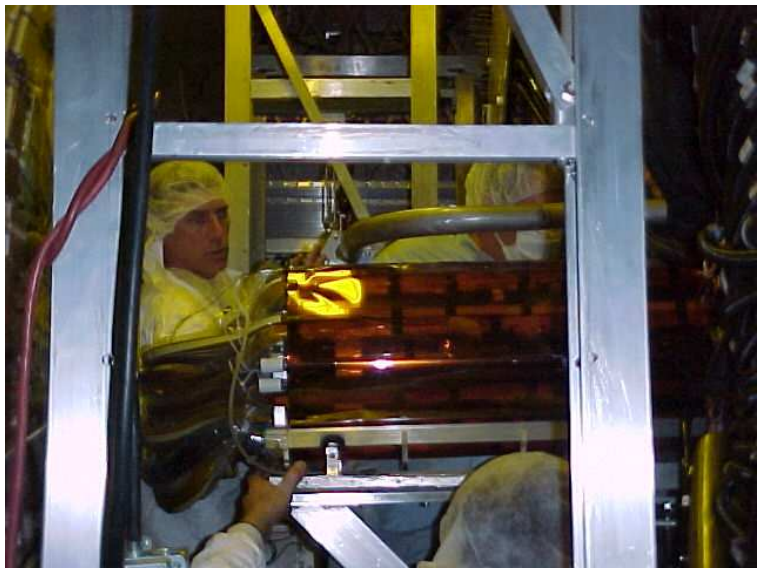} 
\caption{Installation of the silicon detector in the CFT.}
\label{installation}
\end{center}
\end{figure}

The SMT detector was built at the Fermilab Silicon Detector Facility
and then transported about 3\,km to the \D0 collision hall. The installation 
of the barrels and F-disks within the bore of the CFT, shown in 
Fig.~\ref{installation}, was completed in December 2000. The final
H-disk was installed in early February 2001, and about 20\% of the 
electronics were in place for the beginning of Run~II in March.

Detailed descriptions of the transportation and installation of the
barrel detector are given in the following section.

\subsection{Transportation and Installation of the Barrel Detector}   
The assembled detector halves were mounted inside transportation boxes 
for the drive to \D0, as well as the trip from the loading bay to the 
center of the \D0 detector.

The cushioned box incorporated several features for various 
portions of the transportation and installation journey.  The main 
components of the box were a base plate with a rail system mounted on it, 
a 4-sided enclosure (cover), a cable tray assembly mounted under the 
base-plate, and a set of rollers to allow the box to be slid along 
a trolley system in the collision hall.  In addition, a cart 
with air suspension was used during the trip from the assembly lab 
to the loading dock at the experimental hall to reduce accelerations 
to less than $5g$.  

\begin{figure}
\begin{center}
\includegraphics[width=100mm]{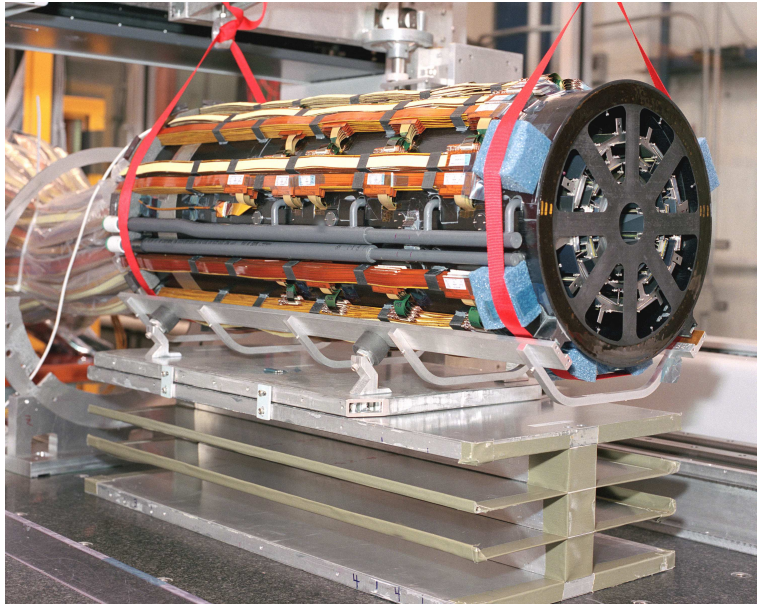} 
\caption{Installation of the silicon detector on the transportation rail.}
\label{rail}
\end{center}
\end{figure}

The first stage of the installation journey involved transferring the 
completed detector assemblies from the rotating fixtures used for 
barrel and disk installation and cabling onto the rail system mounted 
to the base plate of the transportation box.  This was done using the 
slide and ball screw mechanism which was used to install the barrel 
and disk assemblies into the support cylinder.  Straps were run around 
the cylinder and the cylinder was slowly lifted and released from the 
rotating support frame, which was then removed from under the cylinder.  
The transport box base, with cable tray, was then slid under the 
cylinder and the cylinder lowered onto the rails (Fig.~\ref{rail}).  
The cylinder has ``feet'' at the $z$=0 end at $\pm 45^\circ$ which are 
the final supports from the CFT inner barrel.  At the outer end, 
temporary feet were screwed to the outer membrane of the support cylinder.  
Once lowered onto the rails, small clamps were used locally at each foot 
to secure the cylinder to the rail system.  The rail system was coupled to the 
base plate of the box through rubber mounts to provide some isolation 
from impact loading during movement, as well as to allow for small alignment 
corrections when these rails were mated to the pre-existing rail system 
inside the fiber tracker.  

The 4-sided cover was installed next.  This provided mechanical protection 
for the assembly as well as a semi-sealed volume for purge gas.  The outer 
end of the box was left open where the cables exited.  The cables were then 
dressed into the cable tray assembly under the box, then the box assembly 
was sealed from the bottom of the cable tray assembly to the box cover to 
provide a sealed purge gas volume. A nitrogen purge was provided during the 
drive. This was important as the detector was moved during the winter and 
condensation onto the detector was possible after arrival at the collision 
hall. Finally, a set of brackets was installed which coupled to the base 
plate, ran outside the box cover, and terminated above the center of gravity 
of the assembly with rollers for the trolley system used in the collision hall.

The completed package was loaded onto an air-ride cart designed to limit 
impact loading to less than $5g$ during the truck ride to the collision hall.  
Several test runs were made using a dead weight on the cart to simulate the 
silicon load.  Accelerometers attached to fixed points on the cart routinely 
registered loads in excess of $5g$ and occasionally above $10g$ when the truck 
hit small bumps in the road.  However, the mock silicon assembly was insulated 
such that loads were always below the $5g$ threshold.  During the 
transportation of the two detector halves, accelerometers were placed on the 
silicon enclosure to monitor the device in all three directions; none of 
the $5g$ sensors tripped during transportation.  

Upon arrival at the collision hall, the detector halves were lifted by crane 
directly from the rear of the truck to the trolley system.  A length of 
pipe was slid through the trolley system rollers on the transportation box 
and stops were installed to prevent the box from moving along the pipe.  
The pipe was then picked up by the crane and the assembly lifted from the 
air ride cart and ``flown'' to the detector area.  The trolley pipe traveling 
with the silicon was joined to the fixed section running from the outside 
of the detector into the calorimeter gap.  The detector was then rolled along 
the trolley system into the area just outside the calorimeters.  At this 
point, the cables were unpacked and hung from additional carts on the trolley 
system.  The box cover was also removed at this point.  The silicon was  
moved along the remainder of the trolley path into the calorimeter gap where 
the assembly was lowered onto a table mounted onto the calorimeter.  

At this stage, the silicon was still sitting on the rails mounted to the 
transportation box base plate.  The base plate was slid by hand 
until the rails engaged the ends of the rail system inside the inner barrel 
of the fiber tracker.  The final step was to slide the detector along the 
rails into its final position, at which point the outer end of the detector 
was secured in place with a set of nuts engaged on titanium studs at the 
end of the fiber tracker.

After removal of the installation equipment, final connections were made to 
the cooling system and cables. After connection of each cable, a 
functionality test was performed to ensure a good connection.  

The cabling of the more than 15000 connections and the installation 
of all electronics were completely finished in May 2001.  Problems during 
the commissioning mainly concerned power supplies and the distribution of 
the operating and high voltages. 

\newpage
\pagebreak

\section{Performance and Lifetime}

After its installation 2001, the SMT has been successfully operating
for more than eight years. The detector is performing very well, providing 
good tracking and vertexing for the \D0 experiment. 

\begin{figure}[htb]
\vspace{0.5cm}
\centering
\includegraphics[width=5.0in]{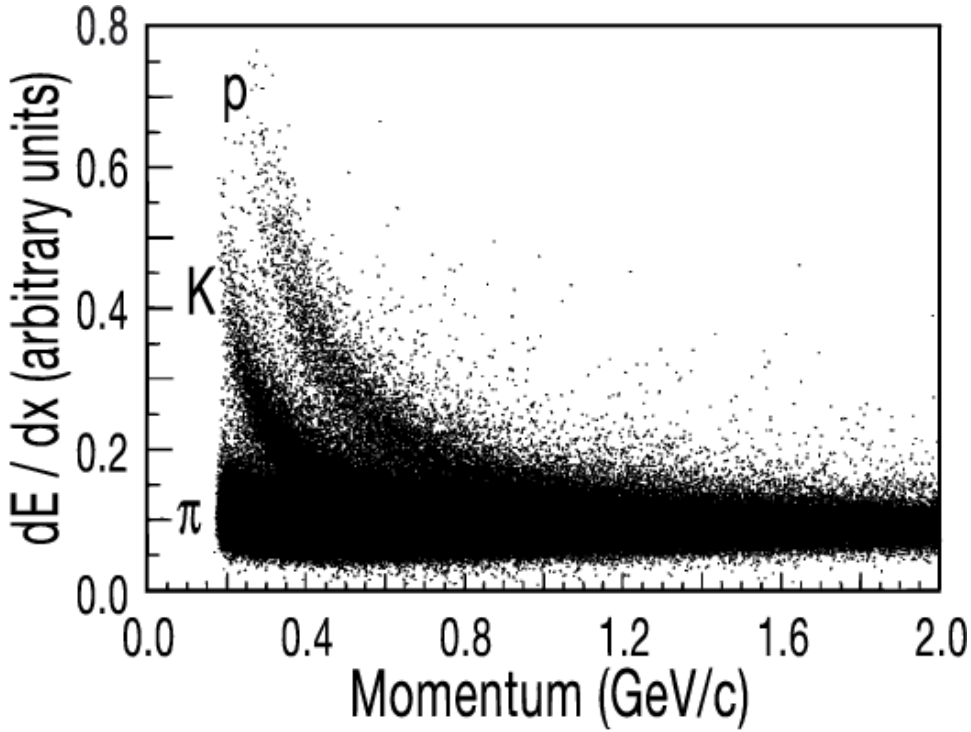}
\caption{Distribution of energy loss for a kaon-enriched 
sample of tracks showing $\pi , K$, and proton bands.}
\label{f:dedx}
\end{figure}

The signal-to-noise ratio was measured to be between 10 and 15 depending on
the module type. The signal is defined as the cluster charge given by a
minimum ionizing particle including a correction for the incident 
angle, and the noise as the rms of the pedestal distributions. The noise 
consists of intrinsic, random noise coming from the front-end of the 
SVXIIe chip, and environmental, coherent noise. The signal-to-noise ratio
calculated only from random noise is between 12 and 18. Gains vary among 
detector types with the n-sides 5--15\% lower than the p-sides due to the 
larger load capacitance. Pulse height information from the SVXIIe is
used to calculate cluster centroids and can also be used for {\it{dE/dx}} 
tagging of low momentum tracks. Figure~\ref{f:dedx} shows {\it{dE/dx}}
distributions after corrections for gain and incident angle are
made.

\begin{figure}[htb]
\centering
\includegraphics[width=5.5 in]{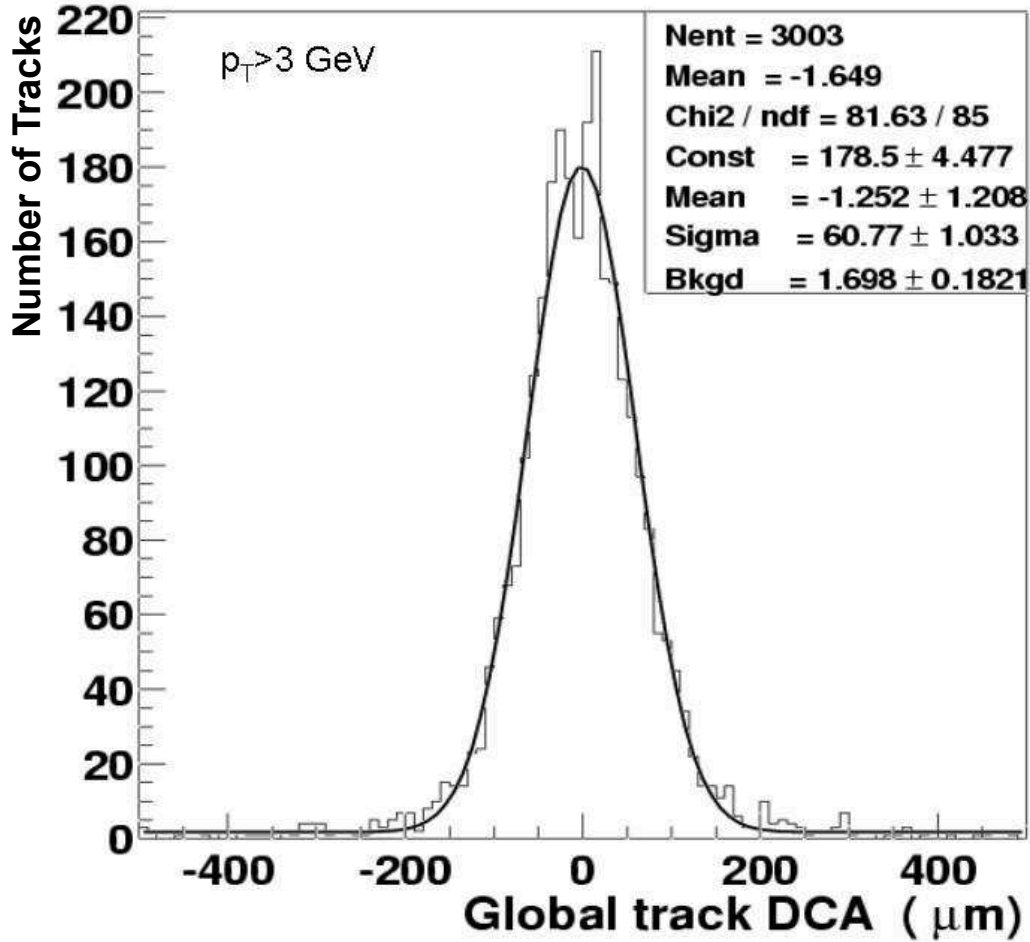}
\caption{The impact parameter of charged tracks 
with transverse momentum large than 3\,GeV/$c$ with respect 
the primary vertex.}
\label{fig:impact}
\end{figure}

\begin{figure}[htb]
\centering
\includegraphics[width=4.0 in]{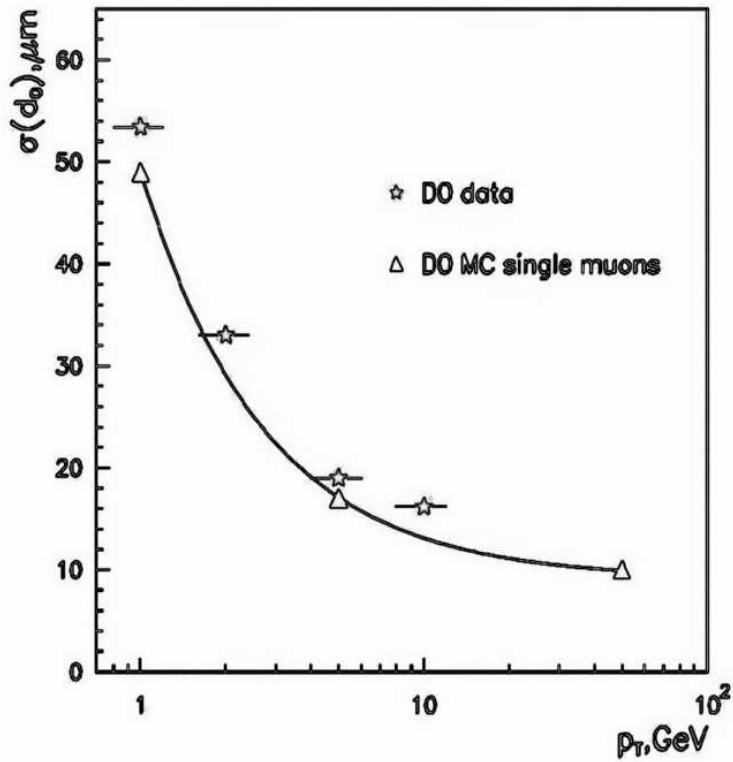}
\caption{The rms of the impact parameter distribution as  
a function of the transverse momentum of charged particles is
shown both for data and simulations.}
\label{fig:impact2}
\end{figure}

\begin{figure}[htb]
\centering
\includegraphics[width=4.0 in]{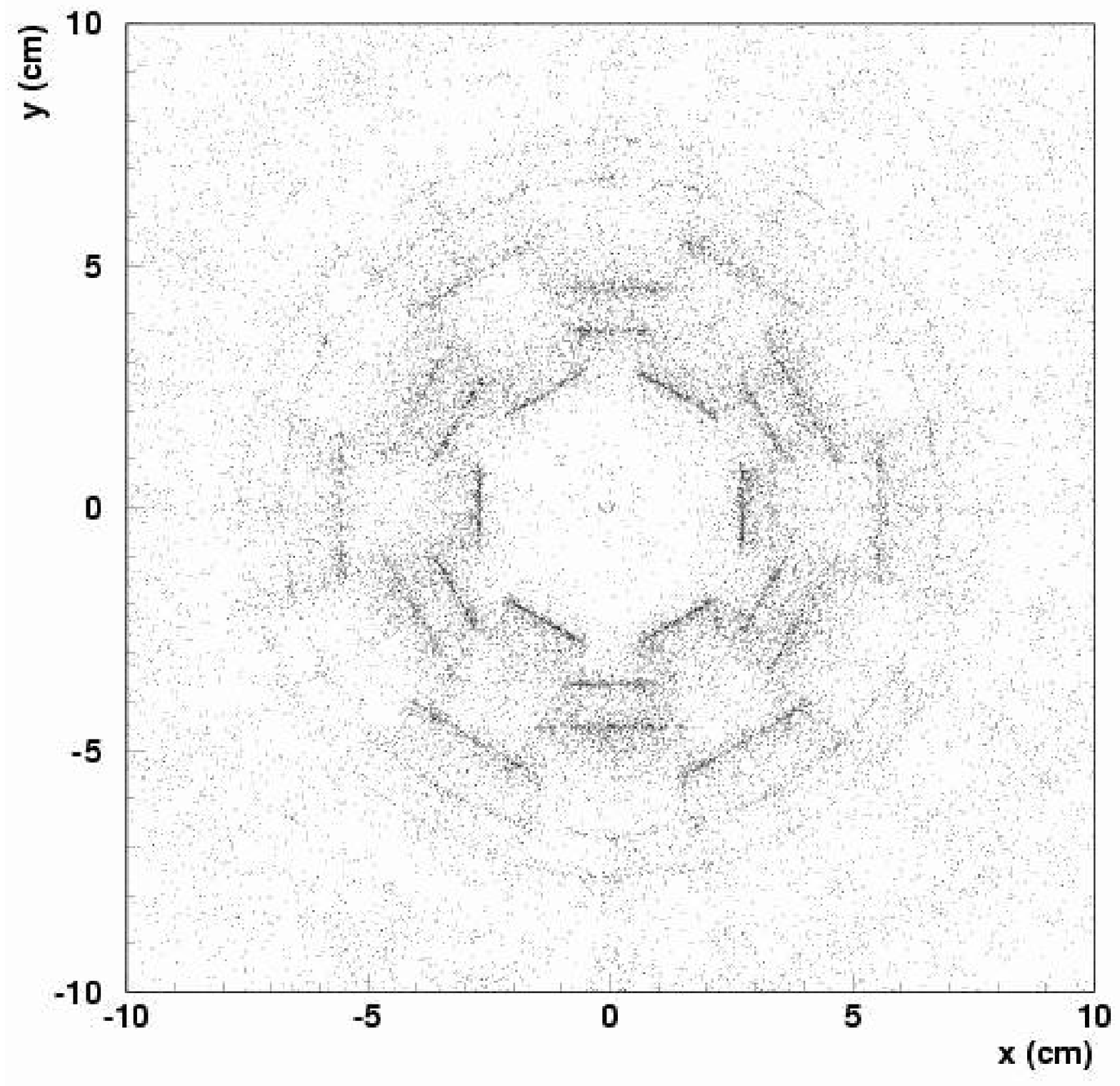}
\caption{A ``photometric'' picture of the SMT produced by plotting 
reconstructed coordinates of  $ \gamma $-conversions.}
\label{fig:gamma}
\end{figure}

Figure~\ref{fig:impact} shows the impact parameter distribution for tracks 
with respect to the primary vertex. Only charged particles with transverse 
momentum larger than 3\,GeV/$c$ are included in the plot. The only a
lignment in this case is the geometrical alignment described in Sec.~8. 
The Gaussian fit has a width of 60\,$\mu$m to which the beam size
contributes 30 to 40\,$\mu$m. After an alignment procedure where both the 
hits in the SMT and CFT and the primary vertex are used to constrain the 
fit, the impact parameter resolution is much improved and corresponds 
roughly to what is expected from simulation studies. The impact parameter as  
a function of transverse momentum of charged particles is shown both for 
data and simulation in Fig.~\ref{fig:impact2}. Figure~\ref{fig:gamma} shows 
a ``photometric'' picture in $x/y$ view of the SMT. The precise tracking 
allows reconstruction of the $xy$ coordinates of photon conversions in the 
detector material. Different layers of the SMT are clearly seen in this 
picture.

Operational difficulties have in many cases been peripheral to the 
silicon detector itself. These include latchup of operational 
amplifiers on the interface boards, low-voltage power supply failures, and 
high leakage currents in the high-voltage distribution boxes. However, the 
detectors themselves have suffered from some important problems and failures 
which are discussed in the following sections. In the next section, some 
very noisy modules are described; in Sec.10.2, problems with the readout 
system are discussed, and finally the aging problem due to radiation damages 
and a life time estimate are presented.
 
\subsection{Noise in the F-wedges}
\begin{figure}[htb]
\centering
\includegraphics[width=5.0in]{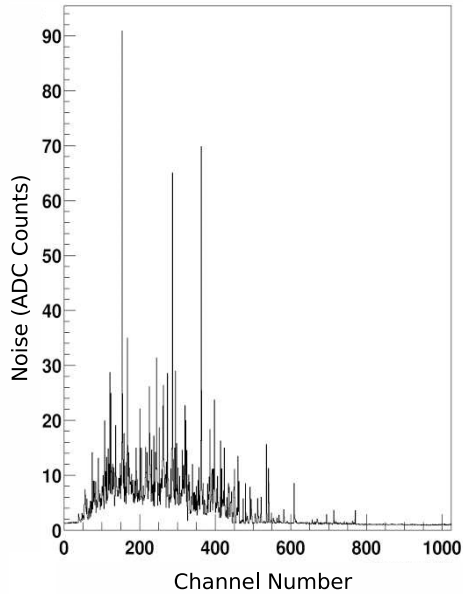}
\caption{Example of the grassy noise seen in the Micron-supplied 
F-disk detectors. Ideally, the entire plot would look like the 
region above channel 800.}
\label{f:noise}
\end{figure}

A serious detector feature is the so-called ``grassy noise'' shown in 
Fig.~\ref{f:noise}, which is confined to the Micron-supplied F-disk 
detectors which correspond to 75\% of the F-disk sensors. This noise is 
characterized by large charge spikes which cover 10--20 strips, and it 
occurs in about 20\% of the events for affected devices. Leakage currents 
typically rise to greater than 100\,mA within 1 h of turn-on at the 
beginning of a store. The only major change in the grassy noise was its onset
about two months after the start of the run. Some of the devices continue 
to slowly increase their current draw. The current increases about 100 mA 
after detectors are biased; these increases for a store which appear to be 
charge-up effects with time constants of tens of minutes.

\subsection{Performance of the Readout System}

\begin{figure}[htb]
\centering
\includegraphics[width=5.3 in]{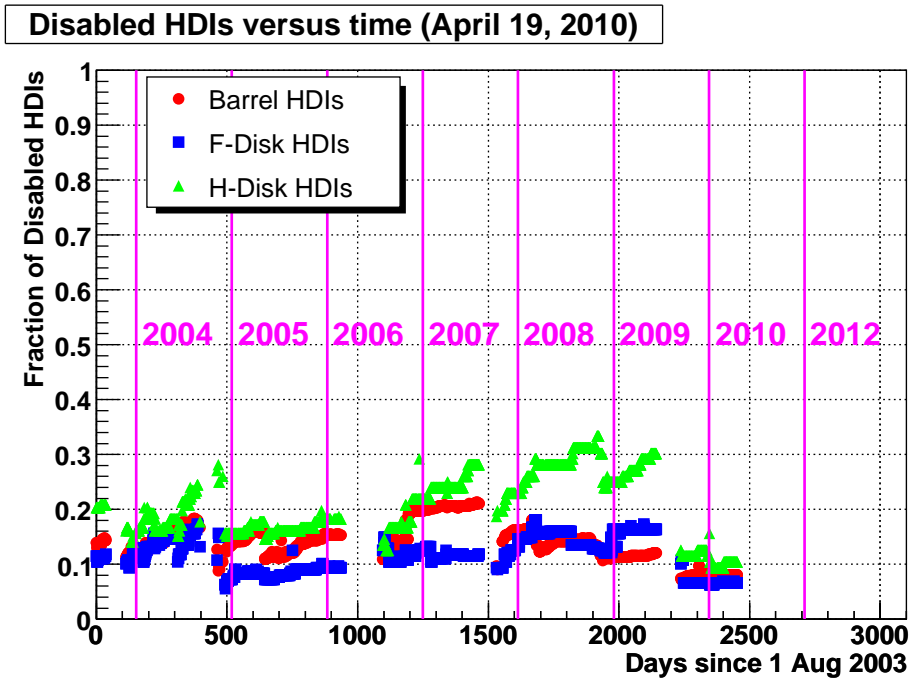}
\caption{The fraction of disabled HDIs as a function of time.}
\label{fig:disable}
\end{figure}

At the start-up in 2001, 95\% of the ladders, 96\% of the F-wedges and 
87\% of the H-wedges were working. Over time, these numbers have varied, 
and  Fig.~\ref{fig:disable} shows the number of disabled modules from 
2003 to 2010. The gaps in the figure correspond to machine shutdowns. 
During each shutdown period, several modules have been recovered by 
repair work to the external electronics and cabling. However, most of 
the disabled modules have problems in a inaccessible area, i.e., in 
the SVXIIe chips, in the HDIs, or in the low mass cables.  

Many of the problems with the chips arise because of an unwanted
increase in the supply currents. This could happen if the SVXIIe 
stands idle for time periods measured in hundreds of seconds; some of the 
supply currents increase by as much as a factor of 5. The cause of this has 
been traced to the use of dynamic CMOS memory in the chip.  
Figure~\ref{f:channel} shows part of the schematic of the SVXIIe. At the 
far right of the schematic, two capacitors are labeled Dout1 and Dout2.  
These capacitors are examples of the dynamic CMOS memory used in the SVXIIe.
In one half of a cycle, the capacitors are either set to a voltage or to 
zero, and they are read in the second half of the cycle.  However, if there is 
no clock signal, leakage currents slowly charge the capacitors. Since 
these devices drive CMOS gates, the  intermediate voltage will turn 
on both the n and p transistors, and the current will increase. This feature 
can lead to excessive chip currents and can evidently cause power wire-bond 
failure. A special signal is installed that cycles the SVXIIe every few 
seconds if there is no beam to avoid the increasing current. However, in 
one incident, the loss of the clock signal to the detector apparently caused 
the failure of several HDIs due to power wire-bond failures. After that, 
a special watch-dog timer is under development that will shut off the chip 
power if the special signal fails.

One chip not working on an HDI due to a power wire-bond failure means that 
the whole HDI is not functional and has to be disabled. However, it has been
possible to recover most of the data from these HDIs. Since it is impossible 
to get to the HDIs, the repair was done from outside the detector. The wire 
bond that failed was for the digital power.  The digital bus lines have 
protection diodes to the digital power bus inside the chip, and there is a 
small resistor in series with the diode to limit the current. It turned out 
to be  possible to power the chip from the bus lines. However, the resistor 
prevents enough power to do an entire readout, but with the chip set to a 
very high threshold and in zero-suppressed mode, the entire string of chips 
except for the one that failed was recovered.

Since there is a diode drop in the protection diodes in the chip, the
bus voltage has to run near 5\,V.  This is higher than the output of
the standard bus drivers so new adapter cards that provided the higher 
voltages were produced.  These new adapter cards were installed where 
needed in the 2007 and 2008 shutdowns.

Another major failure appears to be from connector failures. One H-disk 
unit on each side was removed to make room for a new inner layer 
Ref.~\cite{layor0} in 2006.  The two H-disks had numerous failures before 
being removed, and were investigated after the removal. It turned out that 
14 out of 24 bad HDIs had no failures during extensive bench testing.  
The flexible tail is designed to take care of slight cable mis-alignments.  
However, in the case of the H-disk, the tail was over-constrained by a 
carbon fiber structure. If the low mass cable was installed at a 
slight angle in the plane of the cable, a small torque was placed on 
the cable, which was evidently enough to cause the connectors to open. 

Eight of the H-disk HDIs that had actually failed were further inspected.  
Six had a broken wire bond in the priority passing line.  The other two 
had unknown failures.  The cause of the wire bond failures appears to be 
the  encapsulant. The encapsulant had re-liquefied and the re-liquefaction 
started at the surface of the HDI. The H-disks are mounted vertically so 
when the encapsulant released from the HDI, any stress would be split 
between the one bond wire going across the chips and the 19 wires going 
along the chips.  The evidence suggests that the stress on the one wire 
was too great and it failed. The cause of the re-liquefaction of the 
encapsulant is not known. 

In addition to the problems discussed above, the modules suffered from 
some recoverable errors. The most common are a failure in the 
ring counter that controls the pipeline and a failure to download properly.  
The ring counter error is a design oversight in the chip. Its symptom is 
that in digitize mode it connects a DC voltage to the comparator input rather 
than the output of a capacitor.  The only solution to this is to reset 
the pipeline counters.  This error is induced by electrical noise, and 
the noise from a digitization and readout frequently causes this error.  
To overcome this problem, the pipeline counters are reset after every 
readout.  This also resets the pipeline, so any triggers in the pipeline 
are lost.

\subsection{Radiation Damage}

The lifetime of the SMT is expected to be limited by noise due to 
micro-discharge breakdown in the inner layers, as discussed in Sec.~5. 
The micro-discharge effect depends on bias voltage and becomes unacceptable 
at approximately 150\,V.  The bias voltage needed to deplete the detector 
modules changes with the radiation dose they have received. To predict the 
lifetime of the SMT, it is necessary to monitor the depletion voltage and 
the radiation dose. The inner layer of the inner four barrels consists of 
the DSDM silicon modules which were measured to be the most sensitive to 
radiation. In addition, the inner layer receives the highest radiation dose. 
Therefore, this study is concentrated on the DSDM silicon modules in the 
inner layer.

\begin{figure}[htb]
\centering
\includegraphics[width=4.0 in]{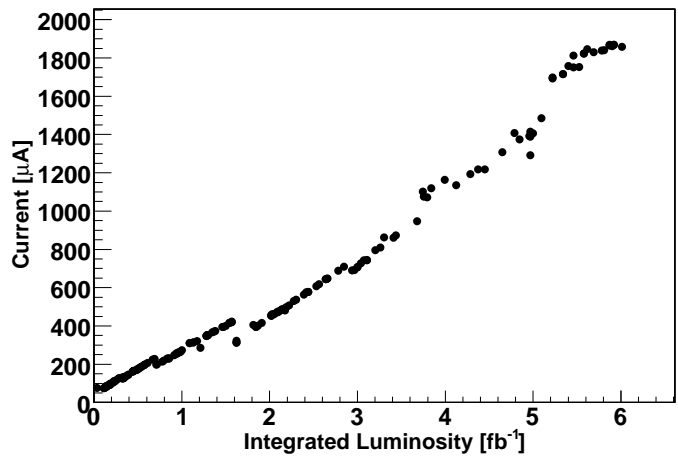}
\caption{Evolution of the leakage current normalized to $20^\circ$C with 
integrated luminosity.}
\label{fig:curr}
\end{figure}
  
The radiation dose received by a silicon module can be measured using the 
leakage current, which depends on flux according to the following formula:
\begin{equation}
I=I_{0} + \alpha \cdot \Phi \cdot V,
\end{equation}
where $I_{0}$ is the leakage current before irradiation, $\alpha$ is the 
radiation damage coefficient, $\Phi$ is the particle flux, and $V$ is the 
detector volume. The coefficient $\alpha$ depends on the particle type, 
temperature, and time~\cite{moll,Wunstorf}. The particle flux is proportional 
to the integrated luminosity delivered by the Tevatron. Therefore the 
dependence of the leakage current on the integrated luminosity is also 
expected to be linear. In Fig.~\ref{fig:curr} the current is plotted versus 
the integrated luminosity for four DSDM silicon modules. It can be seen in 
the figure that there is an almost linear dependence. The drop in the 
leakage current at for example the integrated luminosity $0.6$~fb$^{-1}$ 
occurred because the SMT was warmed up to $15^\circ$C for more than one 
month, and the annealing processes accelerated. The inner silicon layer has 
two sub-layers: the inner sub-layer at radius 27.15\,mm  and the outer 
sub-layer at radius 36.45\,mm. The radiation dose collected  by the outer 
sub-layer is about 1.6 times lower than the one at the inner sub-layer. 
Measurements of several silicon modules in the inner layer gave a 
normalized particle flux from
\mbox{$4.0\times 10^{12}$} to 
\mbox{$5.3\times 10^{12}$~particles$/$cm$^{2}\cdot$fb$^{-1}$}, taking 
annealing parameters from~\cite{Wunstorf} and an asymptotic radiation 
damage constant of \mbox{$3\times 10^{-17}$~A$\cdot $cm$^{-1}$}. 
The spread reflects the accuracy of the method as well as possible 
non-uniformity of the radiation dose.

The silicon detector bulk material at \D0 is slightly n-doped. Under 
radiation, however, donor states are removed and acceptor states created, 
eventually leading to a change from positive to negative space charge, 
i.e. the silicon bulk goes from being n-doped to becoming p-doped. 
This phenomenon is referred to as type inversion and has been confirmed 
by many experiments. It was predicted that the type inversion at the \D0 
silicon bulk material would happen when the integrated normalized particle 
flux reached approximately  $5\times 10^{12}$~particles$/$cm$^{2}$. 
At the point when the Tevatron had delivered about $1$~fb$^{-1}$, the inner 
layer was expected to have reached the type inversion point, where the 
depletion voltage is minimal. The following sections provide a description 
of depletion voltage measurements.  

\subsubsection{Noise Level on the n-side} 
While the silicon bulk material is n-doped, the noise of the n-side strips 
is large and constant at all voltages below the depletion voltage, and is
reduced significantly when depletion is reached. Thus studying the n-side 
noise as a function of bias voltage can be used to determine the
depletion voltage.

In a bias voltage scan, 11 runs are taken, with the high voltage settings 
varied from 0 to 100\% in steps of 10\%. The bias voltage scans were 
performed with and without beams in the Tevatron. The beam presence does 
not affect the measured depletion voltage values.

The DSDM and DS sensors show very different noise behavior as a function 
of bias voltage. For that reason, different procedures are used for DSDM 
and DS devices.

\begin{figure}[htb]
\centering
\includegraphics[width=5.0in]{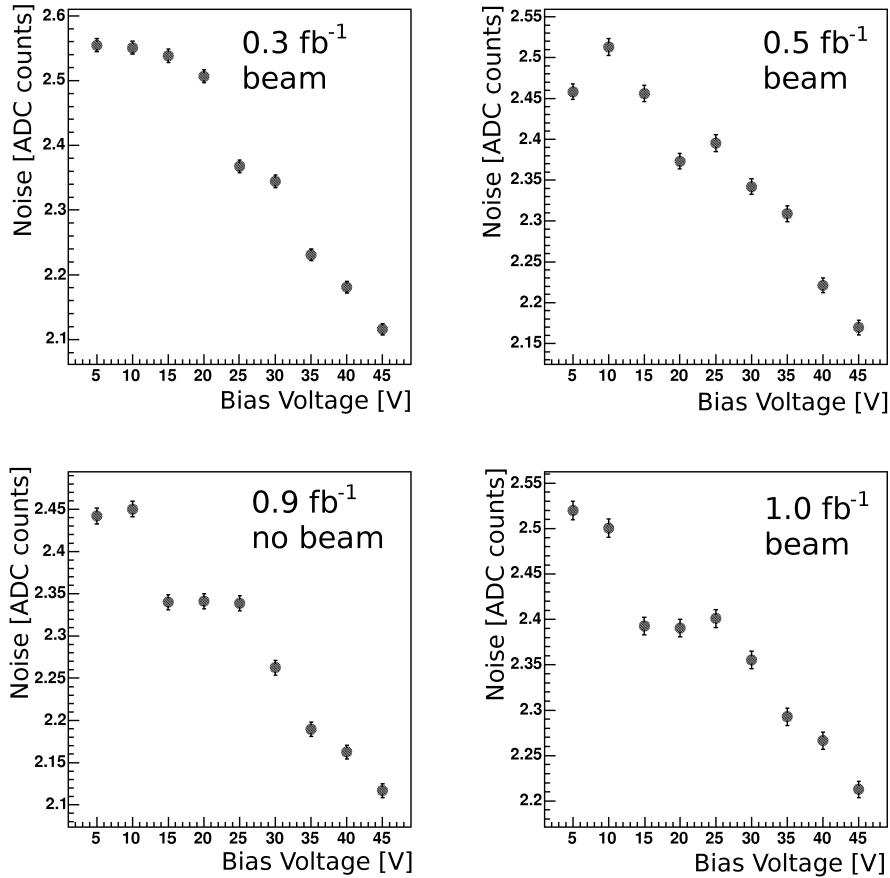}
\caption{Distribution of noise level on the n-side as a function of 
bias voltage for a DSDM silicon module installed 
in the \D0 detector.
The bias voltage scans at different integrated luminosities are shown.
The depletion voltage at 0.3\,fb$^{-1}$ is 25\,V, at 0.5\,fb$^{-1}$ 20\,V,
and at 0.9 and 1.0\,fb$^{-1}$ 15\,V. }
\label{fig:dsdm_irradiated}
\end{figure}

The DSDM devices show a rather unexpected noise behavior as a function of 
bias voltage (see Fig.~\ref{fig:dsdm_irradiated}). There is no abrupt 
decrease in the noise level on the n-side when the depletion voltage is 
reached. Instead, the noise is decreasing rather monotonically with increasing 
bias voltage. For some HDIs, a small kink in the noise can be seen at a 
certain bias voltage, and the position of this kink changes as a 
function of the radiation dose. This kink is interpreted as  an indication of 
the depletion voltage. 

The n-side noise was measured with a non-irradiated test DSDM module and 
abnormal behavior was not observed. This measurement indicates that the 
abnormal noise behavior is caused by the irradiation, and the comparison with 
the DS modules shows that the radiation changes some properties of the 
PECVD layer.

\begin{figure}[htb]
\centering
\includegraphics[width=5.0in]{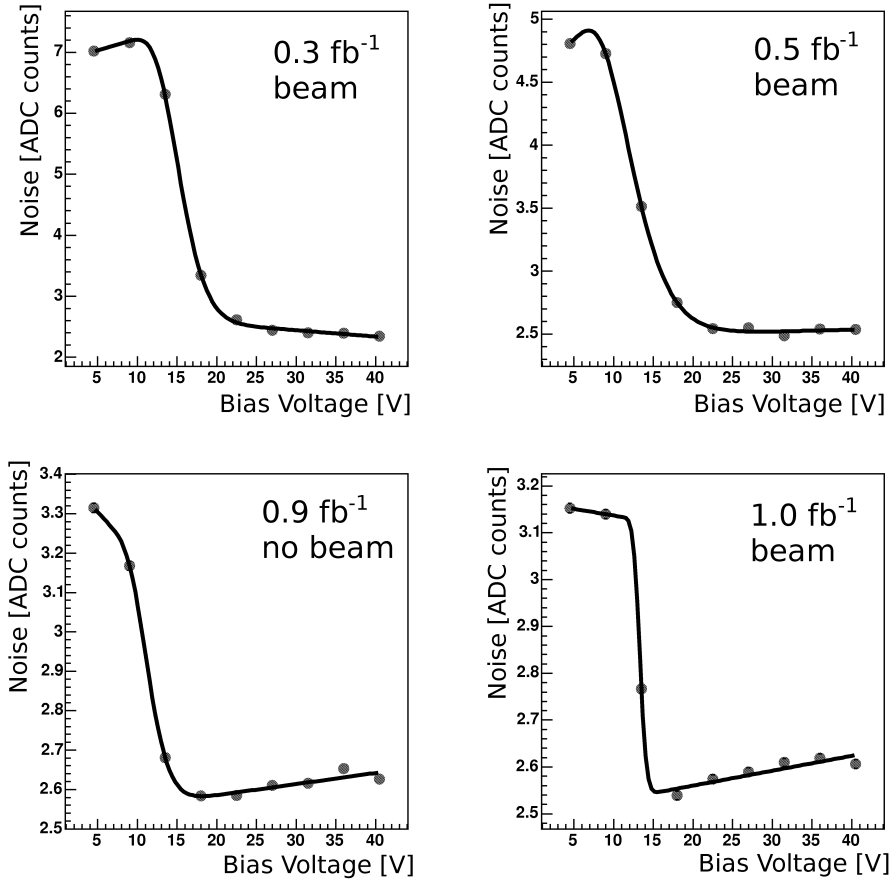}
\caption{Distribution of noise level on the n-side as a function of bias 
voltage for a DS silicon module installed in the \D0 detector. 
The bias voltage scans at different integrated luminosities 
are shown.}
\label{fig:ds_irradiated}
\end{figure}

Most of the DS devices show the expected noise behavior as a function of 
bias voltage (see Fig.~\ref{fig:ds_irradiated}).  At low voltages the noise 
is large, and rapidly decreases to a stable and lower level as soon as the 
bias voltage reaches the depletion voltage. There is some indication of 
noise increase at bias voltages higher than the depletion voltage after 
higher radiation doses.

\subsubsection{Charge Collection Efficiency} 

Another method for determining the depletion voltage uses the 
dependence of charge collection efficiency on bias voltage. For the 
silicon detectors with n-doped bulk material, the charge collection 
efficiency measured on the p-side increases with bias voltage and reaches 
its maximum at full depletion. It is possible that there is some additional 
small increase after that, for example, due to change of  the charge 
collection time.

This method requires bias voltage scans with tracks in the detector. 
A special algorithm has been developed for cluster reconstruction on the 
ladder under study in the vicinity of the expected track position. To 
determine the charge collection efficiency, the cluster charge is measured. 
The bias voltage where the charge collection efficiency reaches 95\% of 
its asymptotic value has been identified as the depletion voltage. 

The depletion voltage measured using the charge collection efficiency 
shows agreement with the noise measurements. 

\begin{figure*}[htb]
\centering
\includegraphics[width=5.9in]{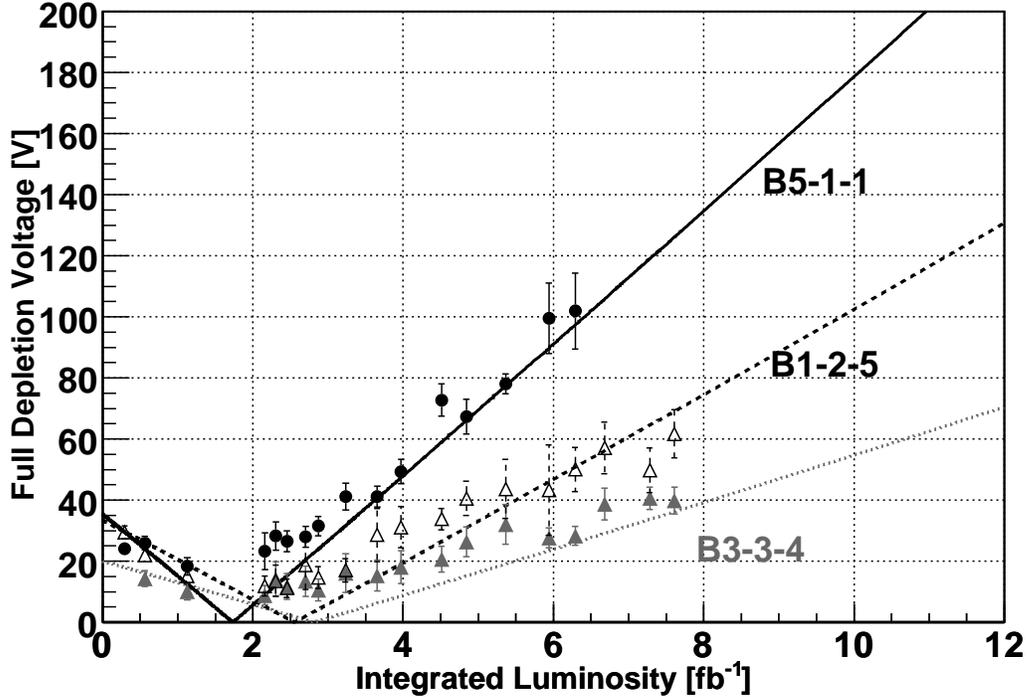}
\caption{Depletion voltage for silicon modules installed in the 
\D0 detector for different radiation doses. 
The DSDM module is located at a radial distance from the beam of 
2.7\,cm, the SS module at 3.6\,cm and the DS module at a 
radial distance of 4.4\,cm.
The lines represent theory predictions according to Ref.~\cite{moll}.
}
\label{fig:booster_comparison}
\end{figure*}

\subsubsection{Lifetime Estimate}

The measurements of the radiation dose and depletion voltage allow us 
to compare the behavior of the silicon sensors installed in SMT
with those irradiated at the Radiation Damage Facility (RDF) at Fermilab. 
A comparison shows that the DSDM silicon modules irradiated at the RDF and 
those installed in the SMT have a different dependence of the depletion 
voltage on the radiation dose. The depletion voltage for the DSDM irradiated 
sensors increases much faster than for the  DS silicon detectors tested at 
the RDF (see Fig~\ref{fig:sens4}). From our measurement see 
Fig.~\ref{fig:booster_comparison}, the behavior of depletion voltage for the 
DSDM silicon modules installed in the SMT is similar to the  behavior of 
the DS modules and is well described by standard 
parameterizations~\cite{moll}. Assuming validity of DSDM detectors tests at 
RDF the discrepancy is probably due to slow annealing of charge trapped in the 
insulator layer between metal layers which was not measured in the RDF 
studies. Assuming this ``normal'' behavior continues, the depletion voltage 
for the DSDM silicon sensors at the inner layer is predicted reach values 
around 150\,V  at a delivered luminosity of \mbox{$8$--$9$~fb$^{-1}$}. It is  
seen that the micro-discharge for these ladders increase at 110\,V with the 
luminosity of 6.5~fb$^{-1}$, which could limit maximum operating voltage for 
these ladders to less than 150 V.

\newpage
\pagebreak

\section{Conclusion}

The SMT contains a large variety of component types.  It holds six different 
types of sensors, one sensor type is double sided, double metal; two are 
double sided; and three are single sided. Nine different types of flex 
circuit carrying the readout chips and other passive electronic components 
were needed to accommodate the various sensors and readout geometries. This 
variety of components made the production of the detector challenging.
Despite many difficulties, the  SMT was successfully assembled and installed 
at the start of Tevatron Run~II. The detector has worked well since then. 
However, due to radiation damage, the detector is aging and in order to
compensate for this anticipated performance degradation and to improve the 
momentum resolution, a new inner barrel layer, called Layer 0 was 
installed in 2006. A description of layer 0 can be found in 
Ref.~\cite{layor0}.  

The SMT detector has been part of the D0 readout chain since the beginning 
of data taking in Run II, contributing to the almost two hundred interesting 
physics publications by the D0 collaboration.

{\bf{Acknowledgements}}

We thank the staff at Fermilab and in particular at SiDet, the D0 
mechanical and electrical support personnel and all our d0 collaborators 
for there support. Especially we would like thank George Ginther and 
Susan Blessing for their thorough reading of the draft and for giving 
countless useful comments and suggestions.  We also acknowledge 
support from the
DOE and NSF (USA);
CEA and CNRS/IN2P3 (France);
FASI, Rosatom and RFBR (Russia);
CNPq, FAPERJ, FAPESP and FUNDUNESP (Brazil);
DAE and DST (India);
Colciencias (Colombia);
CONACyT (Mexico);
KRF and KOSEF (Korea);
CONICET and UBACyT (Argentina);
FOM (The Netherlands);
STFC and the Royal Society (United Kingdom);
MSMT and GACR (Czech Republic);
CRC Program and NSERC (Canada);
BMBF and DFG (Germany);
SFI (Ireland);
The Swedish Research Council (Sweden);
and CAS and CNSF (China).

\newpage
\pagebreak


\begin{thebibliography}{Longauthor99}


\bibitem{d0nim-old}
S.~Abachi \etal, Nucl. Instr. and Meth. A 338 (1994) 185.

\bibitem{top}
S.~Abachi \etal, Phys. Rev. Lett. {74} 2632 (1995).

\bibitem{manyph}
See http;//www-d0.fnal.gov/d0$\_$publications/   
for a complete list of \D0 publications.

\bibitem{d0nim}
V.~M.~Abazov \etal, Nucl. Instr. and Meth. Phys. Res. A 565 (2006) 463.

\bibitem{muon}
V.~M.~Abazov \etal, Nucl. Instr. and Meth. Phys. Res. A 552 (2005) 372.

\bibitem{STT}
T. ~Adams \etal, {\it The D0 Run II impact parameter trigger}, 
arXiv:physics/0701195. 

\bibitem{layor0}
R.Angstadt et al,{\it The Layer 0 Inner Silicon Detector of the D0 
Experiment},
arXiv:0911.2522.

\bibitem{svx2}
The SVXII design Group,{\it A Beginners Guide to the SVXII},
FERMILAB-TM-1892.



\bibitem{1553}
The MIL-1553B standard.\\
http://www.interfacebus.com/Design$\_$Connector$\_$1553.html

\bibitem{vrb}
Fermilab Document ESE-SVX-950719, October 2001.

\bibitem{honey}
Honeywell, Kansas City Plant.\\ 
http://www.honeywell.com/sites/kcp/ 

\bibitem{finisar}
Finisar.\\ 
http://finisar.com// 




\bibitem{micron}
Micron Semiconductor Ltd., 1 Royal Buildings, Marlborough Road, Lancing,
Sussex, BN15 8UN, England.\\
URL http://www.micronsemiconductor.co.uk/.

\bibitem{elma}
ELMA:Research and Production Organization for Electronic Materials
(NPO Elma), 103460 Moscow, Russia.

\bibitem{eurisys}
Canberra Eurisys, 4 Avenue des Frenses, 78067 St. Quentin Yvelines, 
CEDEX, France. \\
URL http://www.eurisysmesures.com/.


\bibitem{md}
T. Ohsugi, al., 
{\em Micro-Discharges of AC-Coupled Silicon Strip Sensors}
(International Symposium on Development and Application of
Semiconductor Tracking Detectors, Hiroshima, Japan, 1993).

\bibitem{moll}
M.~Moll, PhD Thesis, {\it Radiation damage in silicon particle detectors: 
Microscopic defects and macroscopic properties}, DESY-THESIS-1999-040.



\bibitem{brush} 
Brush-Wellman, Cleveland, OH, U.S.A

\bibitem{ablefilm} 
Ablefilm 563 K Ablestik, USA



\bibitem{d0note_3841} The \D0 SMT Production Testing Group, 
{\it Electrical Production Testing of the \D0 Silicon Microstrip 
Tracker Detector Modules}, FERMILAB-TM-2348-E. 

\bibitem{lpc}
LPC, Litchfield, MN, USA.

\bibitem{dyconex}
Dyconex AG, Bassersdorf, Switzerland.\\ 
URL http://www.dyconex.com/.

\bibitem{speedy}
Speedy Circuits, Huntington Beach, CA, USA.\\ 
URL http://www.speedycircuits.com/.

\bibitem{compunetics}
Compunetics Inc., Monroeville, PA, USA.\\ 
URL http://www.compunetics.com/. 

\bibitem{vme4877ps}
User Manual, BiRa Systems, "Model VME 4877PS High Voltage
Power Supply System Manual 2nd Edition March 1998 \\
URL http://www.dsp-fpga.com/products/search/fm/id/?13899

\bibitem{marvin}
L. Bagby et. al, {\it SVX/silicon detector studies.} \\ 
FERMILAB-CONF-95-351


\bibitem{phoenix}
Phoenix Precision, Phoenix, AZ, USA.



\bibitem{Wunstorf} R. Wunstorf, {\it Radiation Hardness of Silicon Detectors: 
Current Status},
IEEE Transactions on Nuclear Science, Vol. 44, No. 3, June 1997.



\end{thebibliography}
\end{document}